\documentclass[longauth]{aa} 
\usepackage{graphicx}
\usepackage{txfonts}
\usepackage{threeparttable}
\usepackage{amsmath}	
\usepackage{amssymb}	
\usepackage{dirtytalk}
\usepackage{multirow}
\usepackage{gensymb}
\usepackage{color}
\usepackage[singlelinecheck=off]{subcaption}

\begin{document}

   \title{SN\,2018bsz: a Type I superluminous supernova with aspherical circumstellar material}
    \titlerunning{SN\,2018bsz: a SLSN-I with aspherical circumstellar material}

   \author{ M. Pursiainen\inst{1} \and
            G. Leloudas\inst{1} \and
            E. Paraskeva\inst{2,3,4} \and
            A. Cikota\inst{5} \and
            J.~P. Anderson\inst{5} \and
            C.~R. Angus\inst{6} \and
            S. Brennan\inst{7} \and
            M. Bulla\inst{8} \and
            E. Camacho-I\~niguez\inst{9,10}\and
            P. Charalampopoulos\inst{1}\and
            T.-W. Chen\inst{8}\and 
            M. Delgado Manche\~no\inst{11}\and       
            M. Fraser\inst{7}\and
            C. Frohmaier\inst{12}\and
            L. Galbany\inst{13,14}\and
            C.~P. Guti\'errez\inst{15,16}\and
            M. Gromadzki\inst{17}\and
            C. Inserra\inst{18}\and
            J. Maund\inst{19} \and
            T.~E. M\"uller-Bravo\inst{13} \and
            S. Mu\~noz Torres\inst{11} \and
            M. Nicholl\inst{20} \and
            F. Onori\inst{21} \and
            F. Patat\inst{22} \and
            P.~J. Pessi\inst{23} \and
            R. Roy\inst{24} \and
            J. Spyromilio\inst{22} \and
            P. Wiseman\inst{13} \and
            D. R. Young\inst{25}     
            }
   
   \institute{DTU Space, National Space Institute, Technical University of Denmark, Elektrovej 327, 2800 Kgs. Lyngby, Denmark\\
              \email{miipu@space.dtu.dk}
         \and
         IAASARS, Observatory of Athens, 15236, Penteli, Greece
         \and
         Department of Astrophysics, Astronomy \& Mechanics, Faculty of Physics, National and Kapodistrian University of Athens, 15784 Athens, Greece
         \and
         Nordic Optical Telescope, Apartado 474, 38700 Santa Cruz de La Palma, Santa Cruz de Tenerife, Spain
         \and
         European Southern Observatory, Alonso de C\'ordova 3107, Casilla 19, Santiago, Chile    
         \and
         DARK, Niels Bohr Institute, University of Copenhagen, Lyngbyvej 2, DK-2100 Copenhagen, Denmark
         \and
         School of Physics, University College Dublin, Belfield, Dublin 4, Ireland
         \and
         The Oskar Klein Centre, Department of Astronomy, Stockholm University, AlbaNova, SE-10691 Stockholm, Sweden
         \and
         Instituto de Astrofísica, Facultad de Física, Pontificia Universidad Católica de Chile, Casilla 306, Santiago 22, Chile
         \and
         Millennium Institute of Astrophysics (MAS), Nuncio Monseñor Sótero Sanz 100, Providencia, Santiago, Chile
         \and
         Departamento de F\'isica Te\'orica y del Cosmos, Universidad de Granada, E-18071 Granada, Spain
         \and
         School of Physics and Astronomy, University of Southampton, Southampton SO17 1BJ, UK
         \and
         Institute of Space Sciences (ICE, CSIC), Campus UAB, Carrer de Can Magrans, s/n, E-08193 Barcelona, Spain
         \and
         Institut d’Estudis Espacials de Catalunya (IEEC), E-08034 Barcelona, Spain
         \and
         Finnish Centre for Astronomy with ESO (FINCA), FI-20014 University of Turku, Finland
         \and
         Tuorla Observatory, Department of Physics and Astronomy, FI-20014 University of Turku, Finland
         \and
         Astronomical Observatory, University of Warsaw, Al. Ujazdowskie 4, 00-478 Warszawa, Poland
         \and
         Cardiff Hub for Astrophysics Research and Technology, School of Physics \& Astronomy, Cardiff University, Queens Buildings, The Parade, Cardiff, CF24 3AA, UK
         \and
         Department of Physics and Astronomy, University of Sheffield, Hicks Building, Hounsfield Road, Sheffield S3 7RH, UK
        \and
         Birmingham Institute for Gravitational Wave Astronomy and School of Physics and Astronomy, University of Birmingham, Birmingham B15 2TT, UK
         \and
         INAF-Osservatorio Astronomico d'Abruzzo, via M. Maggini snc, I-64100 Teramo, Italy
         \and
         European Southern Observatory, Karl-Schwarzschild-Str. 2, D-85748 Garching b. M\"unchen, Germany
         \and
         Facultad de Ciencias Astron\'omicas y Geof\'isicas, Universidad Nacional de La Plata, Paseo del Bosque S/N, B1900FWA, La Plata, Argentina
         \and
         Inter-University Centre for Astronomy and Astrophysics, Pune - 411007, India
         \and
         Astrophysics Research Centre, School of Mathematics and Physics, Queen's University Belfast, Belfast BT7 1NN, UK
             }

   \date{Received XXX; accepted YYY}

 
  \abstract{We present a spectroscopic analysis of the most nearby Type I superluminous supernova (SLSN-I), SN\,2018bsz. The photometric evolution of SN\,2018bsz has demonstrated several surprising features, including an unusual pre-peak plateau and evidence for rapid formation of dust $\gtrsim200$~d post-peak. We show here that the spectroscopic and polarimetric properties of SN\,2018bsz are also unique. While its spectroscopic evolution closely resembles SLSNe-I, with early \ion{O}{II} absorption and \ion{C}{II} P-Cygni profiles followed by Ca, Mg, Fe and other O features, a multi-component H$\alpha$ profile appearing at $\sim30$~d post-maximum is the most atypical. The H$\alpha$ is at first characterised by two emission components, one at $\sim+3000$~km/s and a second at $\sim-7500$~km/s, with a third, near-zero velocity component appearing after a delay. The blue and central components can be described by Gaussian profiles of intermediate width ($\mathrm{FWHM}\sim2000$ -- $6000$~km/s), but the red component is significantly broader ($\mathrm{FWHM}\gtrsim10000$~km/s) and Lorentzian. The blue H$\alpha$ component evolves towards lower velocity offset before abruptly fading at $\sim+100$~d post-maximum brightness, concurrently with a light curve break. Multi-component profiles are observed in other hydrogen lines including Pa$\beta$, and in lines of \ion{Ca}{II} and \ion{He}{I}. Spectropolarimetry obtained before (10.2~d) and after (38.4~d) the appearance of the H lines show a large shift on the Stokes $Q$~--~$U$ plane consistent with SN\,2018bsz undergoing radical changes in its projected geometry. Assuming the SN is almost unpolarised at 10.2~d, the continuum polarisation at 38.4~d reaches $P \sim1.8\%$ implying an aspherical configuration. We propose that the observed evolution of SN\,2018bsz can be explained by highly aspherical, possibly disk-like, CSM with several emitting regions. After the SN explosion, the CSM is quickly overtaken by the ejecta, but as the photosphere starts to recede, the different CSM regions re-emerge producing the peculiar line profiles. Based on the first appearance of H$\alpha$, we can constrain the distance of the CSM to be less than $\sim6.5\times10^{15}$~cm ($430$~AU), or even lower ($\lesssim87$~AU) if the pre-peak plateau is related to an eruption that created the CSM. The presence of CSM has been inferred previously for other SLSNe-I, both directly and indirectly. However, it is not clear whether the rare properties of SN\,2018bsz can be generalised for SLSNe-I, for example in the context of pulsational pair instability, or whether they are the result of an uncommon evolutionary path, possibly involving a binary companion.
   }
  


   \maketitle
%

\section{Introduction}

\label{sec:introduction}

Superluminous supernovae (SLSNe) are stellar explosions characterised by exceptionally bright, often long-lived light curves \citep[e.g.][]{Gal-Yam2009, Pastorello2010, Chomiuk2011, Quimby2011}. Initial classification scheme labelled all SNe brighter than a threshold of $M=-21$ in optical bands as superluminous \citep{Gal-Yam2012}. However, recent sample studies of SLSNe have shown that their populations might extend down to lower luminosities  \citep[e.g.][]{DeCia2018, Angus2019}, demonstrating that such a threshold is somewhat arbitrary. Therefore, SLSNe are currently classified based on morphological similarities to previously discovered SLSNe in addition to the observed brightnesses \citep[see e.g. ][for review]{Gal-Yam2019, Inserra2019}. While SLSNe are intrinsically rare \citep[e.g.][]{Quimby2013, McCrum2015, Prajs2017, Frohmaier2021}, it is possible to discover them at great distances due to their extreme luminosities \citep[e.g.][]{Moriya2019, Inserra2018b, Inserra2021}.

Spectroscopically SLSNe can be divided into two categories: those that do not exhibit hydrogen features (SLSN-I) and those that do (SLSN-II)  \citep[see e.g. ][for review]{Gal-Yam2017}. The more numerous SLSNe-I show spectral similarity to the hydrogen and helium poor Type Ic SNe after maximum brightness \citep[e.g.][]{Pastorello2010, Liu2017}, but they are very diverse based on both their spectroscopic and photometric properties \citep[e.g.][]{Nicholl2015a,Quimby2018, DeCia2018, Inserra2018a, Lunnan2018a, Angus2019}. On the other hand, the rarer H-rich SLSNe-II can be divided into the events similar to SN\,2006gy \citep{Smith2007, Ofek2007} characterised by narrow hydrogen emission lines (also known as SLSNe-IIn in analogue to Type IIn SNe) and to the few events similar to SN\,2008es \citep{Gezari2009, Miller2009, Inserra2018} that exhibit broad hydrogen features instead. 

Due to the long-lived, extremely bright light curves it is clear that SLSNe require a powerful energy source. While normal Type I SNe (both Ia and Ibc) are assumed to be powered by the decay of radioactive nickel, SLSNe would require several solar masses of \element[][56]{Ni} synthesised in the explosion. Only the Pair-Instability SN (PISN) explosions of extremely massive stars ($M\gtrsim140M_\odot$) are thought to be capable of producing sufficient $^{56}$Ni \citep[see e.g.][]{Heger2002, Gal-Yam2009}.  Other scenarios include a rapidly rotating, highly magnetised neutron star (a magnetar) formed in the core-collapse of the progenitor star \citep{Kasen2010, Woosley2010}. First suggested to explain the evolution of peculiar Type Ib SN\,2005bf \citep{Maeda2007}, the rotational-decay of the magnetar is utilised to provide a sufficient energy source. A similar central-engine scenario invoking fallback accretion on to a newborn black hole \citep{Dexter2013, Moriya2018a} has also been discussed in the context of SLSNe.

Lastly, interaction of the SN ejecta with surrounding circumstellar material (CSM) is an efficient mechanism to convert kinetic energy of the ejecta into radiation and is a proposed mechanism for powering SLSNe. Such CSM interaction is commonly observed in various kinds of SNe. Type IIn \citep{Schlegel1990}, Ibn \citep{Foley2007, Pastorello2007} and Icn SNe \citep{Fraser2021,Gal-Yam2022,Perley2022} are assumed to be completely enshrouded by CSM. Moreover, several H-poor SNe, such as Ia-CSM \citep[e.g.][]{Hamuy2003a} and Type Ic SNe \citep[e.g.][]{Chen2018, Kuncarayakti2018}, show strong signs of interaction with H-rich CSM. CSM interaction is already considered to be relevant for SLSNe-IIn due to their spectral similarity with the fainter Type IIn SNe. However, the discovery of a few SLSNe-I with late H emission \citep[see e.g.][]{Yan2015, Yan2017} and the presence of CSM shell around iPTF16eh \citep{Lunnan2018} suggests that CSM interaction can be relevant for this class of objects as well. 

Interaction of SN ejecta with an aspherical CSM has been used to explain peculiar observables of individual SNe. In particular a disk-like CSM has been attributed to be the cause of multi-component H$\alpha$ emission lines seen in the Type IIn SNe 1998S \citep{Gerardy2000, Leonard2000, Fassia2000, Pozzo2004} and PTF11iqb \citep{Smith2015} as well as the late H$\alpha$ emission seen in IIb SN~1993J \citep{Matheson2000b, Matheson2000}. Highly aspherical CSM has also been identified in several type II SNe. Most famously the near-by SN~1987A is surrounded by three CSM rings in an hourglass-shaped structure \citep[see e.g.][for review]{McCray2016}. Finally, the Homunculus Nebula surrounding $\eta$~Car (yet to explode), clearly demonstrate that real CSM surrounding massive stars is often aspherical. 

SN\,2018bsz was first discovered by the  All Sky Automated Survey for SuperNovae \citep[ASAS-SN;][]{Shappee2014} as ASASSN-18km on May 17th 2018 \citep{Stanek2018, Brimacombe2018}. A few days later on May 21st, the event was independently detected by the Asteroid Terrestrial-impact Last Alert System (ATLAS) survey \citep{Tonry2018} as ATLAS18pny. While the early reports classified the event as Type II SN due to its apparent H$\alpha$ P~Cygni profile \citep{Hiramatsu2018, Clark2018}, it was quickly reclassified as SLSN-I after the feature was re-interpreted as  \ion{C}{II} $\lambda6580$ \citep{Anderson2018a}. The host galaxy is  2MASX~J16093905-3203443 at $z=0.0267$ \citep{Jones2009}, making SN\,2018bsz the closest SLSN-I discovered to date.

\citet{Anderson2018} presented a study on the early spectral and photometric properties of SN\,2018bsz, noting some uncharacteristic behaviour even within  the diverse class of SLSNe-I. SN\,2018bsz exhibited a long $>26$~d slowly-rising \say{plateau} before a steeper rise to the maximum brightness. Similar long-lived, red pre-maximum evolution has been seen in SLSN-I DES15C3hav \citep{Angus2019}. The early spectra of SN\,2018bsz were characterised by strong \ion{C}{II} features along the typical \ion{O}{II} absorption features. While \ion{C}{II} features have been identified in the early spectra of several SLSNe-I such as PTF09cnd and PTF12dam \citep{Quimby2018}, the features seen in SN\,2018bsz are visually stronger \citep{Anderson2018}. Furthermore, \citet{Chen2021} analysed late optical and near infrared (NIR) and mid infrared (MIR) photometry of SN\,2018bsz and concluded that a significant amount of dust was formed around the SN $\gtrsim200$~d post-maximum \citep[see also][]{Sun2022}. Attributing the Balmer lines to CSM, \citet{Chen2021} concluded that the dust must have formed in a region of CSM interaction at a time the CSM had cooled below the dust sublimation temperature.

In this paper we focus on the spectroscopic evolution of SN\,2018bsz and present an in depth analysis of the spectra extending to $\sim120$~d post-maximum brightness, when the SN went behind the Sun. Our analysis also includes two epochs of spectropolarimetry. The extensive and high-quality dataset of 16 optical spectra allows us to identify when the hydrogen emission first appears and characterise its evolution in comparison to the rest of the spectral features. Thanks to our concurrent  spectropolarimetric observations we can also infer how the shape of the photosphere changes at the time of the appearance of the strong hydrogen emission. 

The paper is structured as follows: In Sect. \ref{sec:observations} we present our dataset.  In Sect. \ref{sec:spectroscopy} we focus on the analysis of the spectroscopic evolution, followed by comparison of SN\,2018bsz to SLSNe-I and Type Ic SNe, SLSNe-I with hydrogen emission as well as to selected Type IIn SNe in Sect. \ref{sec:SN_comparison}. In Sect. \ref{sec:spectropolarimetry} we analyse the two epochs of spectropolarimetry. In Sect. \ref{sec:discussion} we discuss the implications of the analysis and present a physical scenario to explain the peculiar observables of SN\,2018bsz. Finally, in Sect. \ref{sec:conclusions} we present our summary and conclusions.

\section{Observations}
\label{sec:observations}

\begin{table*}
    \def\arraystretch{1.1}%
    \setlength\tabcolsep{10pt}
    \centering
    \fontsize{9}{11}\selectfont
    \caption{The optical spectroscopic and spectropolarimetric observations of SN\,2018bsz analysed in this paper.}
    \begin{threeparttable}
    
    \begin{tabular}{c c c c c c c c}
    \hline
    \hline
	Date & MJD	& Phase (d)\tnote{a}	& Telescope	&	Instrument & Grism/Arm & $R$ ($\lambda/\Delta\lambda$) & Range     \\
    \hline
2018-05-21\tnote{b} &  58259.2 &     -8.1 & NTT   & EFOSC2       & Gr\#13           & 355             &     3685 --     9315 \\
2018-05-23\tnote{b} &  58261.3 &     -6.0 & NTT   & EFOSC2       & Gr\#11/Gr\#16    & 390/595         &     3380 --    10320 \\
2018-06-01\tnote{b} &  58270.3 &      2.8 & VLT   & X-Shooter    & UVB/VIS/NIR      & 5400/8900/5600  &     3000 --    24800 \\
2018-06-09\tnote{c} &  58278.0 &     10.2 & VLT   & FORS2        & GRIS\_300V       & 440             &     3300 --     9325 \\
2018-06-09          &  58279.0 &     11.2 & VLT   & X-Shooter    & UVB/VIS/NIR      & 5400/8900/5600  &     3000 --    24800 \\
2018-06-16          &  58285.2 &     17.2 & VLT   & X-Shooter    & UVB/VIS/NIR      & 5400/8900/5600  &     3000 --    24800 \\
2018-06-23          &  58292.1 &     23.9 & VLT   & X-Shooter    & UVB/VIS/NIR      & 5400/8900/5600  &     3000 --    24800 \\
2018-07-02          &  58301.1 &     32.7 & VLT   & X-Shooter    & UVB/VIS/NIR      & 5400/8900/5600  &     3000 --    24800 \\
2018-07-08\tnote{c} &  58307.0 &     38.4 & VLT   & FORS2        & GRIS\_300V       & 440             &     3300 --     9325 \\
2018-07-08          &  58307.1 &     38.5 & VLT   & X-Shooter    & UVB/VIS/NIR      & 5400/8900/5600  &     3000 --    24800 \\
2018-08-03          &  58333.6 &     64.4 & NTT   & EFOSC2       & Gr\#11/Gr\#16    & 390/595         &     3380 --    10320 \\
2018-08-13          &  58343.1 &     73.6 & NTT   & EFOSC2       & Gr\#11/Gr\#16    & 390/595         &     3380 --    10320 \\
2018-08-14          &  58344.0 &     74.5 & VLT   & X-Shooter    & UVB/VIS/NIR      & 5400/8900/5600  &     3000 --    24800 \\
2018-09-02          &  58363.1 &     93.1 & NTT   & EFOSC2       & Gr\#11           & 390             &     3380 --     7520 \\
2018-09-17          &  58378.0 &    107.6 & NTT   & EFOSC2       & Gr\#11/Gr\#16    & 390/595         &     3380 --    10320 \\
2018-10-01          &  58392.0 &    121.3 & VLT   & X-Shooter    & UVB/VIS/NIR      & 5400/8900/5600  &     3000 --    24800 \\
    \hline
    \hline
    \end{tabular}
	
	\begin{tablenotes}
        \item[a] With respect to maximum brightness \citep[MJD = 58267.5 in $r$;][]{Anderson2018}.
        \item[b] Previously published in \citet{Anderson2018}.
        \item[c] Spectropolarimetric observation.
    \end{tablenotes}
    
    \end{threeparttable}    
\label{tab:spec_log}
\end{table*}

SN\,2018bsz was intensively followed-up with European Southern Observatory (ESO) facilities. The Director's Discretionary Time (DDT) program (PI: G. Leloudas) was especially critical covering the seasonal gap of the extended Public ESO Spectroscopic Survey for Transient Objects \citep[ePESSTO;][]{Smartt2015}. The primary instruments used were the ESO Faint Object Spectrograph and Camera \citep[EFOSC2;][]{Buzzoni1984} on the New Technology Telescope (NTT) at the ESO La Silla observatory, Chile and X-Shooter \citep{Vernet2011} at the Very Large Telescope (VLT) Melipal unit (UT3) of the ESO Paranal observatory, Chile. The NTT spectra were reduced with the PESSTO pipeline \citep{Smartt2015} and the X-Shooter spectra as described in \citet{Selsing2019}.

In addition to the spectroscopic data, two epochs of spectropolarimetry were obtained with the FOcal Reducer/low dispersion Spectrograph \citep[FORS2;][]{Appenzeller1998} at the VLT Antu unit (UT1). The spectropolarimetry was reduced with a series of \texttt{IRAF} \citep{Tody1986} tasks. The frames were bias subtracted and cosmic rays were removed using \texttt{L.A.Cosmic} \citep{VanDOkkum2001}. Wavelength calibration was applied on the 2D frames with the aid of arc frames and the two beams (ordinary and extraordinary) were extracted in an identical manner with the task \texttt{apall}. Subsequently, we followed \citet{Patat2006} in order to obtain the normalised Stokes parameters ($Q$ and $U$) and their errors through the normalised flux differences. A small correction was applied to correct for the chromatic rotation of FORS2, following the values tabulated in the instrument web page\footnote{\url{www.eso.org/sci/facilities/paranal/instruments/fors}}. The polarisation degree $P$ and the polarisation angle $\theta$ were computed by $Q$ and $U$ and a polarisation bias correction was applied using the Heaviside function approach of \citet{Wang1997}. The flux spectra were derived by summing the ordinary and extraordinary beams and using an archival flux calibration from the observations of the spectroscopic standards with the polarisation units. 

The details of the spectroscopic and spectropolarimetric observations analysed in this paper are presented in Table \ref{tab:spec_log}.  The spectra presented in this paper have been scaled to match the $grizJHK$ photometry of the closest epoch presented in \citet{Chen2021}. The host galaxy of SN\,2018bsz exhibits strong emission lines \citep{Chen2021} as is typical for hosts of SLSNe \citep[e.g.][]{Leloudas2015}. In order to focus on the transient, the wavelength ranges of the known strong host lines have been cut out in the optical spectra presented in this paper. While the high-resolution X-Shooter spectra are not affected as the host lines are very narrow in comparison to the broad transient features, the host line clipping does affect the lower resolution spectra (EFOSC2 and FORS2). Due to the presence of the host galaxy emission lines, narrow emission lines as would be typically be observed in interacting SNe are difficult to investigate.

In addition to the optical ground-based dataset we also analyse two epochs of publicly available spectra taken by Hubble Space Telescope (HST). At $\mathrm{MJD}=58294.5$ ($+26.5$~d) SN\,2018bsz was observed with HST (Program 15488, PI: P. Blanchard) using the Cosmic Origins Spectrograph (COS) and the Space Telescope Imaging Spectrograph (STIS) and at $\mathrm{MJD}=58319.1$ ($+50.2$~d) with COS only (Program 15489, PI: R. Quimby). The reduced HST data were retrieved using the MAST archive\footnote{\url{https://archive.stsci.edu/}}.

\section{Spectroscopy}
\label{sec:spectroscopy}

\begin{figure*}
    \centering
    \includegraphics[width=1.00\textwidth]{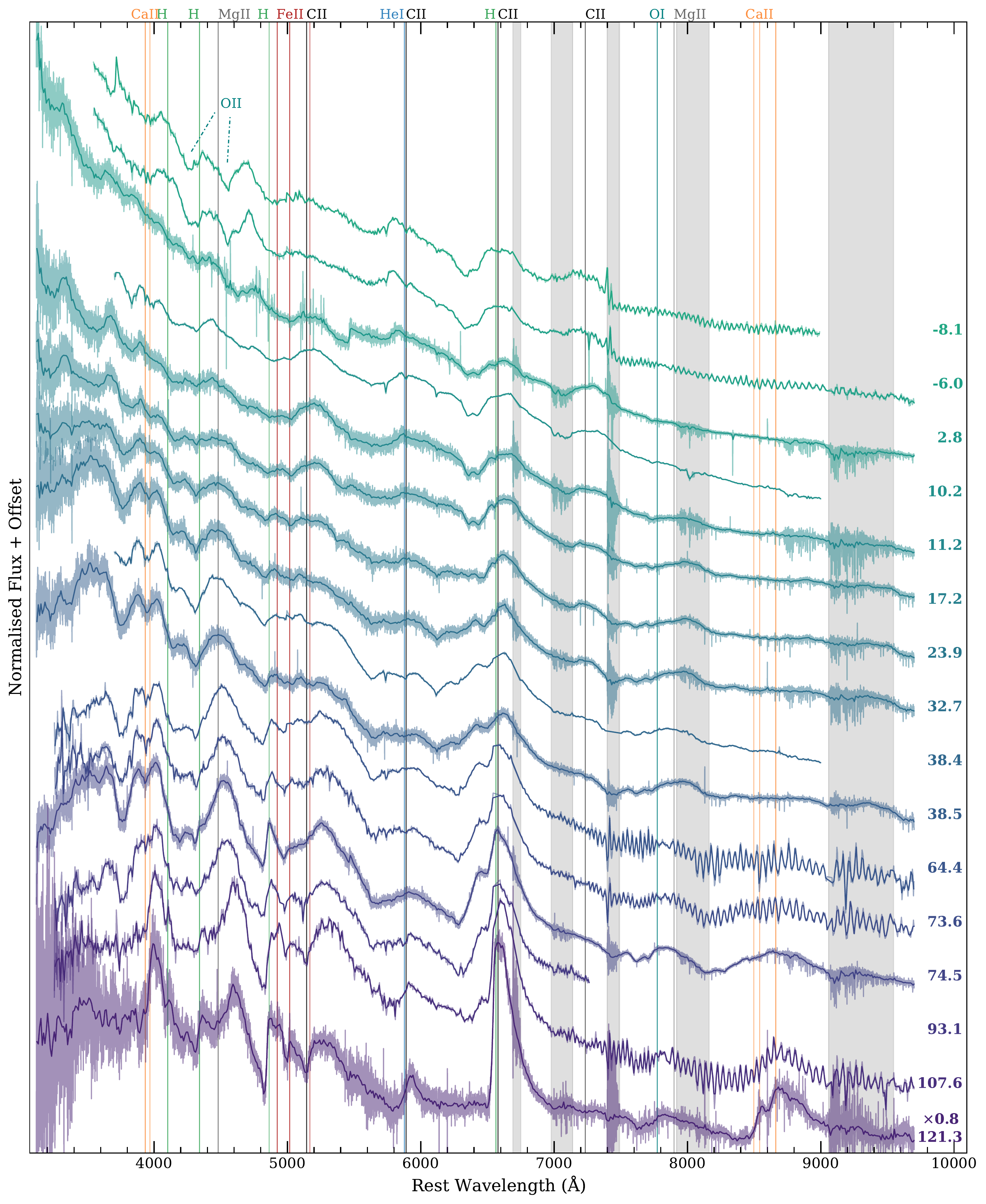}
    \caption{Spectral timeseries of SN\,2018bsz. Both the original spectra (lighter shade) and spectra binned to 10~Å (darker shade) are shown. Significant emission line features have been highlighted with vertical lines. The most significant transition in the spectral sequence occurs at $\sim30$~d when the \ion{C}{II} features have faded and the Balmer emission lines start to appear. Regions of strong telluric absorption are highlighted with grey bands. The host galaxy lines have been clipped in order to focus on the transient. The spectra have been normalised by the average flux of each spectra. Note that the spectrum at 121.3~d has been multiplied by 0.8 for clarity.}
    \label{fig:spectral_timeseries}
\end{figure*}

In Fig. \ref{fig:spectral_timeseries} we present the timeseries of optical spectra of SN\,2018bsz analysed in this paper (see Table \ref{tab:spec_log}). The spectra are described by an underlying blue continuum and an increasing number of emerging absorption and emission features. In the early spectra the most notable line features are those of \ion{O}{II} and \ion{C}{II}. The \say{w} feature created by several overlaying narrow \ion{O}{II} absorption lines is clearly present in the first two spectra but it is no longer visible after the maximum brightness. While this feature is common for SLSNe-I \citep{Quimby2011} it has also been seen in other SN types e.g Type Ib SN\,2008d \citep{Soderberg2008}, Type Ibn SN OGLE-2012-SN-006 \citep{Pastorello2015b} and Type II SN\,2019hcc \citep{Parrag2021}. The \ion{C}{II} $\lambda\lambda5890,6580,7234$ emission lines identified by \citet{Anderson2018} persist until $\sim20$~d post-maximum. Common stripped-envelope SNe (SESNe) and SLSNe-I ejecta lines appear at about $\sim30$~d post-maximum -- broadly at the same time as the hydrogen features. Most notable emerging features are the  \ion{Ca}{II} H\&K, \ion{Mg}{II} $\lambda4481$ and \ion{Fe}{II}  lines around 5000~Å. From 74.5~d onward we can also see the \ion{Ca}{II} NIR triplet and \ion{O}{I} $\lambda7774$ line. At $\sim30$~d the \ion{C}{II} $\lambda6580$ lines have been replaced by an H$\alpha$ emission with multiple components -- uniquely observed in SN\,2018bsz. In this section we investigate the individual line features identified in the spectra. 

\subsection{\ion{C}{II} lines}
\label{subsec:CII_lines}

\begin{figure*}
    \centering
    \includegraphics[width=1.00\textwidth]{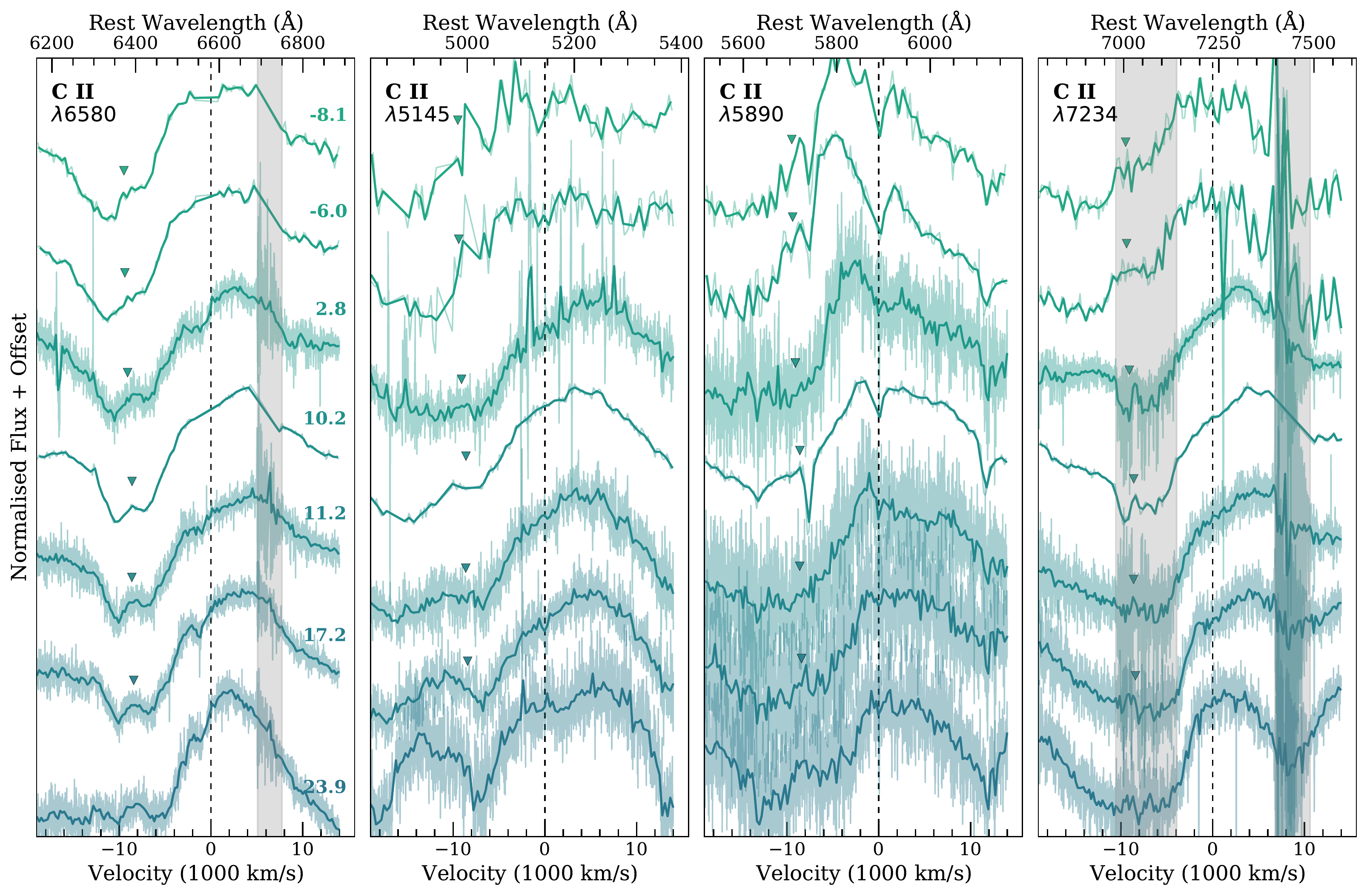}
    \caption{Spectral timeseries of \ion{C}{II} $\lambda\lambda6580,5145,5890,7234$ up to 24~d post-maximum. Both the original spectra (lighter shade) and spectra binned to 5~Å (darker shade) are shown. The location of the high-velocity (HV) emission feature next to \ion{C}{II} $\lambda6580$ and the corresponding locations for $\lambda\lambda5145,5890,7234$ have been marked with triangles. The feature appears to be present only for  \ion{C}{II} $\lambda6580$.  The line strengths from minimum to maximum are scaled to be equal to investigate the evolution of the line profiles. Regions of strong telluric absorption are indicated with grey bands. Note that a linear continuum subtraction has been applied to the displayed spectra to highlight the similarity of the profiles.}
    \label{fig:CII_timeseries}
\end{figure*}

As mentioned above, we confirm the presence of \ion{C}{II} $\lambda\lambda5890,6580,7234$ lines, and we further identify a fourth \ion{C}{II} line at $5145$~Å. The evolution of these \ion{C}{II} features is presented in Fig. \ref{fig:CII_timeseries}. After applying a simple linear continuum subtraction, the line profiles are similar to each other in velocity space throughout their evolution until they are barely detected by 23.9~d (see Fig. \ref{fig:spectral_timeseries}). As the Balmer emission lines become prominent only after $\sim20$~d (see Section \ref{subsec:H_lines}), H$\alpha$ unlikely contributes significantly to the emission of the feature identified as \ion{C}{II} $\lambda6580$, further supporting the \ion{C}{II} identification. The most notable difference between the profiles is that only $\lambda6580$ seems to show prominent P~Cygni absorption. For $\lambda7234$ the absorption interval is strongly influenced by telluric absorption and for $\lambda5145$ several \ion{Fe}{II} lines are likely to contribute in the same wavelength range so we cannot conclude on the presence of absorption components. However, the $\lambda5890$ absorption should have been visible if it was present.  

\citet{Anderson2018} suggested that high-velocity (HV) \ion{C}{II} emission was visible for $\lambda6580$ and $\lambda7234$. Based on Fig. \ref{fig:CII_timeseries} we can confirm the presence of an emission feature in the absorption trough of $\lambda6580$. Adopting the interpretation that the feature is HV \ion{C}{II} emission then the feature is found at $\sim-9000$~km/s. However, the emission feature seen next to \ion{C}{II} $\lambda7234$ is likely related to the telluric absorption affecting this  wavelength range. Furthermore, no emission feature is evident bluewards of $\lambda5145$ or $\lambda5890$. Thus we cannot confirm that the emission is related to \ion{C}{II}. Instead we favour an interpretation of the feature as HV H$\alpha$ emission becoming more prominent at later epochs as discussed in Sect. \ref{subsec:H_lines}.

\subsection{Balmer and Paschen lines}
\label{subsec:H_lines}
\begin{figure*}
    \centering
    \includegraphics[width=0.98\textwidth]{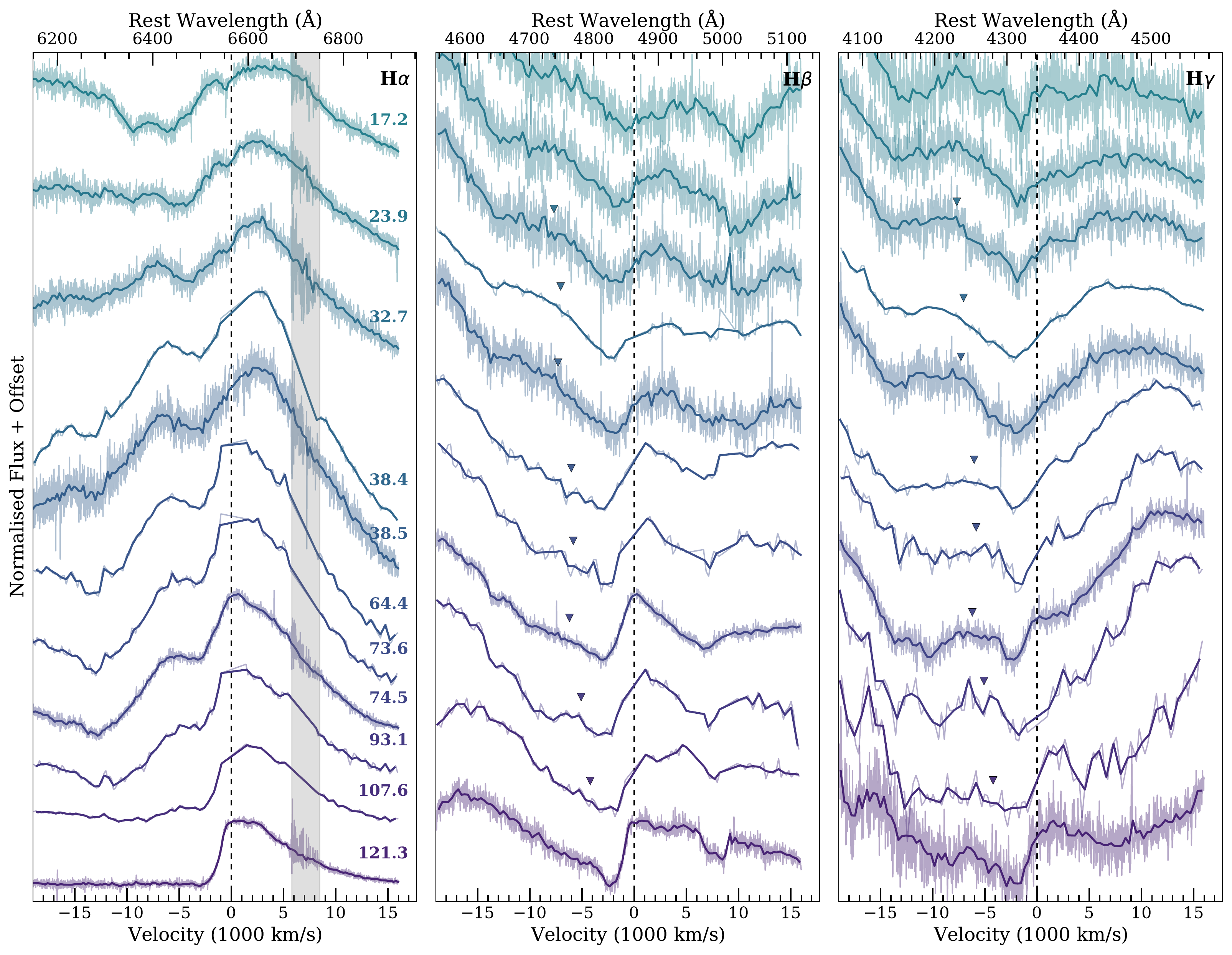}
    \caption{Same as \ref{fig:CII_timeseries} for H$\alpha$, H$\beta$ and H$\gamma$ starting from 17.2~d post-maximum. The corresponding location of the blue H$\alpha$ component has been marked for H$\beta$ and H$\gamma$ with triangles. Note that continuum subtraction has not been applied to any of the displayed line profiles due to the presence of the strong \ion{Mg}{II} $\lambda4481$ line affecting both H$\beta$ and H$\gamma$.}
    \label{fig:Balmer_timeseries}
\end{figure*}

\begin{figure*}
    \centering
    \includegraphics[width=0.98\textwidth]{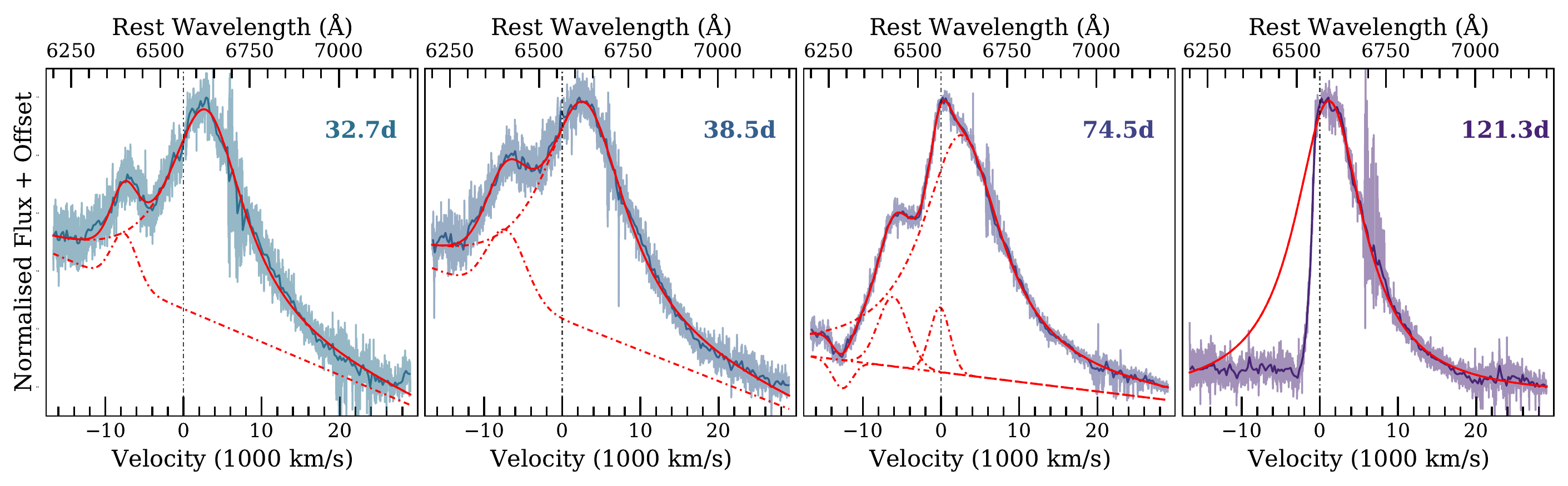}
    \caption{Multi-component H$\alpha$ line profile fits at 32.7~d, 38.5~d, 74.5~d and 121.3~d. The strong, red component is described by a Lorentzian profile but for the other components a Gaussian profile provides a decent fit. The fits are shown with solid red lines and each of the individual components with dashed red lines. Spectra binned to 5~Å are shown with darker shade and unbinned with lighter one. For the first three epochs the fits were performed on the shown data, but for the last epoch only data redward of rest frame H$\alpha$ were used due to the highly asymmetric line profile.}
    \label{fig:Ha_fits}
\end{figure*}

\begin{figure}
    \centering
    \includegraphics[width=0.48\textwidth]{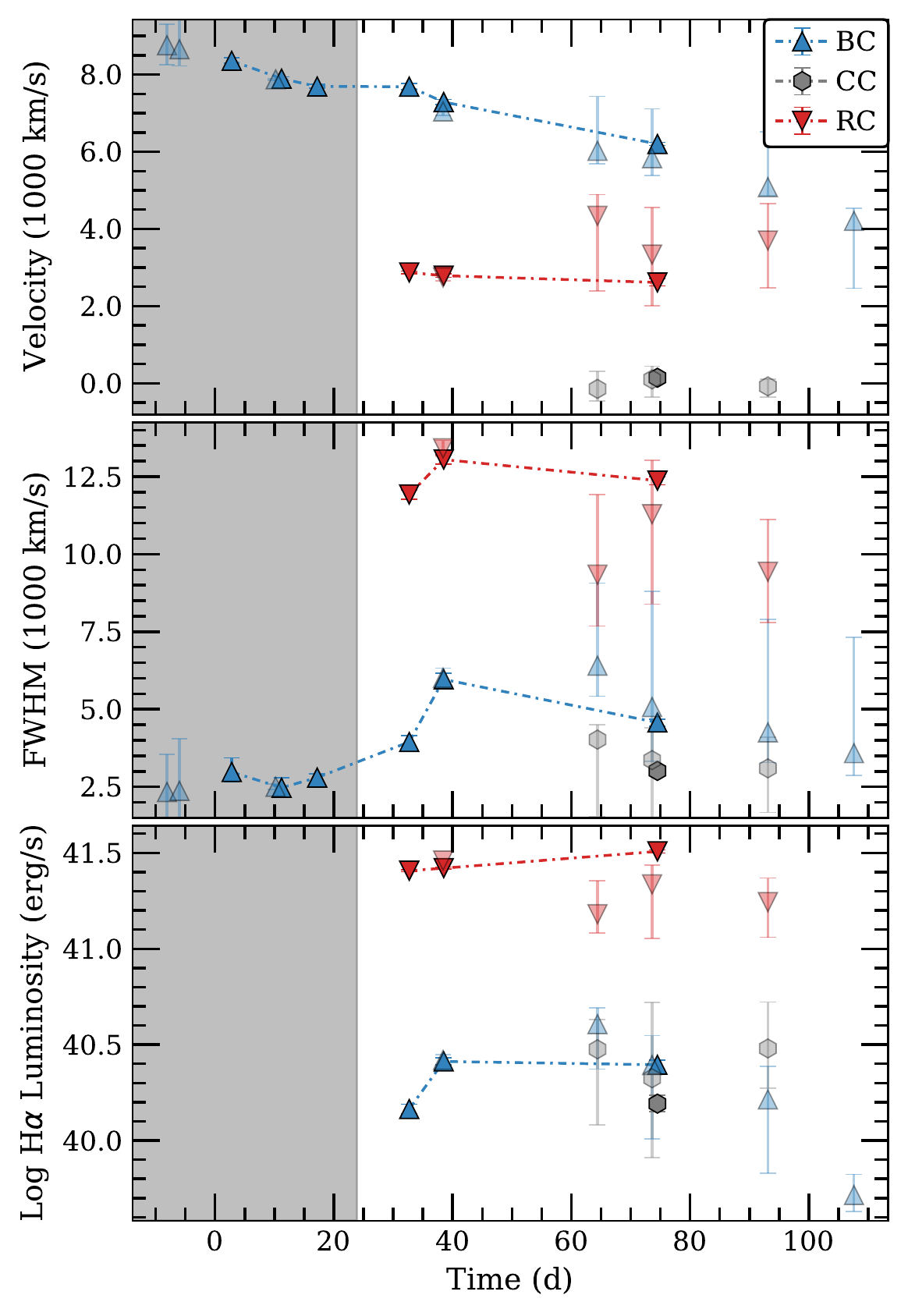}
    \caption{Velocity offset, FWHM and luminosity evolution of the blue component (BC), central component (CC) and red component (RC) of H$\alpha$. The fits of the X-Shooter spectra are shown in darker shade and are connected with dot-dashed lines while the fits to lower resolution spectra are shown with lighter markers. The red component is more luminous by a factor of 10 and does not evolve in velocity or FWHM, while the blue component moves towards the red and eventually disappears. Note that the data points in the grey-shaded region are measured for the HV emission feature at the absorption trough of \ion{C}{II} $\lambda6580$ assuming they are related to the blue component of H$\alpha$. The 107.6~d spectrum was fit only for the blue component as the red component was not describable with a symmetric line profile (as at 121.3~d, see Fig. \ref{fig:Ha_fits}). The shown uncertainties are $2\sigma$ estimated by Markov Chain Monte Carlo \texttt{emcee} package \citep{Foreman-Mackey2012} as implemented by \texttt{LMFIT}.}
    \label{fig:Ha_v_fwhm_evo}
\end{figure}

The most noteworthy transition in our spectral series occurs at $\sim30$~d. The prominent P~Cygni absorption component of \ion{C}{II} $\lambda6580$ has completely vanished by 23.9~d only to be replaced by an emission feature by 32.7~d. At the same time the emission near rest frame \ion{C}{II} $\lambda6580$ persists strong but appears to be changing shape. As \ion{C}{II} $\lambda\lambda5145,7234$ are greatly weakened by $\sim30$~d, it is likely that \ion{C}{II} $\lambda6580$ does not contribute significantly to the emission and that H$\alpha$ now dominates the profile.

In Fig. \ref{fig:Balmer_timeseries} we present the line evolution of H$\alpha$, H$\beta$ and H$\gamma$ lines starting from 17.2~d. As \ion{C}{II} $\lambda6580$ emission profile is present before $\sim30$~d, it is not possible to determine when the H$\alpha$ actually appears. However, the H$\beta$ emission appears to be visible for the first time at 23.9~d. While some excess might also be present at $17.2$~d, it is offset in velocity space with respect to the H$\beta$ seen in the later spectra. As such we adopt 23.9~d as the first epoch the Balmer lines are detected. The emission lines appear to be redshifted by $\sim3000$~km/s. The shift is especially clear for H$\alpha$ and H$\beta$ at 38.5~d, while for H$\gamma$ the emission component is merged with the strong \ion{Mg}{II} $\lambda4481$ and thus is barely visible. By 74.5~d the peaks are observed at the rest frame wavelength, demonstrating a rapid change in the velocity of the emitting material. Given that the H$\alpha$ profile at this epoch appears to show small amount of excess emission at $v\sim0$~km/s, we interpret the drastic velocity change as being caused by an emerging zero-velocity component. A two component model is also possibly needed to explain the \say{flat top} emission profile of H$\alpha$ at 121.3~d.

In addition to the redshifted Balmer emission lines, we also identify blueshifted components. However, the blue emission appears to be present only for H$\alpha$ where it is found at velocity of $\sim-8000$~km/s to begin with -- significantly higher than the redshifted component. The feature moves redward during its evolution and by the last epoch it is visible at 107.6~d it is found at $\sim-4500$~km/s. The location of the blue component in H$\alpha$ has been marked on top of each H$\beta$ and H$\gamma$ profile with triangles in Fig. \ref{fig:Balmer_timeseries}. While no corresponding emission is clearly visible for H$\beta$, the location would coincide with the strong \ion{Mg}{II} $\lambda4481$ line, making it more difficult to identify. On closer examination there does seem to be a \say{bump} in the spectra at a similar velocity as the blue component. The feature is especially clear in our high S/N ratio flux spectrum from spectropolarimetry at 38.4~d. Furthermore, not only is the excess centred at similar velocities, it also extends to $\sim-13000$~km/s similarly to H$\alpha$. While one-to-one comparison of the profiles is difficult due to the strong magnesium line, we consider it likely that the observed excess is caused by the blue emission component of H$\beta$. A similar blueshifted excess appears to be present for H$\gamma$ but significantly stronger than for H$\beta$, and thus it is unclear if it is related to H$\gamma$.

The presence of the multiple H$\alpha$ components can be seen in the line profile fits at 32.7~d, 38.5~d, 74.5~d and 121.3~d in Fig.~\ref{fig:Ha_fits}. At the first two epochs the H$\alpha$ is described by a combination of two emission components, a strong and broad redshifted Lorentzian and a fainter, narrower blueshifted Gaussian superimposed on a linear continuum. By 74.5~d an additional central Gaussian emission component is necessary to achieve a satisfactory fit. Furthermore, the data show the presence of an absorption component, that we fit with a Gaussian centred at $\sim-12000$~km/s ($\sim6300$~Å), but it is unclear if it is related to H$\alpha$ or possibly some other line i.e. \ion{Si}{II} $\lambda6355$. For the final H$\alpha$ profile at 121.3~d we provide a fit with a single Lorentzian component using only the redshifted data ($v>0$~km/s). At this epoch the red side of the profile is well described by the Lorentzian, while the blue side of the profile is absent. No combination of emission and absorption components we attempted provided a reasonable fit, let alone offering a physical explanation for the skewed profile. While we did not achieve an acceptable fit by adding a central component, we consider it likely that the component still persists due to the flat-top shape of the line. Note that while we have presented fits with Gaussian blue and central components, we also attempted fits with Lorentzian profiles instead. The fits were visibly as decent as the ones shown in Fig.~\ref{fig:Ha_fits} and thus we cannot distinguish which profile is preferable. However, for the red component we prefer Lorentzian profile. As shown in Table \ref{tab:Ha_chi2_nu}, Lorentzian profile provides consistently better fits than Gaussian especially for the high-quality X-Shooter spectra. While at the the early epochs the $\chi^2_\nu$ values are similar, the last few require Lorentzian to describe the pronounced red tail of the profile for the fit setups described above. This is well demonstrated by the single-component fits to the last epoch: while the Lorentzian provides $\chi^2_\nu=2.8$, a Gaussian profile results in $\chi^2_\nu=3.3$. The fits were performed using \texttt{LMFIT}\footnote{\url{https://lmfit.github.io/lmfit-py/}} package for \texttt{Python} \citep{Newville2014}.


\begin{table}
    \def\arraystretch{1.1}%
    \setlength\tabcolsep{10pt}
    \centering
    \fontsize{10}{12}\selectfont
    \caption{$\chi^2_\nu$ values for the H$\alpha$ profile fits using Lorentzian (L) and Gaussian (G) red components with otherwise the same fit setups (see text).}
    \begin{threeparttable}
    
    \begin{tabular}{c c c c}
    \hline
    \hline
	Phase (d) & Instrument & $\chi^2_\nu$ (L)	& $\chi^2_\nu$ (G)	\\
    \hline
    32.7	& X-Shooter &	1.95	&	2.17 \\
    38.4	& FORS2     &	3.43	&	4.91 \\
    38.5	& X-Shooter &	1.38	&	1.41 \\
    64.4	& EFOSC2    &	0.56	&	0.57 \\
    73.6	& EFOSC2    &	0.58	&	0.57 \\
    74.5	& X-Shooter &	2.31	&	2.83 \\
    93.1	& EFOSC2    &	0.89	&	0.90 \\
    121.3	& X-Shooter &	2.77	&	3.29 \\
    \hline
    \end{tabular}
    \end{threeparttable}    
\label{tab:Ha_chi2_nu}
\end{table}

While the Balmer lines become visible at about $\sim25$~d, the aforementioned HV feature seen next to \ion{C}{II} $\lambda6580$ is found in a very similar wavelength range as the blue component of H$\alpha$ (see Fig. \ref{fig:spectral_timeseries}). To further investigate this we have shown the continuous velocity evolution -- assuming they are both related to H$\alpha$ -- in Fig. \ref{fig:Ha_v_fwhm_evo}. For epochs up to 17.2~d we fit the \ion{C}{II} P~Cygni profile with Gaussian emission and absorption components and added a single Gaussian emission profile for the HV feature. Starting from 32.7~d, the profile is H$\alpha$ dominated and the fits were performed with the same setup as in Fig. \ref{fig:Ha_fits}. The central component was added to the fits after 60~d. As the evolution of the velocity is effectively linear from $-8.1$~d to 107.6~d it is likely that the emission feature seen in the early spectra is HV H$\alpha$ emission that later develops into the blue component.  The HV H$\alpha$ emission is not very strong in the early spectra and thus detecting an equivalent HV feature in other Balmer lines is difficult, although in the case of H$\beta$ an emission feature appears to be present at 10.2~d at the same velocity. The feature can also be tentatively seen in the following X-Shooter spectra until 32.7~d when it has grown into the blue excess next to H$\beta$ mentioned earlier.

\begin{figure*}
    \centering
    \includegraphics[width=0.98\textwidth]{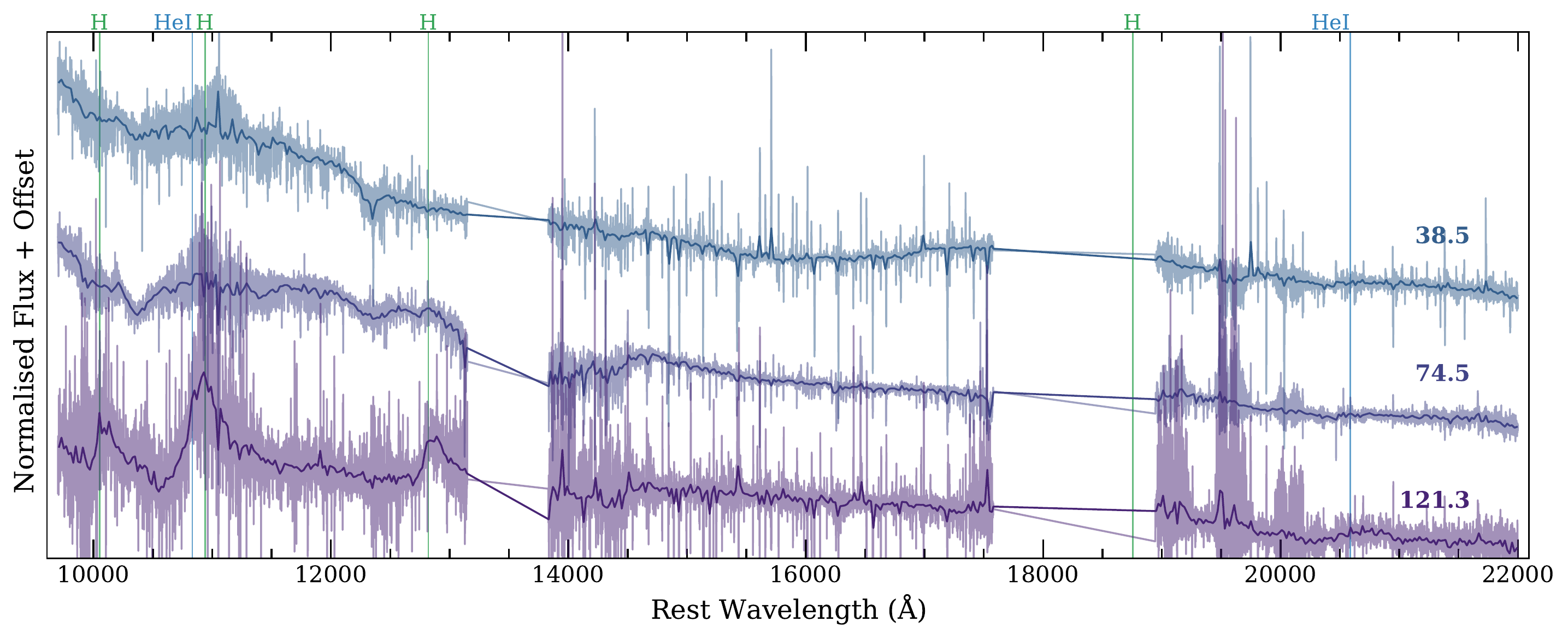}
    \caption{The X-shooter NIR spectra at the last three epochs (lighter shade) and spectra binned to 20~Å (darker shade). Paschen lines are marked with green lines and \ion{He}{I} with blue. As no clear line features were present at the earlier epochs of NIR spectroscopy, the spectra have been excluded from the figure.}
    \label{fig:nir_timeseries}
\end{figure*}

\begin{figure}
    \centering
    \includegraphics[width=0.48\textwidth]{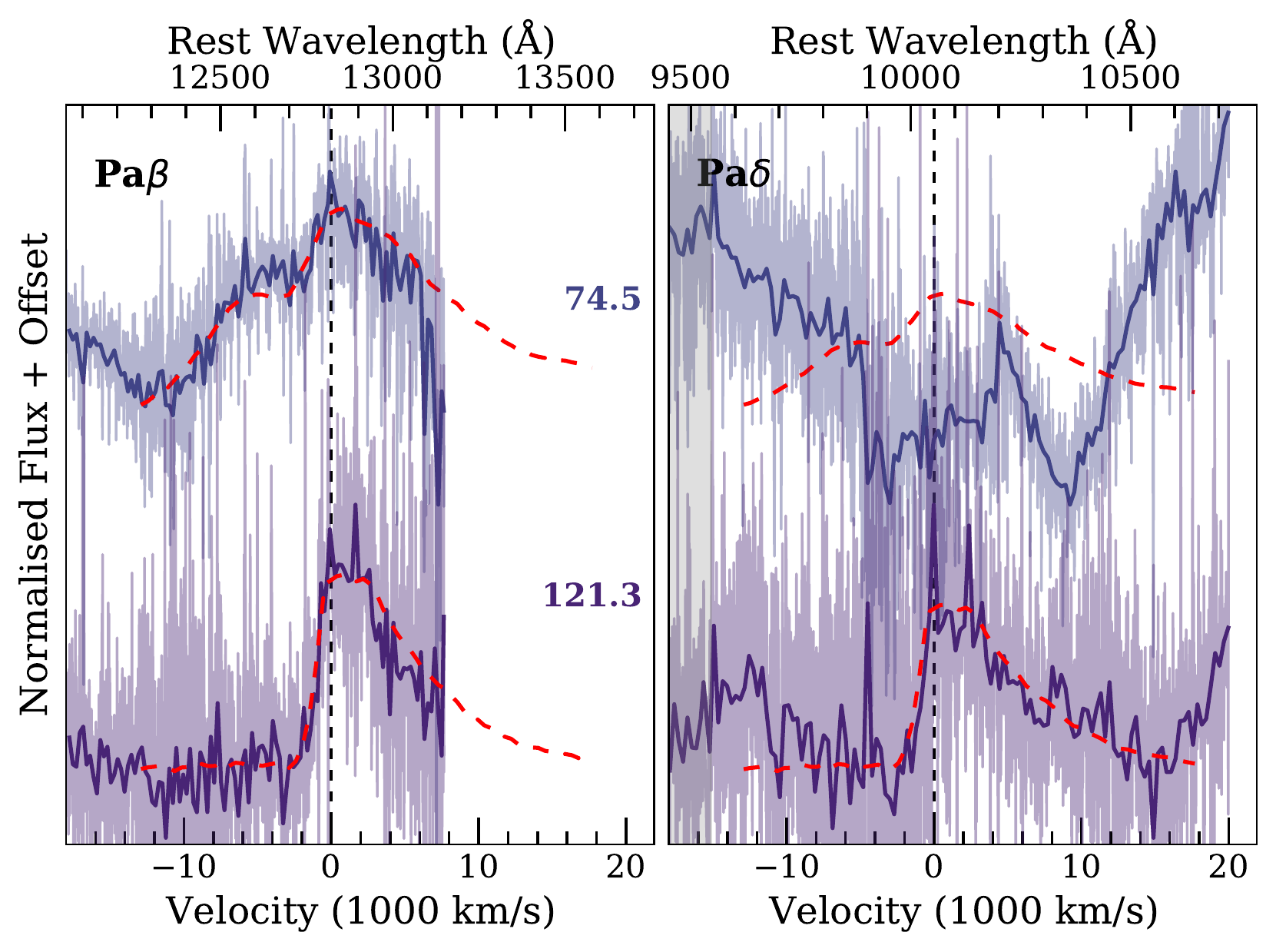}
    \caption{Same as Fig. \ref{fig:CII_timeseries} for Pa$\beta$ and Pa$\delta$ lines at 74.5~d and 121.3~d. A scaled H$\alpha$ profile has been shown over the data highlighting the similarity. Pa$\delta$ is not detected at 74.5 days. The spectra have been binned to $10$~Å.}
    \label{fig:Ha_Pbd}
\end{figure}

\begin{figure}
    \centering
    \includegraphics[width=0.48\textwidth]{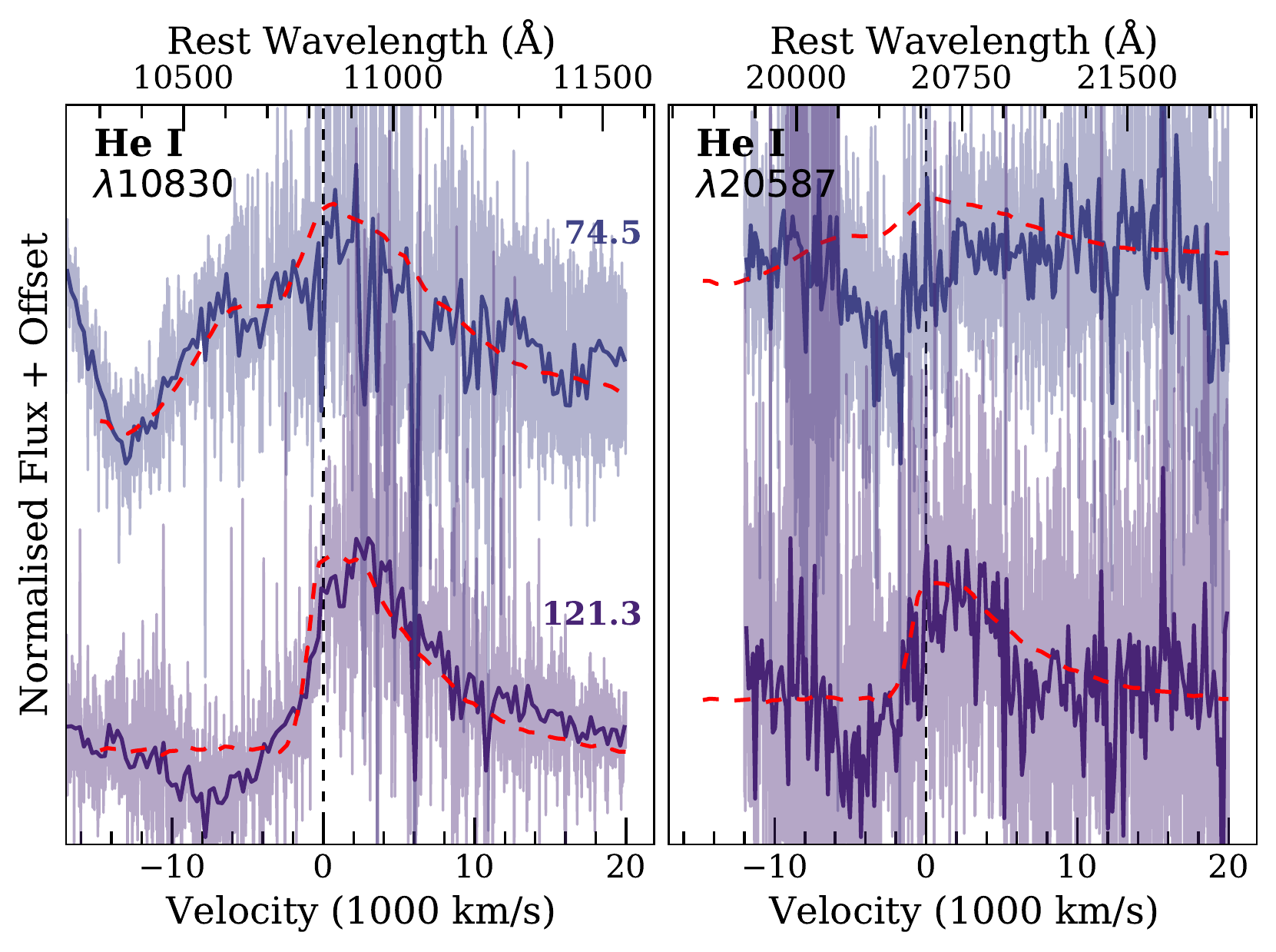}
    \caption{Same as Fig. \ref{fig:Ha_Pbd} for \ion{He}{I} $\lambda\lambda10830, 20587$ profiles at 74.5~d and 121.3~d. \ion{He}{I} $\lambda20587$ is not detected at 74.5 days, but $\lambda10830$ shows remarkable similarity to H$\alpha$ despite the high level of noise.}
    \label{fig:HeI_NIR_timeseries}
\end{figure}

We also show the velocity offset evolution for the red and central components of H$\alpha$ and the evolution of the FWHM and luminosity for all three components in Fig.~\ref{fig:Ha_v_fwhm_evo}. The red component is found at significantly lower velocity than the blue throughout the evolution, but the value also does not clearly decrease in time. The FWHM of the red component is consistently above $10000$~km/s. Given the component is best described by a Lorentzian profile it is very likely that the line has undergone significant amount of electron scattering \citep[e.g.][]{Chugai2001}. In comparison the blue and central components have values around $\mathrm{FWHM}\lesssim6000$~km/s and they can be described by Gaussian profiles. Finally, the central component is consistently found close to rest frame wavelength (i.e. zero-velocity). In the luminosity evolution it is clear that the red component drives the luminosity of the line. While the blue and central component are found below values of $\sim3\times10^{40}$~erg/s the red component is consistently around $\sim3\times10^{41}$~erg/s. Note that the X-shooter spectra at 23.9~d and 121.3~d have been excluded from the figure. At 23.9~d some excess emission appears to be present at the location of the blue component, but as it is very tentative and it is difficult to be certain if it is real. On the other hand, at 121.3~d no successful fit was found for the highly skewed profile (see Fig.~\ref{fig:Ha_fits}.) Additionally the spectrum at 107.6~d was fit only for the blue component as the red part of the profile was likewise not describable with symmetric line profiles.

The higher resolution of the X-Shooter spectra as compared to the NTT and FORS2 spectra, provides for more reliable fits -- in particular due to the removal of the narrow host galaxy emission lines. For the lower resolution spectra the affected region around H$\alpha$ is broad ($\sim80$~\AA) due to multiple host lines (H$\alpha$ and [\ion{N}{II}] $\lambda\lambda6548, 6584$). As a result the fits to the red and central components are less reliable. In Fig. \ref{fig:Ha_v_fwhm_evo} the dot-dashed lines have been drawn through the X-Shooter epochs and the measured quantities from the other spectra are shown with lighter colours.

In Fig.~\ref{fig:nir_timeseries} we present the last three X-Shooter NIR spectra (38.5~d, 74.5~d and 121.3~d). The strongest line features are observed at 121.3~d. Most prominent features coincide with Paschen Pa$\beta$, Pa$\gamma$ and Pa$\delta$ lines. As Pa$\gamma$ is the strongest of the three, its strength is likely affected by the nearby \ion{He}{I} $\lambda10830$. The Pa$\alpha$ line is found at a region of strong telluric absorption. In Fig. \ref{fig:Ha_Pbd} we compare the Pa$\beta$ and Pa$\delta$ lines to H$\alpha$. While the spectrum is noisy, the lines have asymmetric profiles visibly similar to H$\alpha$ at 121.3~d. At 74.5~d the Pa$\beta$ emission appears to have a very similar shape to H$\alpha$ but emission is not detected for Pa$\delta$. No clear hydrogen features are visible at 38.5~d (or before) as can be seen in Fig.~\ref{fig:nir_timeseries}. This is likely a result of high level of continuum emission at the earlier phases, diminishing the emission lines to a degree they are no longer clearly visible over it. The same effect is also visible for the H$\alpha$ line: while the luminosity of the profile remains roughly constant in time (see Fig. \ref{fig:Ha_v_fwhm_evo}), the line becomes visually stronger in comparison to the continuum and other line features as can be seen in Fig. \ref{fig:spectral_timeseries}.

As the Pa$\beta$ and H$\alpha$ have similar profiles at 74.5~d and 121.3~d, it is unlikely that the changes in the profiles are caused by dust. The effect of dust is strongly wavelength dependent and a NIR line should be affected significantly less than an optical one. Therefore, for the dust to explain the disappearance of the blue component at $\sim100$~d, the effect would have to be negligible at 74.5~d but by 121.3~d the dust would have to be responsible for hiding the blue component of Pa$\beta$ as well as H$\alpha$. As this would require a significant increase in the dust mass in a short amount of time during the photospheric phase, it seems unlikely.

\subsection{\ion{He}{I} lines}
\label{subsec:HeI_lines}

\begin{figure}
    \centering
    \includegraphics[width=0.48\textwidth]{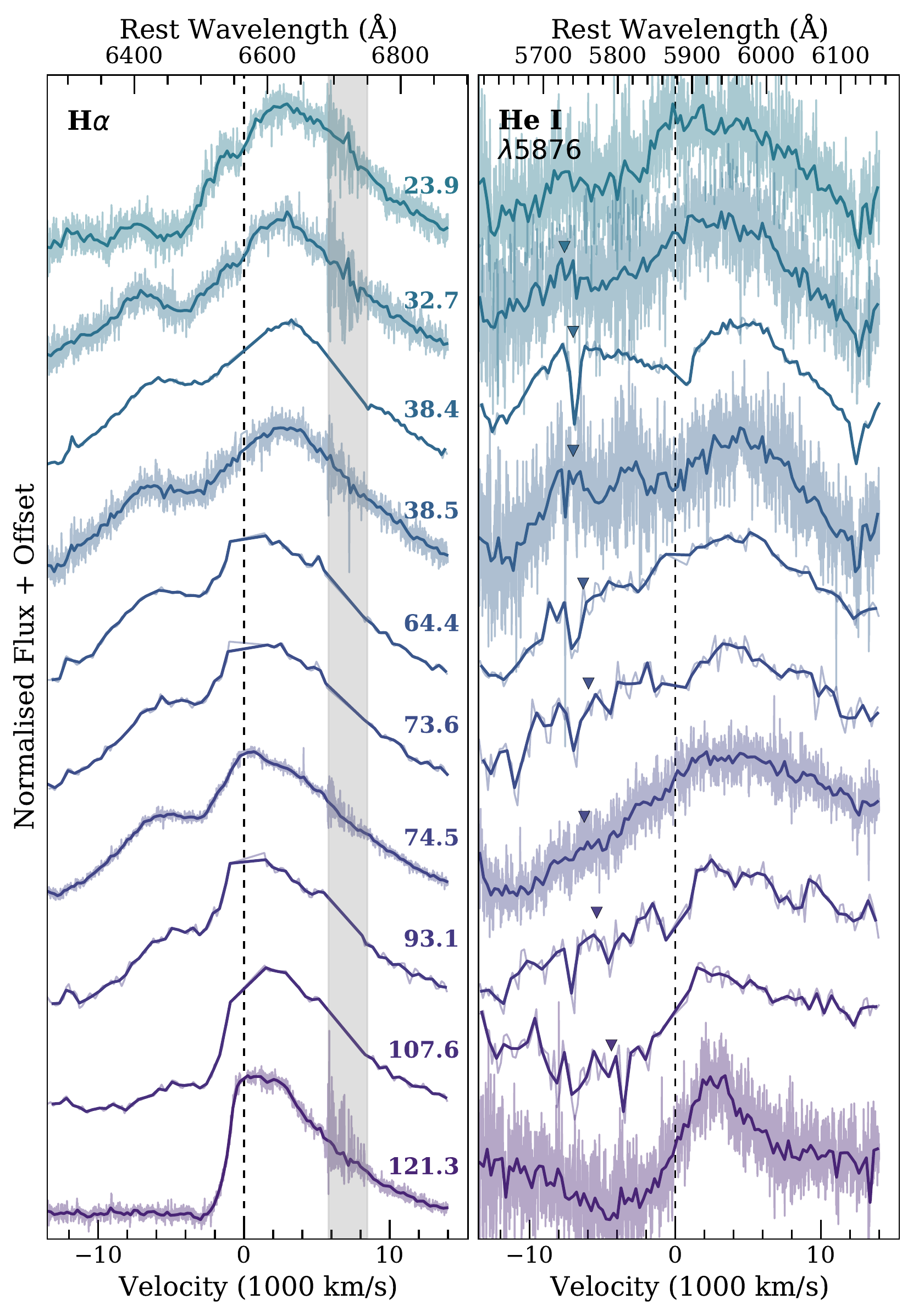}
    \caption{Same as Fig. \ref{fig:CII_timeseries} for H$\alpha$ and \ion{He}{I} $\lambda5876$. The location of the blue H$\alpha$ component is marked with triangles for $\lambda5876$. \ion{He}{I} line is similar to H$\alpha$, but the central component appears to be absent. Note that the regions of strong telluric absorption at $\sim13500$~Å and $\sim18000$~Å have been removed.}
    \label{fig:HeI_5876_timeseries}
\end{figure}

\begin{figure}
    \centering
    \includegraphics[width=0.48\textwidth]{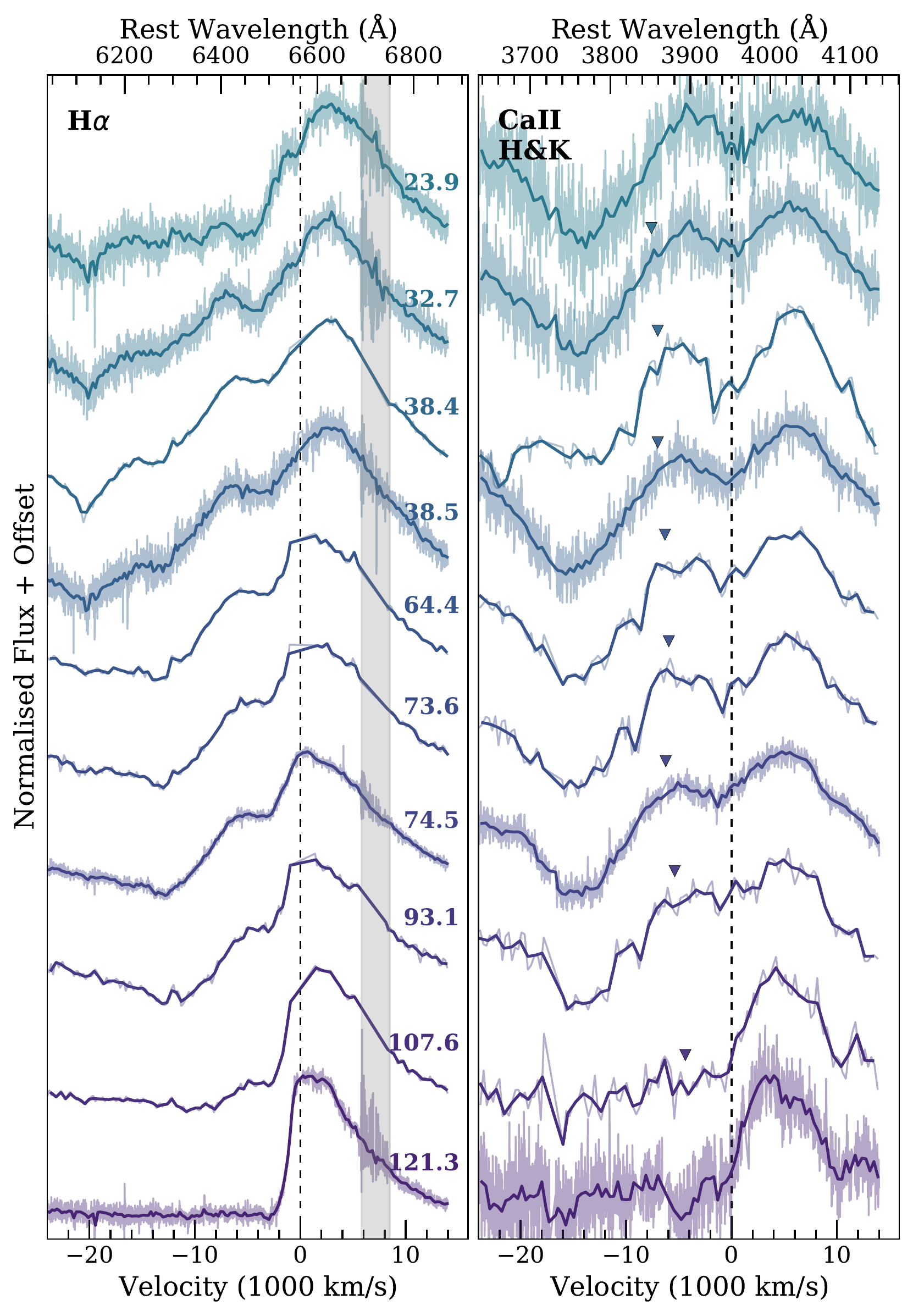}
    \caption{Same as Fig. \ref{fig:HeI_5876_timeseries} for H$\alpha$ and \ion{Ca}{II} H\&K. The \ion{Ca}{II} feature is centred at 3951.5~Å, i.e. the average wavelength between the H and K components. The \ion{Ca}{II} is similar to H$\alpha$, but lacks the central component.}
    \label{fig:CaII_H_timeseries}
\end{figure}

\begin{figure}
    \centering
    \includegraphics[width=0.48\textwidth]{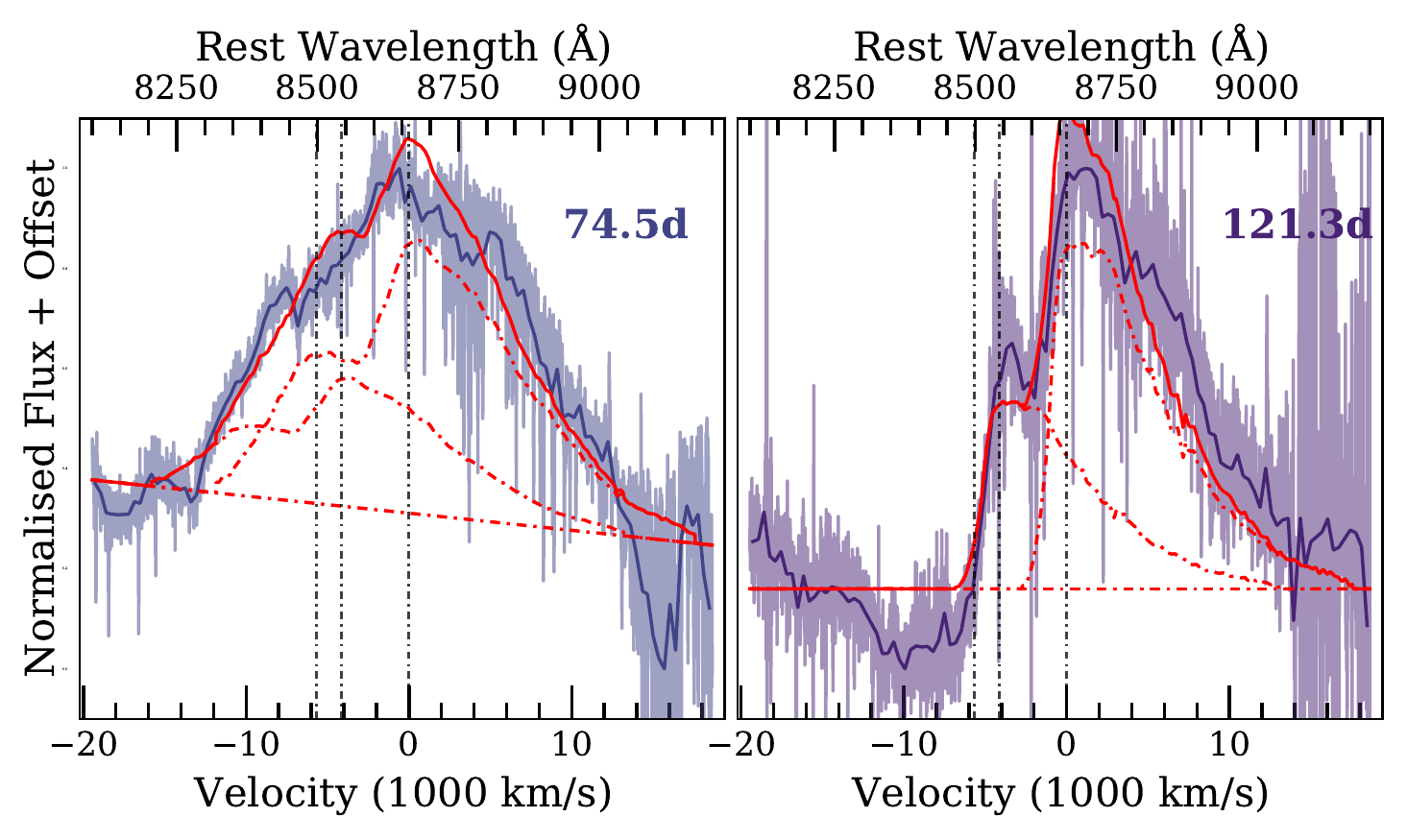}
    \caption{H$\alpha$ line profile fits to \ion{Ca}{II} $\lambda\lambda8498, 8542, 8662$ NIR triplet at 74.5~d and 121.3~d. In addition to linear background, each \ion{Ca}{II} line is fitted assuming it has the shape of H$\alpha$ profile at the same epoch. At 74.5~d the general shape of the profile is described by the fit, but at 121.3~d the asymmetric profile is well matched by the two redder \ion{Ca}{II} lines. Velocity scale is measured from $\lambda8662$. The three \ion{Ca}{II} lines are marked with dashed vertical lines.}
    \label{fig:CaII_NIR_fits}
\end{figure}

\begin{figure*}
    \centering
    \includegraphics[width=0.98\textwidth]{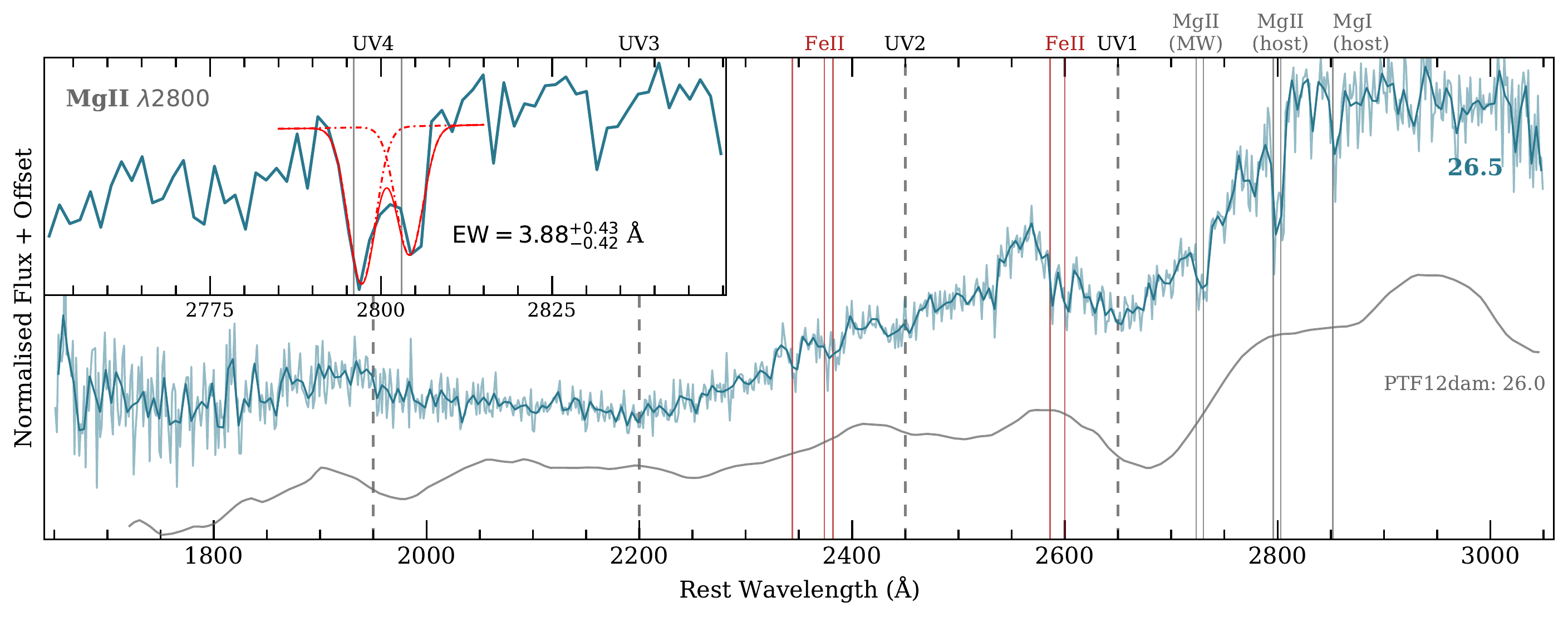}
    \caption{Normalised HST/STIS spectrum of SN\,2018bsz taken at 26.5~d post-maximum brightness (lighter shade) and the spectrum binned to 5~Å (darker shade). Several narrow host galaxy absorption lines of Fe~II, Mg~I and \ion{Mg}{II} as well as Milky Way \ion{Mg}{II} $\lambda\lambda2796,2803$ doublet have been highlighted. Approximate locations of four SLSNe-I absorption bands \citep[UV1: 2650~Å, UV2: 2450~Å, UV3: 2200~Å and UV4: 1950~Å;][]{Quimby2018} have been marked, but only UV1 is clearly present for SN\,2018bsz. In the inset we show a close up of the host galaxy \ion{Mg}{II} $\lambda\lambda2796,2803$ doublet region including a double Gaussian fit to the line profile. The combined equivalent width of the lines is $3.88^{+0.43}_{-0.42}$~Å. We also show spectrum of PTF12dam for comparison as it is one of the few high quality HST spectra of SLSNe-I taken at similar epoch to SN\,2018bsz. The spectrum was first presented in \citet{Quimby2018} and downloaded from the \textit{Open Supernova Catalog} \citep{Guillochon2017}.}
    \label{fig:STIS_spec}
\end{figure*}

\begin{figure*}
    \centering
    \includegraphics[width=0.98\textwidth]{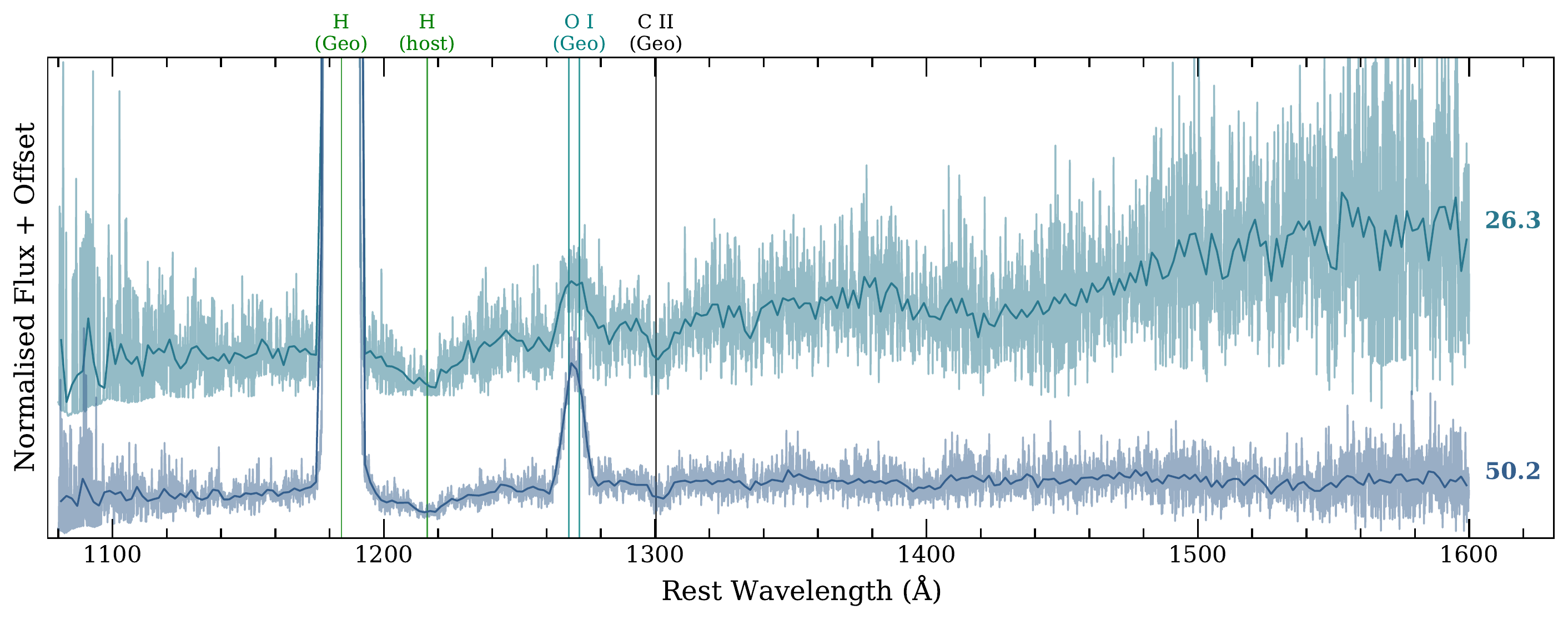}
    \caption{The two epochs of HST/COS spectroscopy taken at 26.3~d and 50.2~d post-maximum brightness (lighter shade) and the  spectra binned to 2~Å (darker shade). We identify several geocoronal lines (Geo) as well as possible Ly$\alpha$ absorption from the host galaxy. No clear transient features are present in the spectra.}
    \label{fig:COS_spec}
\end{figure*}

In the 121.3~d X-Shooter NIR spectra we identify emission by \ion{He}{I} $\lambda10830$ and $\lambda20587$ as presented in Fig.~\ref{fig:HeI_NIR_timeseries}. While the spectrum is very noisy at the location of the latter line, the detected feature resembles the one seen around 10830~Å. Both of the profiles also appear to be similar to the over-plotted H$\alpha$ profile, except they also seem to show blueshifted absorption. For $\lambda20587$ the velocity is $\sim-5000$~km/s but for the $\lambda10830$ it is found to be $\sim-7500$~km/s, as measured from the absorption trough. Neither of the \ion{He}{I} lines are clearly present in the earlier spectra, but at 74.5~d the spectrum around $\lambda10830$ line appears to have a very similar shape to H$\alpha$. However, as the wavelength range appears to be very noisy drawing any definite conclusions is difficult.  

We also identify the commonly observed optical \ion{He}{I} $\lambda5876$ in our spectra and in Fig.~\ref{fig:HeI_5876_timeseries} we present the timeseries of the spectral region in comparison to the H$\alpha$ starting from 23.9~d post-maximum. While the \ion{He}{I} line is significantly fainter than H$\alpha$ and thus spectra are noisier in comparison, it does appear to have a blue emission component at $<40$~d. The feature is most notable in the high S/N ratio spectrum at 38.4~d, when the line profile appears to be distinctly similar to that of H$\alpha$. However, at later phases ($\gtrsim60$~d) the blue component is no longer identifiable and instead the \ion{He}{I} feature appears to consist of a single broad component with a redshifted peak that has a similar width as the whole H$\alpha$ profile at every epoch. In the last three spectra, the emission bluewards of the rest wavelength seems to become weaker in parallel with the disappearance of the blue H$\alpha$ component. At 121.3~d the peak of the profile appears to be redshifted by $\sim2000$~km/s unlike for the NIR lines or the H$\alpha$, but the profile also exhibits blueshifted absorption at $\sim-5000$~km/s similarly to the NIR features. Finally, due to the presence of NIR \ion{He}{I} lines we are convinced the emission line feature around $\lambda5876$ is truly \ion{He}{I} rather than the nearby, common Na~I D ($\lambda\lambda5890,5896$) or \ion{C}{II} $\lambda5890$ which was prominent in the early ($\lesssim25$~d) spectra of SN\,2018bsz.

\subsection{\ion{Ca}{II} lines}
\label{subsec:CaII_lines}

In addition to the hydrogen and helium emission exhibiting peculiar broad profiles with several components, we note that also \ion{Ca}{II} lines appear to be similar in shape. In Fig.~\ref{fig:CaII_H_timeseries} we show the similarity of \ion{Ca}{II} H\&K emission profile with the H$\alpha$ line during the whole spectral timeseries. Notably the blue and red components are found at comparable velocities. While the figure is centred at the middle of the two features (3951.5~Å) the effect remains if centred at either H ($\lambda3969$) or K ($\lambda3934$). The blue component also disappears at the same time as in H$\alpha$. While the blue component is distinct throughout the timeseries -- unlike for \ion{He}{I} $\lambda5876$ -- the evolution resembles the helium line in that the peak of the line profile is redshifted by $\sim2000$~km/s at the last epoch. The \ion{Ca}{II} line also shows clear absorption, but at different epochs than helium -- the absorption is visible only at $<100$~d and it is found at $\sim-15000$~km/s as is typical for SESNe and SLSNe. 

The \ion{Ca}{II} NIR $\lambda\lambda8498, 8542, 8662$ triplet also appears to be remarkably similar to the H$\alpha$ profile -- at least at the last epoch. In Fig. \ref{fig:CaII_NIR_fits} we show the best fits to the \ion{Ca}{II} NIR triplet lines at 74.5~d and 121.3~d using the H$\alpha$ profile at the respective epoch as a template. At 74.5~d only the width of the \ion{Ca}{II} NIR is similar to a combination of H$\alpha$ line profiles, but at 121.3~d the similarity is significant. The combination of the two redder \ion{Ca}{II} lines provides a nearly perfect fit with the highly asymmetric H$\alpha$ profiles. Similarly to the \ion{He}{I} NIR line, the line peaks are found at the rest frame of the respective lines -- unlike for \ion{Ca}{II} H\&K and \ion{He}{I} $\lambda5876$. \ion{Ca}{II} NIR triplet shows clear absorption at both epochs: at 74.5~d the absorption is found at $\sim-15000$~km/s while at 121.3~d it is found at $\sim-9000$~km/s as measured from \ion{Ca}{II} $\lambda8662$. Note that \ion{Ca}{II} $\lambda8498$ does not appear to contribute to the emission at either of the epochs.

\subsection{UV features}

In Fig. \ref{fig:STIS_spec} we present the HST/STIS spectrum taken at 26.5~d post-maximum. We show identifications for narrow host galaxy absorption lines of \ion{Fe}{II}  ($\lambda\lambda2344,2374,2382,2586,2600$), \ion{Mg}{II} ($\lambda\lambda2796,2803$ doublet) and Mg~I ($\lambda2852$) as well as the \ion{Mg}{II} absorption from the Milky Way. In the inset we show a double Gaussian fit to the host \ion{Mg}{II} $\lambda\lambda2796,2803$ doublet line that results in a combined Equivalent Width (EW) of $3.88^{+0.43}_{-0.42}$~Å ($1\sigma$). The value is on the high end of the distribution of SLSNe-I hosts \citep[$2.6\pm1.2$;][]{Vreeswijk2014}. 

In the figure we also mark the locations of four known transient absorption bands often seen in UV spectra of SLSNe-I with black dashed lines. The features are found roughly at UV1: 2650~Å, UV2: 2450~Å, UV3: 2200~Å and UV4: 1950~Å \citep{Quimby2018}. While these absorption features are typically fairly strong, only UV1 is clearly detected for SN\,2018bsz, with possible detections of UV2 and UV3. In this regard the spectrum looks very similar to the HST spectrum of PTF12dam \citep{Quimby2018} -- albeit the features are found to be bluer in SN\,2018bsz. The nature of the marked UV features is still under debate and several combinations of ions have been suggested to be the cause. The commonly discussed identifications are \ion{Mg}{II} (UV1), \ion{Si}{III} (UV2) and \ion{C}{II} (UV3) suggested by \citet{Quimby2011} and \ion{C}{II} + \ion{Mg}{II} (UV1), \ion{C}{II} (UV2); \ion{C}{III} + \ion{C}{II} (UV3) and Fe~III (UV4) presented by \citet{Howell2013}. As discussed above, SN\,2018bsz has strong \ion{C}{II} features in the optical so it would not be surprising to see them in the NUV as well. The UV2 and UV3 lines have been at least partially attributed to \ion{C}{II} by either \citet{Quimby2011} or \citet{Howell2013} but these features are faint in SN\,2018bsz. This could imply that \ion{C}{II} does not contribute significantly to these features, possibly promoting the alternative identifications. This could also imply that UV1 is caused by \ion{Mg}{II} rather than a blend with C~II. However, the STIS spectrum was obtained at a relatively late epoch, when lines of \ion{C}{II} were also weak in the optical. Not many NUV spectra of SLSNe-I are available at later epochs \citep[see e.g. ][]{Quimby2018}. In fact, the shown spectrum of PTF12dam is one of the latest ones and at earlier phases the four UV absorption features were clearly visible for PTF12dam \citep{Quimby2018}. It is therefore possible that SN\,2018bsz exhibits \say{typical} evolution for SLSNe-I and that all four UV dips were present at earlier phases. 

In Fig. \ref{fig:COS_spec} we show the two epochs of HST/COS spectra taken at 26.3~d and 50.2~d post-maximum. Dominant features in this wavelength ranges are geocoronal (airglow) lines, most notably Ly$\alpha$ $\lambda1216$ and \ion{O}{I} $\lambda1302,1306$. We also identify a faint Ly$\alpha$ absorption feature at the redshift of the host galaxy present at both epochs. As such we associate it with the galaxy. No Ly$\alpha$ emission is visible in either of the spectra despite the prominent Balmer lines at a comparable epochs.

\section{Comparison to known SNe}
\label{sec:SN_comparison}
\subsection{SESNe \& SLSNe}

\label{subsec:spec_comp_SLSN}

\begin{figure*}
     \centering
     \begin{subfigure}[b]{0.492\textwidth}
         \centering
         \includegraphics[width=\textwidth]{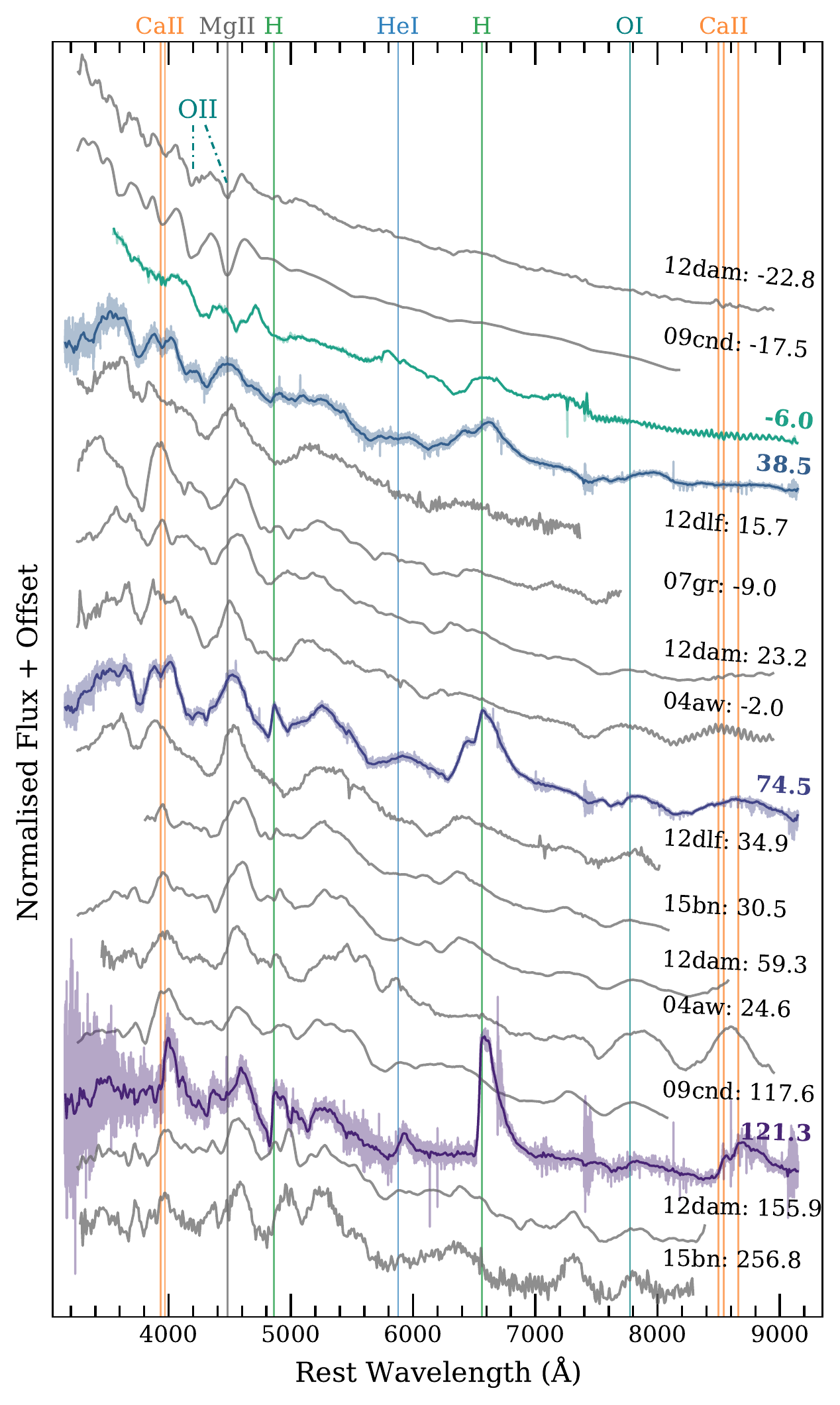}
     \end{subfigure} %
    ~
     \begin{subfigure}[b]{0.492\textwidth}
         \centering
         \includegraphics[width=\textwidth]{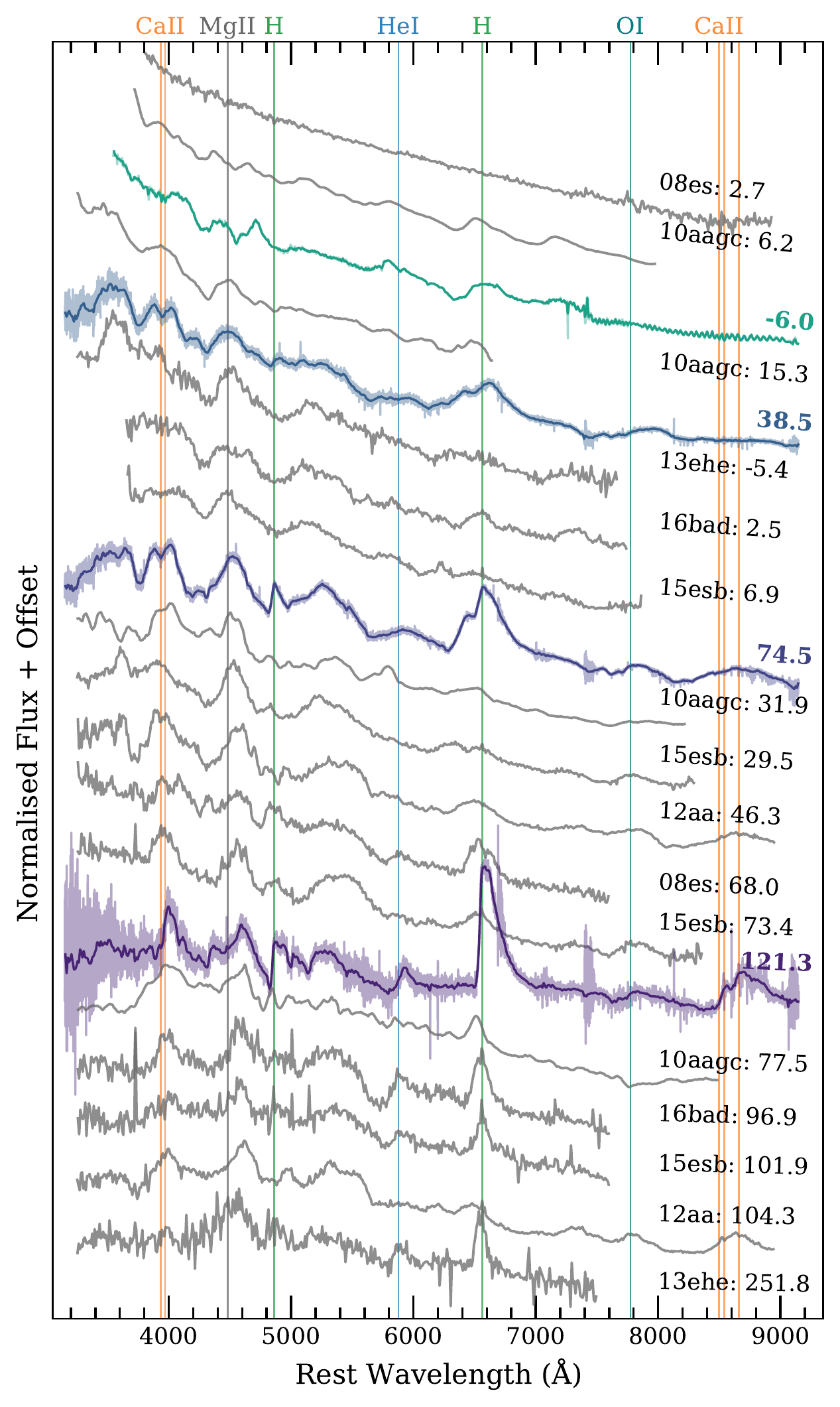}
    \end{subfigure}
     \vspace*{-10pt}
    \caption{SN\,2018bsz spectral evolution in comparison to literature SESNe and SLSNe-I (left) and in comparison to SLSNe-I with detected H$\alpha$ emission and SLSN-II 2008es (right). All spectra have been binned to 10~Å, but unbinned spectra are shown for SN\,2018bsz. SN\,2018bsz is spectroscopically similar to SESNe and SLSN-I, but the Balmer lines are unique even for SLSNe-I with late H emission. Note that the literature spectra have been mangled to have the same color as SN\,2018bsz at a relevant epoch to ease the comparison. The literature spectra were first published in the following papers: SN\,2004aw \citep{Taubenberger2006a}, SN\,2007gr \citep{Valenti2008}, SN\,2008es \citep{Miller2009},PTF09cnd \citep[][see also \citet{Quimby2018}]{Quimby2011}, PTF10aagc \citep{Quimby2018}, PTF12dam \citep{Quimby2018}, LSQ12dlf \citep{Nicholl2014},  SN\,2012aa \citep{Roy2016}, iPTF13ehe \citep{Yan2015}, SN\,2015bn \citep{Nicholl2016a} and iPTF15esb and iPTF16bad \citep{Yan2017}, and the data was downloaded from the \textit{Open Supernova Catalog} \citep{Guillochon2017} and WISeREP \citep{Yaron2012}.}
    \label{fig:SLSN_spectral_comp}
\end{figure*}

As discussed in Sect. \ref{sec:spectroscopy}, the spectra of SN\,2018bsz exhibit several features commonly seen in SESNe and SLSNe. The similarity has been further highlighted in Fig. \ref{fig:SLSN_spectral_comp} where four epochs of SN\,2018bsz are shown with selected SESNe and SLSNe-I demonstrating typical photospheric evolution for the classes. At early epochs SN\,2018bsz resembles SLSNe-I with prominent \ion{O}{II} and \ion{C}{II} features. As mentioned by \citet{Anderson2018}, the \ion{O}{II} features in most SLSNe -- such as PTF09cnd and PTF12dam -- are found at higher velocity when compared to SN\,2018bsz. On the other hand, while \ion{C}{II} lines are not seen in all Type~I SLSNe, they have been reported PT09cnd and PTF12dam shown in the figure \citep{Quimby2018}. 

The spectra of SN\,2018bsz at 38.5~d and 74.5~d are very similar to SESNe pre-peak spectra -- apart from the prominent Balmer lines. \ion{Ca}{II} H\&K absorption, \ion{Mg}{II} $\lambda4481$, \ion{O}{I} $\lambda7774$ and \ion{Fe}{II}  emission centred at $\sim5200$~Å seen in SN\,2018bsz are typical in Type~Ic SNe as demonstrated with SN\,2007gr \citep{Valenti2008} and SN\,2004aw \citep{Taubenberger2006a}. Due to these dominant ejecta lines SN\,2018bsz resembles Type Ic SNe but with a delay as is typical for Type~I SLSNe \citep{Pastorello2010}. Thus the spectra of SN\,2015bn, LSQ12dlf, PTF12dam and PT09cnd all demonstrate remarkable similarity to SN\,2018bsz at comparable epochs as expected. A notable feature to highlight is the emission line near the blue H$\alpha$ component of SN~2018bsz seen in the SLSNe spectra at 30 -- 60~d post-peak. In LSQ12dlf the feature was identified as \ion{Si}{II} $\lambda6355$ \citep{Nicholl2014} while for SN~2015bn it has been discussed as  [\ion{O}{I}] $\lambda\lambda6300, 6364$ \citep{Nicholl2016a}. PTF12dam also exhibits a very similar feature but \citet{Quimby2018} did not  provide an identification after excluding H$\alpha$ due to its apparent blueshift ($\sim-6000$ km/s) and [\ion{O}{I}] as the feature was redshifted relative to that line. Regardless of the nature of the feature, it bears striking similarity to the blue shoulder of H$\alpha$ in SN\,2018bsz. However, the shoulder is centred at $\sim6400$~Å at the first epoch it is visible (32.7~d). This corresponds to a redshift of $\sim2800$~km/s (as measured from the \ion{Si}{II} line) and the line moves redder in time until at 107.6~d it is redshifted by $\sim5700$~km/s. Thus it does not seem plausible to assume that the shoulder is related to either \ion{Si}{II} or [\ion{O}{I}] emission, despite the  similarity. 

Based on the spectral evolution of the ejecta lines we confirm that SN\,2018bsz is a Type~I SLSN, but with strong hydrogen emission. While the presence of hydrogen signatures typically excludes a SN from being classified as Type I, four similar SLSNe-I with late hydrogen emission attributed to CSM interaction have been detected before. In Fig. \ref{fig:SLSN_spectral_comp} we compare the spectral timeseries of these four SNe to SN\,2018bsz. First in PTF10aagc the symmetric Balmer lines first appear at $77.5$~d post-peak and they are found to be blueshifted by $\sim-2000$~km/s \citep{Quimby2018}. Furthermore, \citet{Yan2017} presented spectroscopic data of three Type~I SLSNe -- iPTF13ehe (published earlier in \citealt{Yan2015}, iPTF15esb and iPTF16bad -- with broad, late-time H$\alpha$ first detected at 251~d, 73~d and 97~d from peak, respectively. We note that for iPTF13ehe and iPTF16bad no spectra were reported between peak brightness and the detection of H$\alpha$ so the time of appearance is unconstrained. As can be seen in the figure, the H$\alpha$ line profiles in these SNe were symmetric and seemed to consist of a single component close to rest frame H$\alpha$. However, in all three SNe the H$\alpha$ appeared to be slightly blueshifted $\lesssim-1000$~km/s and for at least two of them the line became slightly redshifted ($\lesssim500$~km/s) in time \citep{Yan2017}.

These four SNe evolve in a similar manner resembling SLSNe-I but it seems that PTF10aagc has the most in common with SN\,2018bsz. First, the early spectra of PTF10aagc show \ion{O}{II} absorption at similar velocities to SN\,2018bsz. The SN also exhibits strong \ion{C}{II} emission lines similar to those seen in SN\,2018bsz. Secondly, while the Balmer emission line do not have similar profiles, the relatively high blueshift seen in PTF10aagc is comparable to the blue H$\alpha$ component of SN\,2018bsz at the later epochs. In comparison, the three iPTF SNe have less in common with SN\,2018bsz. While their evolution is broadly speaking similar to SLSNe-I, at early epochs it appears to be faster. The SNe exhibit clear Type~Ic SN-like spectra at the time of peak brightness -- behaviour not typical for SLSNe-I. Consequently, none of the SNe exhibit \ion{O}{II} absorption and only one of them (iPTF16bad) has clear \ion{C}{II} lines in the early spectra. However, as iPTF15esb and iPTF16bad were discovered close to peak and iPTF13ehe had its first spectrum taken at $-\sim9$~d, it is possible that the lines had simply faded by the time of the first spectra. In the later evolution the three iPTF SLSNe resemble SN\,2018bsz, but the hydrogen emission is found at low velocities in comparison.

In Fig. \ref{fig:SLSN_spectral_comp} we also show two spectra of Type Ic SN\,2012aa discussed exhibiting broad late time H$\alpha$ emission \citep{Roy2016}. While classified as Type Ic, its peak luminosity ($M_V\sim-20$) is similar to SN\,2018bsz \citep[$M\sim-20.5$;][]{Anderson2018} warranting a comparison. In SN\,2012aa the H$\alpha$ emission first appears at $\sim47$~d and it is found at a constant blueshift of $\sim-2000$~km/s \citep{Roy2016}. While thus similar to PTF10aagc, SN\,2012aa does not exhibit either \ion{O}{II} or \ion{C}{II} features in its early spectra albeit the first spectrum is taken at $+8$~d and the features could have already faded.

In addition, SN\,2018bsz shares common characteristics with Type II SLSNe, especially with SN\,2008es \citep{Gezari2009,Miller2009} as shown in Fig. \ref{fig:SLSN_spectral_comp}. While its early spectrum at $\sim3$~d post-max is featureless blue continuum, at $\sim68$~d SN\,2008es is similar to SN\,2018bsz and the other SLSNe-I shown in the figure as the SN exhibits both typical SLSN-I features as well as broad Balmer emission lines. The other members of SLSNe-II, CSS121015:004244+132827 \citep{Benetti2014}, SN\,2013hx and PS15br \citep{Inserra2018}, evolve in a similar manner. The key difference between SLSNe-I with late H emission and SLSNe-II seems to be that for SLSNe-II H emission appears together with the other line features \citep[see e.g.][]{Gezari2008, Miller2009, Benetti2014, Inserra2018}, while for the SLSNe-I there is a definite delay. Furthermore, SN\,2013hx and PS15br show multi-component H$\alpha$ emission at nebular phase \citep{Inserra2018}. For SN\,2013hx three components -- blue, central and red at $-4700$, $-190$ and $+4000$~km/s, respectively -- are identified, while for PS15br only blue ($-4700$~km/s) and central ($-390$~km/s) are seen. Given these similar characteristics it seems possible that SLSNe-II and SLSNe-I with late, broad Balmer emission belong to the same population of stellar explosions as already indicated by similar, extreme host properties \citep{Schulze2018}.

Regardless of the observed differences, SN\,2018bsz, PTF10aagc, iPTF13ehe, iPTF15esb and iPTF16bad are members of rare subclass of SLSNe-I, characterised by broad hydrogen emission after peak brightness.  Given that the hydrogen emission is likely arising in external material, it is plausible that the progenitor systems of the SNe are similar. The differences between the observables could then be explained by differences in the geometry of where the hydrogen is located with respect to the progenitor. The similarities of these SLSNe-I are further discussed in Sect. \ref{subsec:implications}

\subsection{Type IIn SNe}
\label{subsec:spec_comp_IIn}
\begin{figure}
    \centering
    \includegraphics[width=0.48\textwidth]{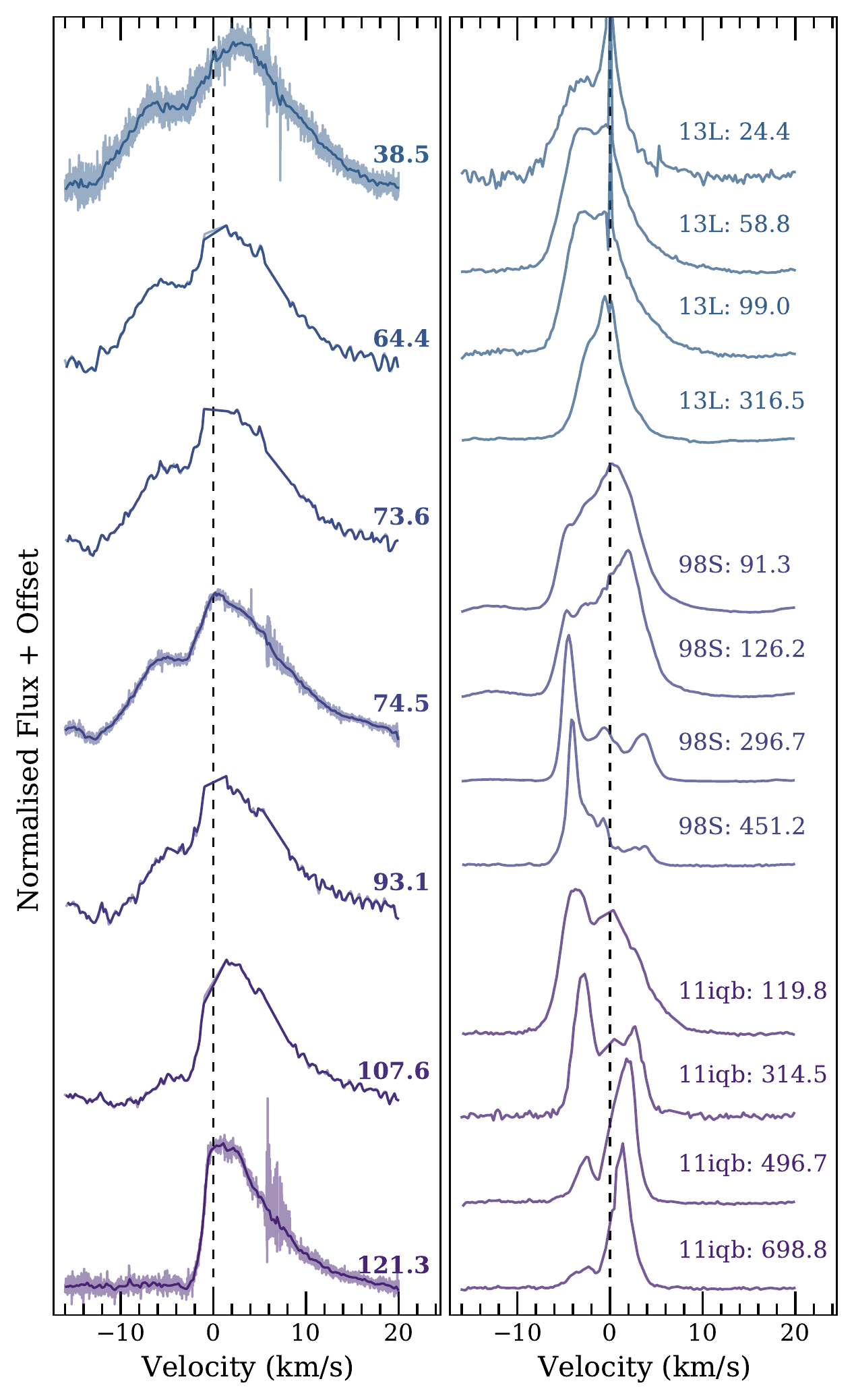}
    \caption{Spectral evolution of the H$\alpha$ region of SN\,2018bsz (left) and the Type IIn SNe 2013L, 1998S and PTF11iqb (right). The H$\alpha$ has a multi-component profile similar to the shown literature SNe IIn. All spectra have been binned to 5~Å, but unbinned spectra are shown for SN\,2018bsz. The literature spectra were first published in the following papers: SN\,1998S \citep{Leonard2000, Pozzo2004}, PTF11iqb \citep{Smith2015} and SN\,2013L \citep{Andrews2017} and the data was downloaded from the \textit{Open Supernova Catalog} \citep{Guillochon2017} and WISeREP \citep{Yaron2012}.}
    \label{fig:Ha_comp}
\end{figure}

While the spectral timeseries of SN\,2018bsz as a whole clearly resembles Type~I SLSNe, the peculiar evolution of H$\alpha$ is not similar to even those few SLSNe-I with hydrogen. Instead, such line evolution has been observed in three Type IIn SNe attributed to CSM interaction. To highlight the similarity, we present the H$\alpha$ evolution seen in SN\,2018bsz in comparison to SN\,2013L \citep{Andrews2017, Taddia2020}, SN\,1998S \citep{Leonard2000, Fassia2000} and PTF11iqb \citep{Smith2015} in Fig. \ref{fig:Ha_comp}. However, we note that similar line profiles have also been seen in other types of SNe e.g. Type IIP SNe 2004dj \citep{Vinko2006, Chugai2007}, 2007od \citep{Andrews2010} and 2011ja \cite{Andrews2016} and Type IIb SN\,1993J \citep{Matheson2000b, Matheson2000} likewise associated with CSM interaction, but the lines are more distinct in the Type IIn SNe.

The three Type IIn SNe show multiple broad components of H$\alpha$. While in SN\,2013L there appears to be only central and blueshifted lines during the evolution, SN\,1998S and PTF11iqb have both blue and redshifted components along with a central one. These components evolve differently. In SN\,2013L the multi-component profile arises early $\sim20$~d after peak brightness. To begin with the two components are roughly equally strong, but in time the blue one becomes weaker and narrower while the line velocity also decreases \citep{Andrews2017, Taddia2020}. In SN\,1998S and PTF11iqb on the other hand, the clear multi-component profile becomes visible only at $\sim100$~d after peak. While only central and blueshifted lines are present to begin with, both SNe develop a distinct redshifted component in time: for SN\,1998S it is clearly visible at $\sim120$~d \citep{Leonard2000} and for PTF11iqb at $\sim200$~--~$300$~d \citep{Smith2015}. The evolution of the multi-component profiles mirror each other. As can be seen in Fig. \ref{fig:Ha_comp}, in SN\,1998S the dominant component changes from red to blue, while in PTF11iqb the change is opposite. Otherwise the profiles appear to be very similar to each other and by the time of the last shown spectra the weaker component is barely visible. 

When the spectral properties of the whole optical range are considered, the evolution of SN~1998S and PTF11iqb appears to be nearly identical apart from the evolution of the Balmer lines as demonstrated by \citet{Smith2015}. Both exhibit only narrow Lorentzian lines over blue continua in the early spectra as expected of Type IIn. After $\sim20$~d post-discovery the spectra evolve to strongly resemble that of normal Type II SNe with no clear CSM signatures. However, at about $\sim100$~d after peak, the multi-component Balmer lines appear. First they are visible together with the Type II ejecta lines, but eventually only the lines arising from the CSM persist. SN\,2013L on the other hand evolves a bit differently. Its spectra exhibit only emission features with profiles similar to that of H$\alpha$. For instance the \ion{Ca}{II} NIR triplet can be nicely explained by a combination of several H$\alpha$ line profiles -- similarly to SN\,2018bsz (see Fig. \ref{fig:CaII_NIR_fits}). 

The overall evolution of the H$\alpha$ in SN\,2018bsz is reminiscent of that in the shown Type IIn SNe, with the exception that in SN\,2018bsz the line evolves faster in time and has broader components found at higher velocities. SN\,2013L has a profile similar to SN\,2018bsz with a blue shoulder becoming weaker and moving redward in time. The main difference is the complete lack of the redshifted component. Similarly, despite the early evolution of PTF11iqb being different, the last two shown epochs have a very similar profile to SN\,2018bsz at $107.6$~d with a strong redshifted peak and weak blueshifted one. Furthermore, both blue and redshifted peaks appear to be shifting to lower velocities during the evolution. The key difference is now the lack of visible central component which instead was prominent at the earlier epochs. On the other hand, SN\,1998S had a similar profile at the early epochs with a strong redshifted peak and a fainter blueshifted one, but at later times the profile evolves very differently to SN\,2018bsz.

Given the common trends between the IIn SNe and SN\,2018bsz it does seem reasonable to assume that the mechanism that generates the profiles is similar in nature. The differences in timescales and velocities can be plausibly explained as higher velocity naturally means faster evolution. The other differences could indicate differences in geometry of the external material. One such difference is that the central component appears to always be present from the beginning in the three literature SNe, but not in SN\,2018bsz. Finally, given the spectral similarity of SN\,2018bsz with SLSNe-I and the three Type IIn SNe, we consider that the spectra of SN\,2018bsz are ejecta-dominated at $<25$~d while after the appearance of the strong H$\alpha$ emission they are CSM-dominated. The resulting implications to the CSM structure will be discussed in Sect. \ref{subsec:CSM_conf}.

\section{Spectropolarimetry}
\label{sec:spectropolarimetry}

\begin{figure*}
    \centering
    \includegraphics[width=0.98\textwidth]{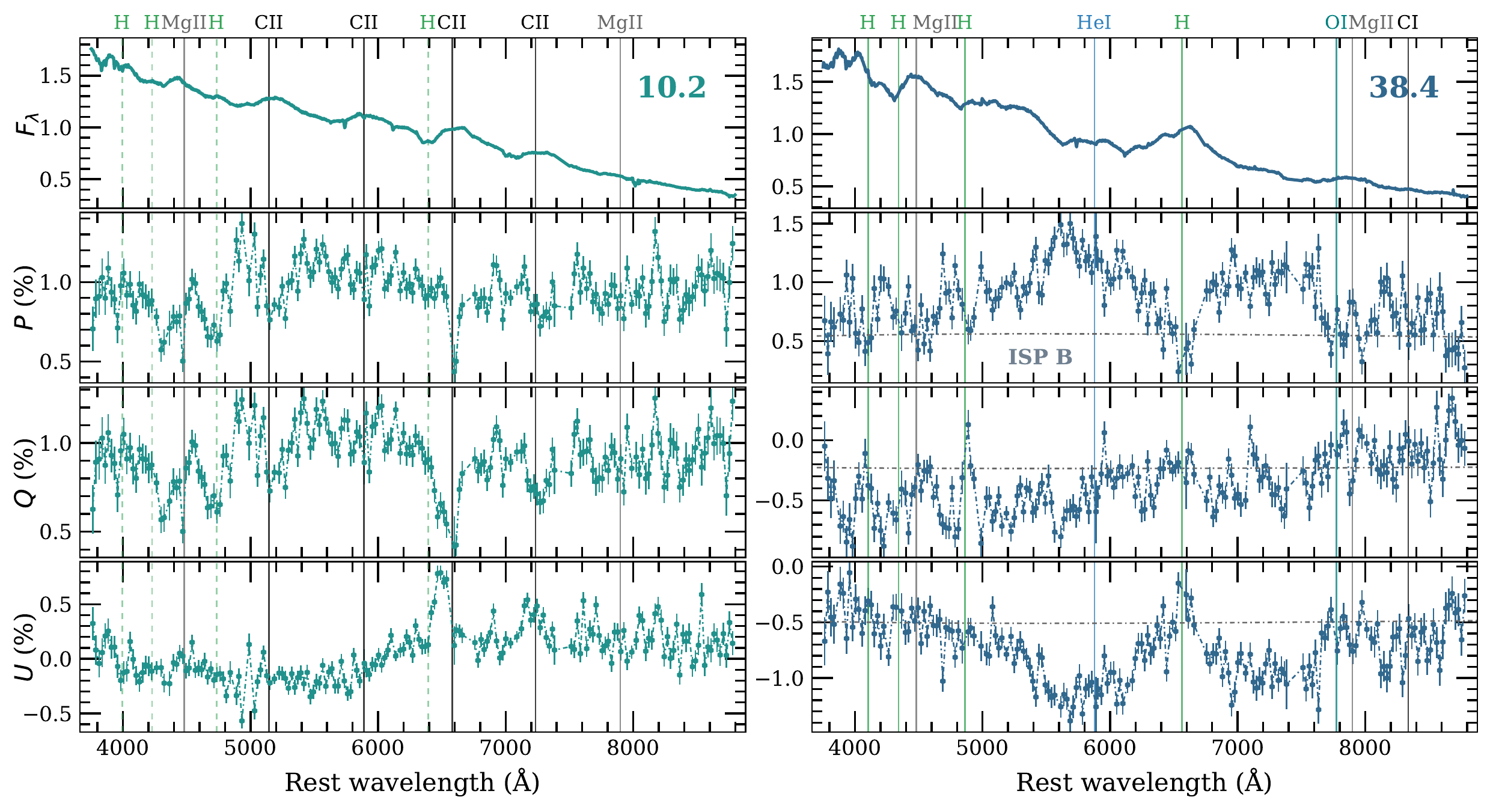}
    \caption{Flux spectrum, polarisation degree $P$, and normalised Stokes $Q$ and $U$ parameters of SN\,2018bsz at 10.2~d (left) and 38.4~d (right). The polarisation and Stokes spectra have been binned to 25~Å but the flux spectra are unbinned. Vertical lines show the location of major spectroscopic features. Note that at $10.2$~d hydrogen lines are shown at blueshift of $-8000$~km/s. The dash-dotted lines in the right panels present the estimation for interstellar polarisation (ISP) in our case B (ISP B), where it has been assumed that the strongest emission lines completely depolarise the spectrum (see Sect. \ref{subsec:ISP}). The data have not been corrected for the ISP.}
    \label{fig:fqupx}
\end{figure*}

\begin{figure}
    \centering
    \includegraphics[width=0.48\textwidth]{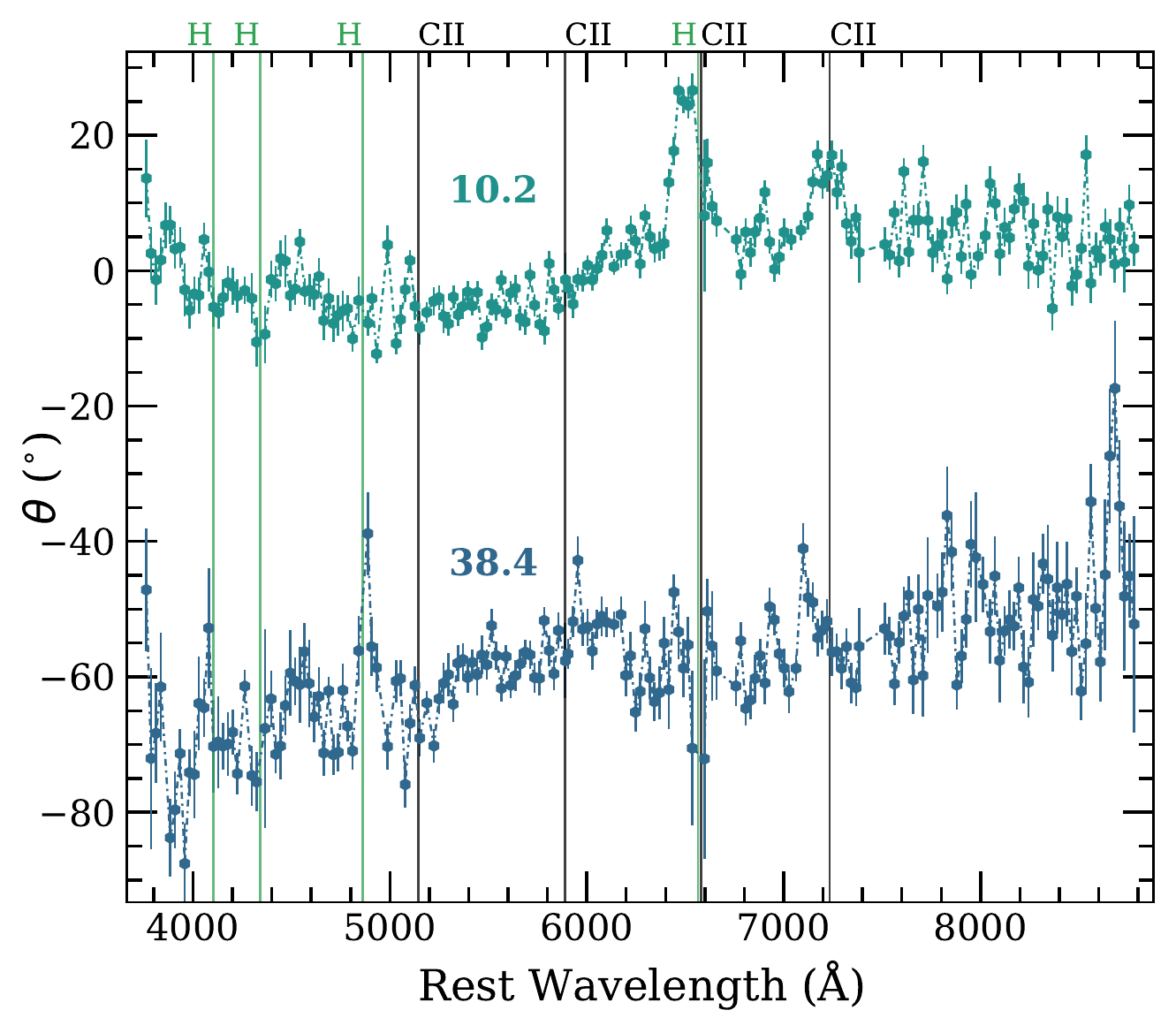}
    \caption{Polarisation angle $\theta$ for SN\,2018bsz at 10.2~d (green) and 38.4~d (blue). The values have been binned to 25~Å. There is a large change in the polarisation angle between the two epochs, corresponding to an average rotation of $\sim 60^{\circ}$. It is worth noting that the polarisation angle at 10.2 days is almost constant with wavelength, with the exception of the \ion{C}{II}/H$\alpha$ region.} 
    \label{fig:theta}
\end{figure}

\begin{figure*}
    \centering
    \includegraphics[width=0.98\textwidth]{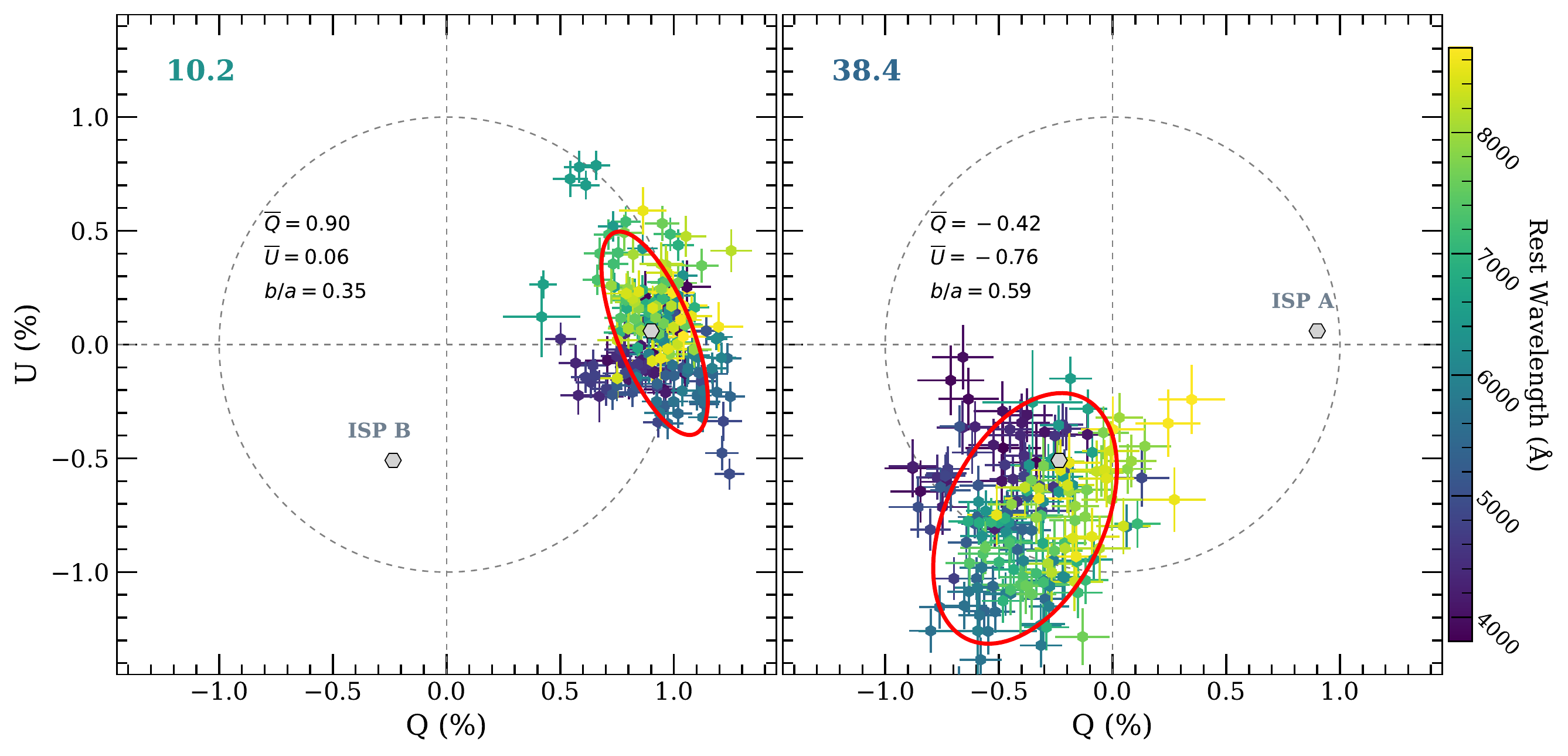}
    \caption{The location of SN\,2018bsz on the $Q$ -- $U$ plane at 10.2~d (left) and 38.4~d (right). 
    The individual points are coloured according to their wavelength as indicated in the colorbar. Thin dashed lines have been drawn at $Q = 0$, $U=0$ and $P=1$ to guide the eye. The two ellipses in red are the outcome of a principal component analysis \citep{Maund2010} where the major axis of the ellipse is aligned with the direction of the maximum variance of the data and the axial ratio $b/a$ parameterizes the ratio of polarisation carried by the orthogonal to the dominant direction (minor to major axis). The locations of two alternative ISP solutions that have been examined (ISP A and ISP B; see Sect. \ref{subsec:ISP}) have been marked with grey hexagons, but the data  has not been corrected for any ISP. There is a large change in the SN polarisation between the two epochs shown the large shift in the barycentre ($\tilde{Q}$, $\tilde{U}$), which is independent of the ISP. Such large changes on the $Q$ -- $U$ plane have rarely been observed for SN explosions. }
    \label{fig:qu_plane}
\end{figure*}

\begin{figure}
    \centering
    \includegraphics[width=0.48\textwidth]{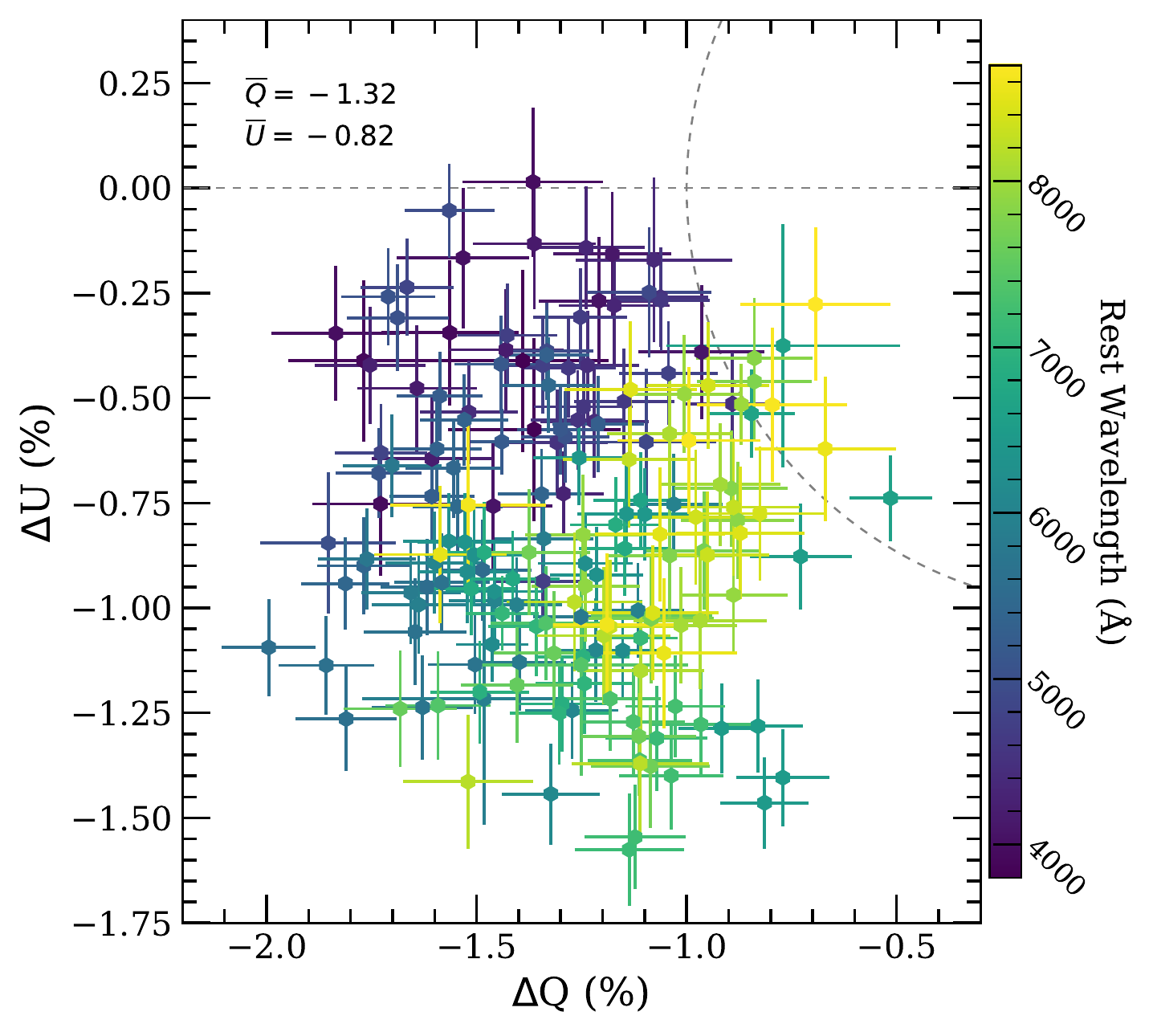}
    \caption{The location of SN\,2018bsz on the $\Delta Q$ -- $\Delta U$ plane, where $\Delta Q = Q_{38.4}-Q_{10.2}$ and $\Delta U = U_{38.4}-U_{10.2}$. As the change is ISP-independent the values demonstrate the true change of the polarisation. Barycenter of the shift is found at $\tilde{Q}=-1.32$ and $\tilde{U}=-0.82$.} 
    \label{fig:dqdu}
\end{figure}

Spectropolarimetry is a powerful tool to investigate the geometric structure of SN explosions. For SNe the source of continuum polarisation is assumed to be Thomson scattering from free electrons abundant in the SN ejecta especially during the photospheric phase \citep[see e.g. ][]{Hoflich1991}. In case of a perfectly spherical photosphere the net polarisation of the SN is zero as the light is linearly polarised equally in all directions. On the contrary, deviations from a spherical photosphere produce a non-zero polarisation. The geometry of many SN explosions has been studied  with the aid of spectropolarimetry -- including the famous Type\,II SN\,1987A \citep[see e.g.][]{Schwarz1987, Jeffery1987, Cropper1988} and Type IIb SN\,1993J \citep{Trammell1993, Tran1997, Stevance2020}. A comprehensive review of SN spectropolarimetry is provided by \cite{Patat2017}.

Two epochs of spectropolarimetry were obtained with FORS2 at 10.2 and 38.4~days post-maximum brightness. We are therefore fortunate to have a snapshot of both the ejecta-dominated (\ion{C}{II}-dominated; $\lesssim25$~d; Fig. \ref{fig:CII_timeseries}) and the CSM-dominated phases (H$\alpha$-dominated; $\gtrsim25$~d; Fig. \ref{fig:Balmer_timeseries}) of SN\,2018bsz. Our spectropolarimetry is shown in Fig.~\ref{fig:fqupx}, where we plot the flux spectrum, the polarisation spectrum, and the normalised Stokes parameters $Q$ and $U$. The polarisation angle $\theta$ has been plotted separately in Fig.~\ref{fig:theta} to facilitate comparison between the two epochs. In addition, Fig. \ref{fig:qu_plane} shows the spectropolarimetric measurements on the Stokes $Q$ -- $U$ plane. 

Figure~\ref{fig:fqupx} shows that there has been a very significant evolution in the polarisation properties of SN\,2018bsz between the two epochs confirming a radical change in the SN and its projected geometry during these four weeks. At $+$10.2 days the barycenter of the data in the Stokes plane is found at $\tilde{Q}=0.90$\%, $\tilde{U}=0.06$\%, where $\tilde{Q}= \sum_{i} (Q_i /\delta Q_i^{2}) / \sum_{i} (1 /\delta Q_i^{2})$ is a weighted mean, with $Q_i$ and error $\delta Q_i$ referring to the i-th wavelength bin  (and similar for $\tilde{U}$). At $+$38.4 days, the barycenter has moved to $\tilde{Q}=-0.42$\%, $\tilde{U}=-0.76$\%, manifesting a large shift. To further quantify the evolution we provide $\Delta Q$ -- $\Delta U$ plane in Fig. \ref{fig:dqdu}, where $\Delta Q = Q_{38.4}-Q_{10.2}$ and $\Delta U = U_{38.4}-U_{10.2}$. The barycenter of the change between the two epochs is found at $\Delta\tilde{Q}=-1.32$ and $\Delta\tilde{U}=-0.82$. As the change is independent of the interstellar polarisation (ISP) contribution, it provides a measurement of the actual polarisation shift.

We have searched the literature for previous core-collapse supernovae with multi-epoch spectropolarimetry and we have only identified SN\,2001ig \citep{Maund2007a}, a Type IIb SN, as potentially showing such a large change in the loci of the data on the $Q$ -- $U$ plane with time. This is also reflected in Fig.~\ref{fig:theta} that shows that the polarisation angle changed by about 60$^\circ$. However, we note $\theta$ depends strongly on the ISP correction and the exact values should be taken with caution. Following the methodology of \citet{Maund2010}, we performed a principal component analysis in order to estimate the direction of maximum variance of the data, illustrated by the direction of the major axis of an ellipse on the $Q$ -- $U$ plane. In addition, the axial ratio $b/a$ of the ellipse (minor over major axis) parameterizes the ratio of polarisation carried by the dominant and the orthogonal direction. A theoretical ratio of $b/a = 0$ would mean that all polarisation is carried by a dominant axis, corresponding to a perfectly axial symmetric geometry \citep{Wang2008}. These ellipses have been drawn on Fig. \ref{fig:qu_plane}, where it can be seen that both their origin, rotation angle and axial ratio has changed.

\begin{figure}
    \centering
    \includegraphics[width=0.48\textwidth]{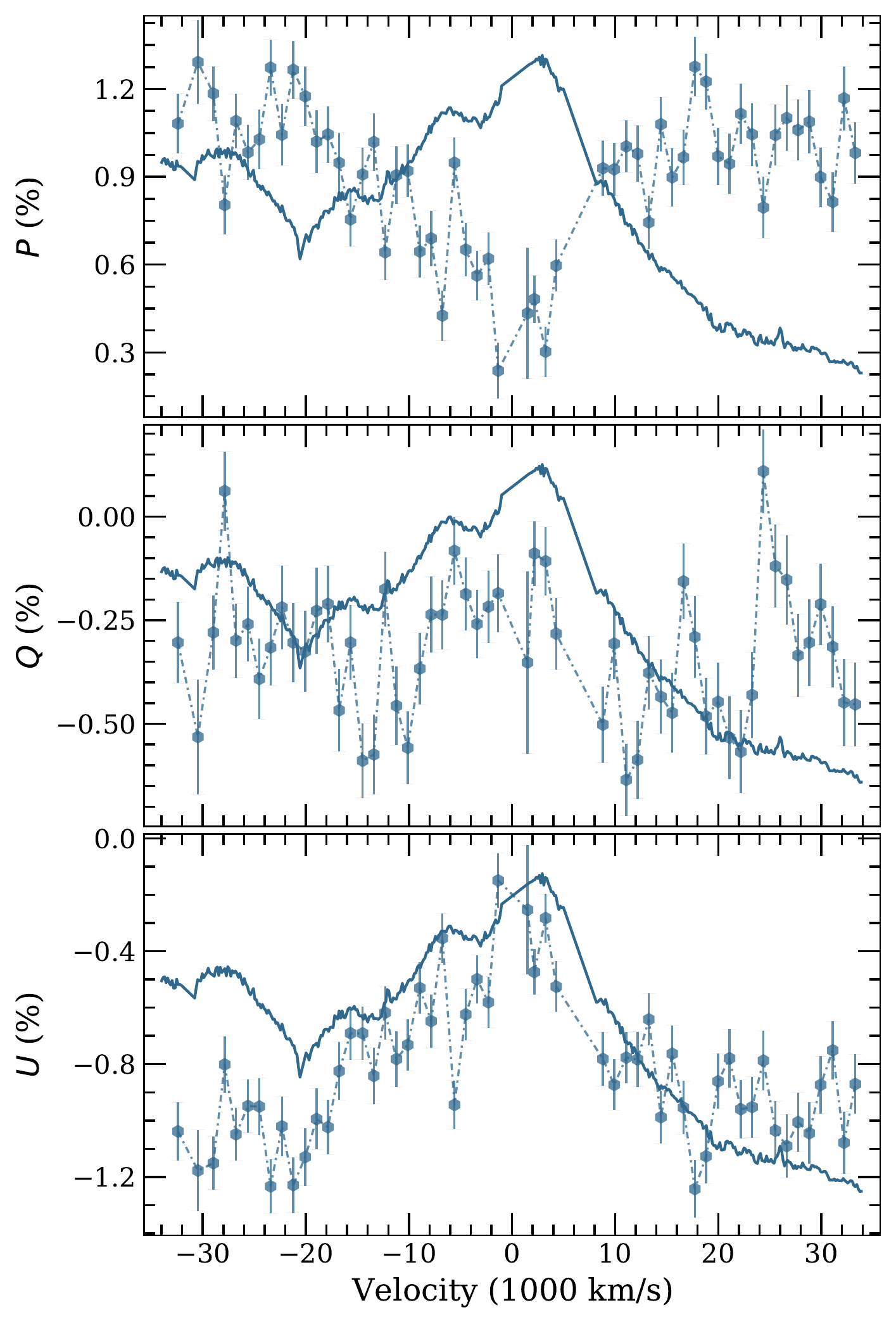}
    \caption{$P$, $Q$ and $U$ for the H$\alpha$ line at 38.4~d in velocity space. Scaled H$\alpha$ profiles are shown for clarity. A strong depolarisation occurs across the emission line.}
    \label{fig:38.4_Ha_pux}
\end{figure}

A more detailed look at Figure \ref{fig:fqupx} reveals further differences between the polarisation properties of the two spectra. The most noticeable is the strong depolarisation at the location of the complex H$\alpha$ line at day $+$38.4. This is typical for strong emission lines, but the effect is only mild at $+$10.2 days, confirming once more that the possible contribution from H$\alpha$ is limited at these phases. In addition, we see a similar depolarising effect at the location of O~I and the developing \ion{Ca}{II} IR triplet at the red edge of the FORS2 spectrum (see Sect.~\ref{subsec:CaII_lines}), which is again not seen at $+$10.2 days. The H$\alpha$ profile at $+$38.4 days is shown in more detail in Fig. \ref{fig:38.4_Ha_pux} in velocity space. The depolarisation effect seems to extend to $v\sim-20000$~km/s, mirroring the emission profile.

The blue part of the spectrum is depolarised in both epochs by the presence of multiple lines. The average continuum polarisation, best measured between 5400 -- 6200~\AA\ and above 7500~\AA\ for the first epoch, is 1.0\% $\pm$ 0.1\% (standard deviation) at $+$10.2 days. It is a bit higher at $+$38.4 days, reaching 1.2\% $\pm$ 0.1\%, although these absolute values depend on the ISP. To demonstrate the effect of the ISP correction, the ISP corrected polarisation spectra, normalised Stokes parameters $Q$ and $U$ as well as the polarisation angles are shown in Fig. \ref{fig:FPQUX_ISP_corr} for both ISP solutions.

\subsection{Interstellar polarisation: alternatives and implications}
\label{subsec:ISP}

Interstellar polarisation (ISP) is caused by dust grains along the line of sight and can significantly affect the polarimetric signature of a SN. Its presence is a constant nuisance for the determination of the intrinsic SN~ polarisation as there is no unique and unambiguous way to estimate and determine it. \citet{Stevance2020} provides an educative summary of methods that have been employed in the past to estimate and remove ISP from observations of SNe. There is no single method that is 100\% reliable in our case. We therefore focus quantitatively on two alternatives for the ISP, neither of which we regard entirely convincing but they can be considered as two limiting cases. These two alternatives are: ISP A) the SN is almost unpolarised (spherical) in the first epoch, which means that the ISP is responsible for the bulk of the observed polarisation at 10.2 days; and ISP B) that the strongest emission line (H$\alpha$) observed in the second epoch completely depolarises the spectrum and the observed polarisation is caused purely by the ISP. First, however, we discuss some general considerations that constrain the ISP. 

SN\,2018bsz is found at a relatively low Galactic latitude of $+14^{\circ}$, resulting at moderate extinction along the line of sight of $E(B-V) = 0.214$ \citep{Schlafly2011}. Using $P_{\mathrm{ISP}} < 9 \times E(B-V)$ \citep{Serkowski1975} we obtain that the maximum ISP contribution from the Galaxy could be up to 1.93\%, i.e. quite significant and not particularly constraining. Unfortunately, there are no sufficient nearby stars in the Heiles catalogue \citep{Heiles2000} that can be used to draw reliable constraints on the Galactic ISP. We only find one star within $3^{\circ}$ with a reported $P \sim 1.2$\% but the angular distance from SN\,2018bsz is already large. A second star is found within $5^{\circ}$ and this time $P \sim 1.9$\%. Statistics only become possible when increasing the search radius to an angular distance of $6^{\circ}$, but now 3/6 stars are consistent with negligible polarisation, while the other 3 present a large spread in their values. Furthermore, all reported polarisation angles are different and we therefore consider this test quite inconclusive. However, it does show that the Galactic ISP could be significant towards SN\,2018bsz, possibly of the same order of magnitude that we measure. In addition, there could be dust within the host galaxy of SN\,2018bsz. \cite{Chen2021} provide an extensive discussion on the subject, examining values ranging from  $E(B-V)_{\mathrm{host}} = 0.04$ from Na I D absorption  \citep{Anderson2018} to $E(B-V)_{\mathrm{host}} = 0.32$ from the Balmer decrement at the host galaxy. \cite{Chen2021} favour the lower values $E(B-V)_{\mathrm{host}} = 0.04 - 0.10$ in their analysis, consistent with the color temperature of the SN compared to other SLSNe. Irrespective, supposing that the Serkowski relation \citep{Serkowski1973} also applies to the host of SN\,2018bsz, there  could also be a significant ISP contribution from the host. Of course, we cannot simply add the polarisation degrees $P_{\mathrm{ISP}}$ from the Milky Way and the host, as they can even cancel out depending on their polarisation angles. 

Another consideration on the ISP results from the fact that the observed polarisation angle at 10.2 days is almost constant with wavelength, if we ignore the \ion{C}{II}\,/\,H$\alpha$ region (Fig.~\ref{fig:theta}). Considering only the blue part of the spectrum ($3800 < \lambda < 6150$) \AA, we get a (weighted) mean polarisation angle of $\theta = -3.5^{\circ} \pm 4.4^{\circ}$ (standard deviation). Considering also the red part of the spectrum (without \ion{C}{II}\,/\,H$\alpha$) we have $\theta = -1.6^{\circ} \pm 6.0^{\circ}$. Since the observed polarisation is the superposition of multiple components (the SN intrinsic polarisation and the ISP, which in turn potentially consists of more components), this can be used to derive some constraints on their relative position on the $Q$ -- $U$ plane. Two possibilities exist for the total ISP: i) the intrinsic polarisation of the SN at this epoch is approximately zero (the SN is almost spherical) and the bulk of the measured polarisation is due to the ISP, which has a location approximately consistent with the barycenter of the data on the $Q$ -- $U$ plane (this is the same as alternative A above); ii) the ISP has to lie on a \say{special} location on the $Q$ -- $U$ plane relative to the SN intrinsic polarisation, as for most random locations the observed polarisation angle would vary with wavelength, unless if the intrinsic polarisation of the SN varied with wavelength in such a way that, when added to the ISP, the wavelength dependence would cancel out. We consider the last combination too contrived \citep[see][for a similar argumentation]{Tanaka2009}. Several such \say{special} locations exist on the $Q$ -- $U$ plane: for example, the ISP could lie along the axis of maximum variance (major axis of the ellipse in Fig. \ref{fig:qu_plane}), or even on the orthogonal direction, as long as it is far enough from the ellipse origin. This argument does not really help us determine the exact value of the ISP, but it does impose some constraints on its expected location. 

We now examine the possibility that the SN is relatively unpolarised at 10.2 days (ISP case A). Except for the fact that the observed polarisation angle is constant with wavelength, this is also motivated by the fact that other SLSNe-I have shown low levels of polarisation around maximum light only increasing later with time \citep{Leloudas2015b,Inserra2016, Leloudas2017a}. For simplicity, to study this case, we set ISP A to be exactly at the location of the barycentre on the Stokes plane ($Q_{\mathrm{ISP}}=0.90$\%, $U_{\mathrm{ISP}}=0.06$\%). This is illustrated with a grey hexagon in Fig. \ref{fig:qu_plane}. Subtracting vectorially this ISP from the data in both epochs, we obtain the intrinsic SN polarisation. Naturally the average $Q$ and $U$ at 10.2 days become equal to zero by construction. The continuum polarisation averaged over 5400~--~6200~\AA\ becomes 0.26\% $\pm$ 0.12\% (a polarisation bias correction has been applied). On the other hand, the distance of the second epoch data to the origin increases, resulting in an increase of the intrinsic polarisation to 1.81\% $\pm$ 0.15\% at 38.4 days. In addition, the polarisation angle is $-73.5^{\circ} \pm 12.5^{\circ}$ (where the large standard deviation is mostly affected by the \ion{C}{II}\,/\,H$\alpha$ region). The spectropolarimetry corrected for ISP A is shown in Fig. \ref{fig:FPQUX_ISP_corr}.

\begin{figure*}
    \centering
    \includegraphics[width=0.98\textwidth]{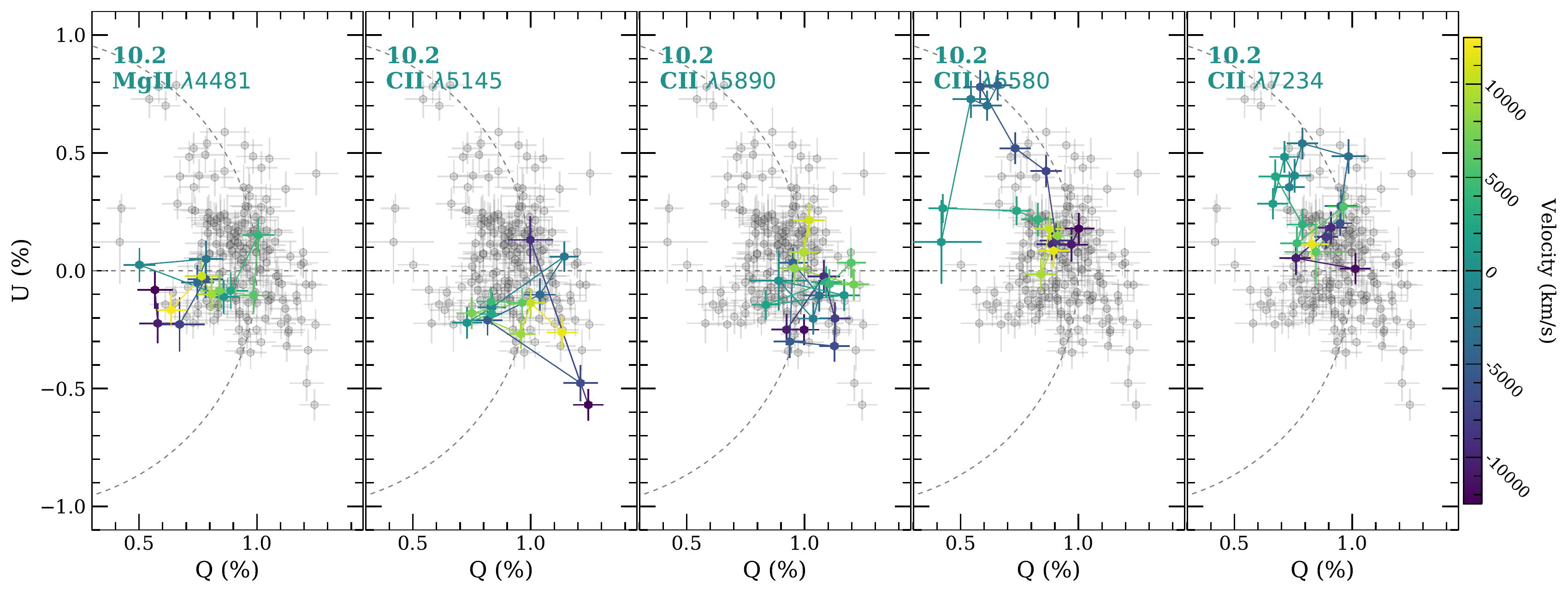}
    \caption{$Q$ -- $U$ plane at 10.2~d for \ion{Mg}{II} $\lambda4481$ and \ion{C}{II} $\lambda5145,\lambda5890,\lambda6580,\lambda7234$. The line regions are highlighted with colors as dictated by the colorbar, while the remaining values are shown in grey.}
    \label{fig:10.2_lines_qu}
\end{figure*}

Another possibility, widely applied in SN observations, is to use the depolarisation of the strongest emission lines and use the minimum as an indication for the ISP. This is what we have done for ISP case B, where we have used the minimum of the H$\alpha$ line at $+$38.4 days to \say{fit} a Serkowski law $p(\lambda)/p_{\mathrm{max}} = \exp{[-K \ln^2{(\lambda_{\mathrm{max}}/\lambda)}]}$, where $p_{\mathrm{max}}$ is the maximum polarisation at wavelength $\lambda_{\mathrm{max}}$ \citep{Serkowski1973}. In practice this is not a real fit as we only consider a very limited wavelength range and we therefore fix $\lambda_{\mathrm{max}} = 5500$ \AA\ and and $K = 1.15$ \citep{Serkowski1975} to obtain $p_{\mathrm{max}} = 0.56$\%. Assuming $\theta_{\mathrm{ISP}} = -57.4^{\circ}$ (determined again from the minimum of H$\alpha$) it is possible to determine $Q_{\mathrm{ISP}}$ and  $U_{\mathrm{ISP}}$ and subtract it from the two datasets to obtain the intrinsic polarisation at the two epochs. The \say{ISP B} solution has been plotted in Fig. \ref{fig:fqupx} together with the data of the second epoch. It can be seen that, although this ISP has been derived solely from H$\alpha$, it is also consistent with the minima of other emission lines, such as $H\beta$ and O~I. It is therefore an acceptable solution. 
In addition, the location for $Q_{\mathrm{ISP}} = -0.23$\% and  $U_{\mathrm{ISP}}= -0.50$\% (referring to 5500 \AA) is also shown in Fig. \ref{fig:qu_plane}. The value can be assumed to be constant since the ISP B solution is almost flat over the wavelength range in Fig. \ref{fig:fqupx} due to the small value of $p_{\mathrm{max}}$. We observe that indeed ISP B does lie in a \say{special} location on the $Q$ -- $U$ plane, i.e. it happens to be along the minor axis of the ellipse describing the data variance, justifying the expectation above. In this second case, however, it is the first epoch that ends up having the largest intrinsic polarisation, i.e.  $P = 1.36\% \pm 0.12$\%, while at 38.4 days we get an average  $P = 0.66\% \pm 0.14$\% (these numbers always refer to the wavelength range 5400 -- 6200~Å and the uncertainty is the standard deviation). In this case, the degree of asymmetry is therefore larger in the first epoch. The ISP B corrected spectropolarimetry is shown in Fig. \ref{fig:FPQUX_ISP_corr}. However, we note that for this ISP determination method to work, the emission lines need to completely depolarise the spectrum. At the relatively early phase this spectrum was obtained (the spectrum is still mostly photospheric and the ejecta optically thick), we have serious doubts on whether this can be the case. Furthermore, from spectroscopy considerations alone (the H$\alpha$ profile), we expect more significant asymmetries in the second epoch. For this reason, we do not consider the ISP B case to be very likely, but it is a viable limiting case, useful for our discussion. The same applies perhaps to ISP case A but this study allowed us  to get an idea of the polarisation levels involved and the possible ranges for both epochs.

\begin{figure*}
    \centering
    \includegraphics[width=0.98\textwidth]{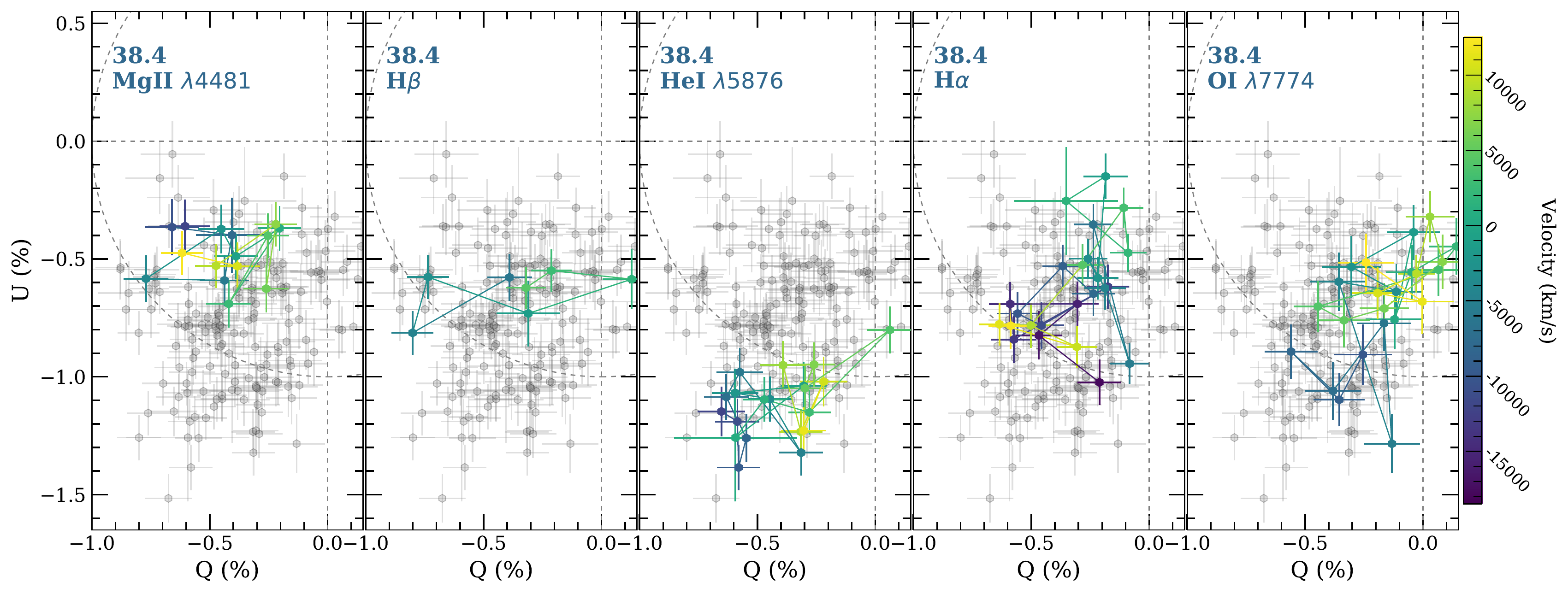}
    \caption{Same as \ref{fig:10.2_lines_qu} for \ion{Mg}{II} $\lambda4481$, H$\beta$, \ion{He}{I} $\lambda5876$, H$\alpha$ and O~I $\lambda7774$ at 38.4~d.}
    \label{fig:38.4_lines_qu}
\end{figure*}

Irrespective of the true value of the ISP, what remains most important is the strong evolution observed in the polarisation properties of SN\,2018bsz between 10.2 and 38.4~days. The ISP does not evolve with time and therefore this evolution must be intrinsic to the SN. Case A ISP corresponds to transitioning from an ejecta-dominated phase, with an almost spherical photosphere, to a CSM-dominated phase with strong asymmetries. Case B ISP corresponds to transitioning from a photosphere that is already highly aspherical \citep[possibly ellipsoidal, e.g.][]{Hoflich1991,Inserra2016} to a CSM-dominated phase that is, however, overall more spherical. As argued in this section we consider the first possibility more reasonable of the two, but the truth may lie somewhere in between.

\subsection{Loops on the $Q$ -- $U$ plane}

We have investigated whether the profiles of different lines present any particular structure on the Stokes $Q$ -- $U$ plane. We have found evidence that the \ion{C}{II} profiles form loops as a function of wavelength, with the strongest being for $\lambda6580,\lambda7234$ (Fig. \ref{fig:10.2_lines_qu}). The presence of such loops is extensively discussed by \citet{Wang2008} and recent modelling has shown that they are a natural product of clumpy ejecta \citep[see e.g.][]{Cikota2019}. We therefore conclude that the \ion{C}{II} lines that dominate the early spectrum of SN\,2018bsz, and by extension of a few other SLSNe, are formed in clumps of material in the outer ejecta. 

There is less evidence for organised structure in the lines dominating the spectrum at $+$38.4 days. The only lines with some possible effects are \ion{Mg}{II} $\lambda4481$ and O~I $\lambda7774$. The profile of these lines, as well as those of other dominant lines at  $+$38.4 days (including H$\alpha$) are shown in Fig. \ref{fig:38.4_lines_qu}. 

\subsection{Comparison with other SLSNe}

There have so far been few polarimetric observations of SLSNe-I. These include LSQ14mo \citep{Leloudas2015b}, SN\,2015bn \citep{Inserra2016,Leloudas2017a}, PS17bek \citep{Cikota2018}, and SN\,2017egm \citep{Bose2018, Maund2019, Saito2020} using linear polarimetry  \citep[circular polarimetry has been obtained for the additional OGLE16dmu by ][]{Cikota2018}. The evidence collected to date can be summarised by the following conclusions: i) SLSNe show typically low levels of polarisation around maximum light consistent with spherical ejecta at these phases; ii) When followed at later phases polarisation seems to increase revealing a higher level of asymmetry in the inner layers. The first fact lends support to the case ISP A versus ISP B for SN\,2018bsz. If true, this would mean that the CSM-dominated photosphere at 38.4~d would be highly asymmetric for SN\,2018bsz ($P \sim 1.8$\%).

Spectropolarimetry has only been collected for SN\,2015bn and SN\,2017egm. For these SNe, although the polarisation degree increased with time, the polarisation angle did not change considerably leading  \cite{Inserra2016} and \cite{Saito2020} to propose an axisymmetric configuration for these SLSN-I, which was retained during their evolution. SN\,2015bn and SN\,2017egm, however, were rather \say{ordinary} SLSN-I. SN\,2018bsz is a unique SN also in its polarimetric properties. The large shift observed in the $Q$~--~$U$ plane for SN\,2018bsz has certainly not been observed for SLSNe-I and highlights a dramatic change between 10.2 and 38.4 days. This change is consistent with the sudden appearance of CSM interaction and change in the sky-projected geometry. In fact, such large changes have rarely been observed for other types of core-collapse SNe, including SESNe. It is not a surprise that one of the few comparable changes we have observed for  past SNe, were for the Type IIb SN\,2001ig, which demonstrated a $\sim40\degree$ rotation attributed to a shift between a nearly spherical hydrogen shell and an aspherical helium core \citep{Maund2007a}.

\section{Discussion}
\label{sec:discussion}

In the previous sections we have presented an in depth analysis to the spectroscopic and spectropolarimetric characteristics of Type I SLSN 2018bsz. The SN appears to be similar to SLSNe-I as its spectral timeseries exhibits several features commonly associated with them. However, the presence of multi-component hydrogen emission lines mirrored by \ion{He}{I} and \ion{Ca}{II} clearly sets it apart from the diverse population and even in comparison to SLSNe-I with late H emission SN\,2018bsz appears unique. Therefore, it seems reasonable to assume that the hydrogen is external to the SN and is located in the CSM. The assumption is further supported by the fact that especially the H$\alpha$ emission appears to be similar to some Type IIn SNe and that late-time NIR entail abundant dust formation \citep{Chen2021} typically assumed to occur in regions of strong CSM interaction \citep[see e.g.][for review]{Matsuura2017}. Therefore, the following discussion will concentrate on the nature, location and geometry of the CSM structure around SN\,2018bsz.

\subsection{Observational Constraints for the CSM}
\label{subsec:CSM_constraints}

The CSM emission appears in form of Balmer emission lines at $\sim25$~d post-maximum. From the beginning, H$\alpha$ exhibits two distinct components, one found at an offset of $\sim-7500$~km/s and the other at $\sim3000$~km/s. The red component is found to be broad ($\mathrm{FWHM}\gtrsim10000$~km/s) with a clear Lorentzian shape, while the blue component is significantly narrower ($\mathrm{FWHM}\sim5000$~km/s) with a profile well matched with both Gaussian and Lorentzian fits. The blue component also appears to be present as faint emission from the time of first spectrum at $-8.1$~d (see Fig.~\ref{fig:Ha_v_fwhm_evo}). The two components likely imply two distinct emitting regions, one moving towards and one moving away from the observer. The clear difference in the line widths implies that emission from the red component undergoes significantly more electron scattering in the line of sight in comparison to the blue component. By 74.5~d post-maximum a third, central component has appeared (albeit the feature could have occurred earlier but could not be probed due to a gap in the X-Shooter observations). Due to reasonably high width ($\mathrm{FWHM}\sim3000$~km/s) the line emitting region either has a moderate velocity dispersion despite the low central velocity (i.e. reasonably high expansion velocity to perpendicular direction from the line of sight) or the line has been scattered significantly.

At around $100$~d post-maximum the multi-component H$\alpha$ profile undergoes several changes. First the blue component starts to fade quickly after 93.1~d. By 107.6~d the component is significantly suppressed and by 121.3~d it is no longer visible. Intriguingly, the loss of the blue component appears to be coupled with the clear skewing of the red component as by 121.3~d the entire blue part of the profile is missing (see Fig.~\ref{fig:Ha_fits}). Intuitively the two must be linked together and explained by the same physical process -- absorption in the line of sight. However, such a configuration might be difficult to create. The components likely originate in two distinct physical locations travelling to opposite directions. For any absorbing body to affect both, it would have to be in front of both emitting regions in a way that only the far-side (i.e. higher velocity) of the red component would be unaffected. Effectively the absorbing body would thus have to cover almost -- but crucially not fully -- the whole SN. At the same time it can not hide the observed central component. Such a combination of observables might be difficult to explain with a single absorbing region.

Furthermore, the nature of the hypothetical absorbing material might also be an issue. The only realistic possibility for suddenly increasing absorption would be the emergence of dust. As discussed by \citet{Chen2021} dust is seen around SN\,2018bsz at  $>200$~d post-maximum, but the 100~d gap in the observations (due to solar occultation) makes it impossible to identify when the dust first appeared. However, for the dust to be the absorber it would have to cause a visible difference in the evolution of H$\alpha$ and Pa$\beta$ when none is seen at 74.5~d or 121.3~d (see Fig.~\ref{fig:Ha_Pbd}). Therefore, the late evolution of the blue component is likely caused by some other physical mechanism than absorption.

\begin{figure}
    \centering
    \includegraphics[width=0.49\textwidth]{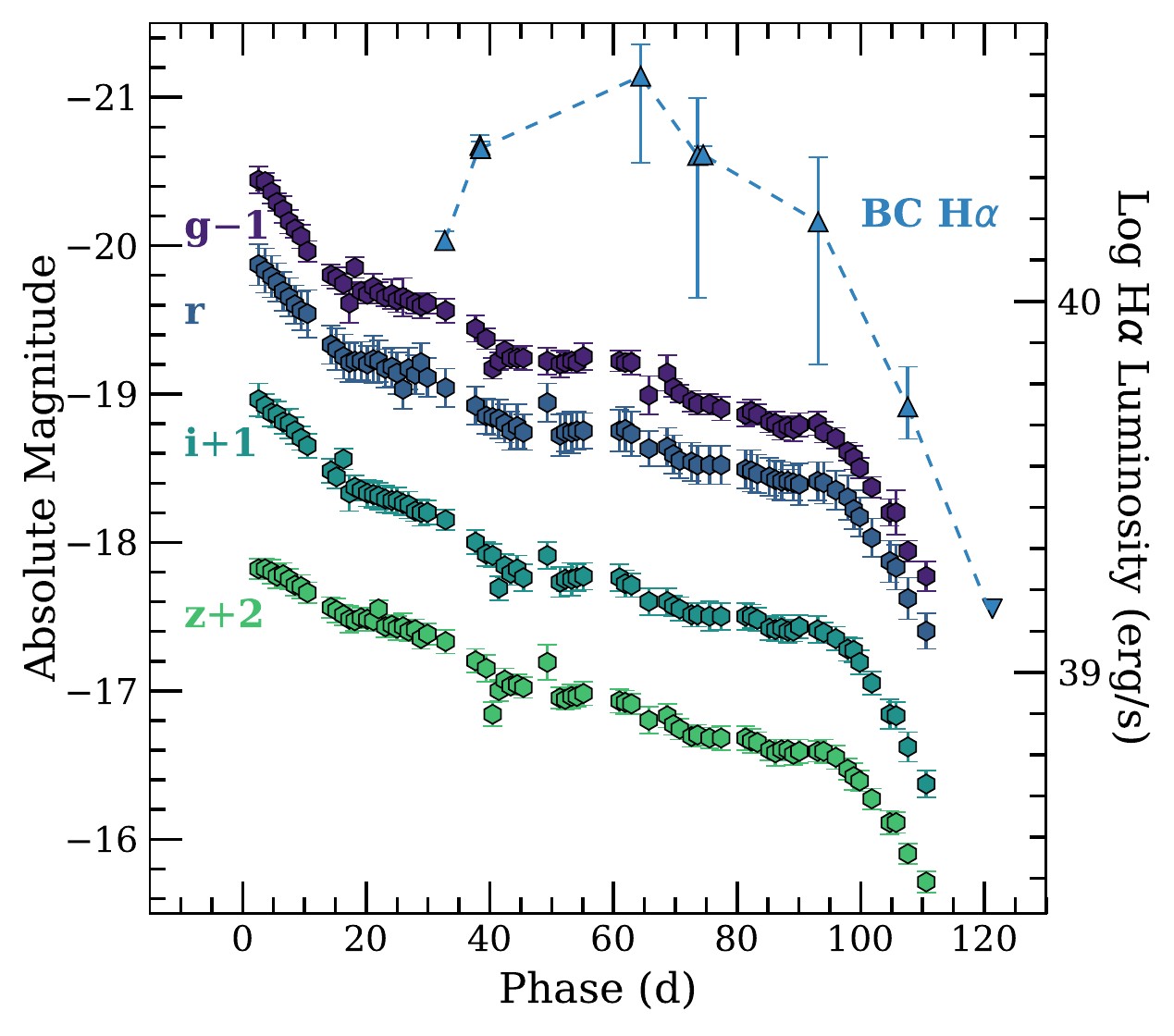}
    \caption{GROND $g$\arcmin, $r$\arcmin, $i$\arcmin and $z$\arcmin~ absolute magnitude light curves of SN\,2018bsz \citep{Chen2021} in comparison to the integrated luminosity of the blue H$\alpha$ component (BC). The blue component becomes fainter simultaneously with the light curve at $\sim100$~d. The H$\alpha$ luminosities were estimated based on the Gaussian fits described in Sect. \ref{subsec:H_lines}. At 121.3~d no emission component was seen and the stated value is a $2\sigma$ upper limit for the non-detection shown with a downward triangle. Note that the axis scales are set so that $2.5~\mathrm{mag}=1 \mathrm{Log}(L)$.}
    \label{fig:Ha_LC_evo}
\end{figure}

On the other hand, the disappearance of the blue component at $\sim100$~d coincides with a break seen in the optical light curves as demonstrated in Fig. \ref{fig:Ha_LC_evo} and the concurrent changes are likely related. The spectra evolve only a little between $\sim25$~d (when the CSM interaction lines appeared) and $\sim90$~d, so the spectral changes between 93.1~d and 121.3~d stand out (see Fig.~\ref{fig:spectral_timeseries}). Not only do the blue emission components of H$\alpha$, \ion{Ca}{II} and \ion{He}{I} lines fade away, but so do also some prominent absorption lines such as \ion{Ca}{II} H\&K. While the spectrum at 121.3~d is clearly not yet nebular, it seems to be evolving towards it. It appears that at $\sim100$~d the photosphere started to recede faster than before, resulting in both the light curve break and in the observed spectral evolution. This is likely caused by a physical mechanism that stopped or at least slowed down the receding photosphere and held it roughly in place radially until $\sim100$~d, thus producing the light curve \say{plateau}.

While the magnetar scenario has often been discussed as the cause of the long-lived light curves of SLSNe-I \citep[see e.g.][]{Kasen2010, Inserra2013, Nicholl2017a}, the evolution of the plateaus seen in Type IIP SNe resemble SN\,2018bsz the most. For a Type IIP SN the plateau occurs as the shocked H-rich ejecta remains optically thick until the hydrogen recombines. As the ejecta expands, each layer eventually passes through the recombination \say{front} becoming optically thin in the process. Once all of the hydrogen has recombined the optical depth of the ejecta suddenly decreases causing the photosphere to recede radially. In an analogue the post-maximum plateau in SN\,2018bsz could be caused by optically thick, H-rich CSM that is located inside the photosphere. In case a significant portion of the CSM has completely emerged from the photosphere by $\sim100$~d, this would result in the photosphere suddenly receding faster explaining the drastic changes in the light curves and the spectra. As the red component still persists at $121.3$~d, the emerged CSM would have to be on the near-side of the SN explaining the sudden disappearance of the blue component. As the emission is originating in optically thick CSM, the component would likely fade quickly after the CSM re-emerged from the photosphere.

While such a scenario gives a reasonable description for the break in the light curve and the sudden fading of the blue component as well as the whole spectral evolution, it does not explain why the blue-side of the red component is suppressed in the late spectra. As discussed above, absorption seems like the only mechanism capable of hiding it with the blue component and -- as the scenario seems unrealistic -- the two features are unlikely related. Therefore, the apparent skewing of the red component must be caused by some other physical mechanism. One possible explanation is related to the significant electron scattering the red component clearly undergoes. In certain conditions an emission from a body of gas can be significantly skewed as the emission from the near-side of the body is scattered less or not at all, while the emission from the far-side has to scatter several times before escaping. As a result the red-end of the profile exhibits an extremely long tail, while the blue-end appears to be significantly narrower in shape -- exactly as seen for SN\,2018bsz at $121.3$~d. Such profiles have been investigated for instance by \citet{Roth2018}, who analysed line shapes arising from emission in hot, outflowing gas optically thick to electron scattering. While their models were created for tidal disruption events (TDEs) they clearly demonstrated that such skewed profiles exist. However, for SN\,2018bsz more appropriate comparison might be the scenario presented by \citet{Taddia2020} fo type IIn SN\,2013L. They demonstrate that emission from a spherical CSM shell naturally generates a similar skewed profile. Under this scenario the far-side of the shell is not visible and the strong scattering is occurring on the sides of the shell. The peaks of the skewed profiles are blueshifted due to the nature of the scattering: emission from the near-side (i.e. higher velocity/blueshift) of the emitting body is scattered less, while emission from further away is both redshifted more intrinsically and scattered more on the way. In SN\,2018bsz the situation is slightly different as we have already assumed that the red component is arising from receding material. As the red component is found at significantly lower velocity than the blue one (see Fig. \ref{fig:Ha_v_fwhm_evo}), the component could easily be blueshifted by such scattering explaining at least some of the velocity difference. Furthermore, as it is clear that the red component is affected by significant electron scattering such a scenario does seem to have merit in explaining why the profile appears to be skewed in the late spectra. However, here it is important to note that at $32.7$~d and $38.5$~d the red component appears to have a symmetric profile with an overlaying, comparatively faint blue component as shown in Fig. \ref{fig:Ha_fits}. If we assume that the component was truly symmetric to begin with, it would have had to become skewed during the evolution suggesting that something changed in the scattering or emitting region. This could be related to an increasing amount of the CSM region becoming geometrically visible as the ejecta-driven photosphere recedes, but determining the cause of the skewing likely requires radiative transfer modelling and is outside the scope of this paper.

Finally, the complex CSM structure cannot block the emission from the ejecta i.e. both the ejecta and the CSM have to be visible at the same time. While some typical ejecta lines (e.g. Ca~II) appear to have similar emission profile to the H$\alpha$, several lines of e.g. Mg and Fe clearly resemble the ones seen in normal Type Ibc SNe and SLSNe-I i.e. they are not significantly scattered or obstructed during the evolution. In the following we discuss some potential configurations of both CSM and ejecta that could naturally explain the observables.

\subsection{CSM configuration}
\label{subsec:CSM_conf}

\subsubsection{Spherical CSM}

Spherical CSM is often discussed in relation to interacting SNe due to its simplicity even if it might not be always realistic. However, for SN\,2018bsz this scenario might be applicable as such a configuration has been suggested for the Type IIn SN\,2013L, which has similar line evolution as SN\,2018bsz (see Fig. \ref{fig:Ha_comp}) by \citet{Taddia2020}. As mentioned above, in their scenario lines emerging from a geometrically thin but optically thick shell behind the SN shock can explain the observables. The main emission component is a skewed one with a blueshifted peak but very extended red tail. The blue-side of the component would be emitted from the near-side of the shell as it is expanding at a reasonable velocity, while the emission from the sides of the shell would undergo more electron scatterings producing the red tail. In their model \citet{Taddia2020} also add a central emission component emitted from unshocked, optically thick CSM outside the shell and a narrow P~Cygni profile from optically thin CSM even further from the SN explosion. However, SN\,2013L did not exhibit a red H$\alpha$ peak -- unlike SN\,2018bsz and the other two IIn SNe presented in Fig. \ref{fig:Ha_comp}. The absence of the red component is an important difference and questions whether such a scenario can be applied directly to SN\,2018bsz. Redshifted emission naturally requires emitting material receding from the observer. But in the \citet{Taddia2020} spherical CSM model, the emitting shell would block the emission from the receding side and the red peak would not be visible. A spherical CSM shell would therefore have difficulties explaining the multi-component lines seen in SN\,2018bsz. 

An alternative scenario could be related to a highly aspherical (bipolar) explosion inside a roughly spherical CSM. In this case, the ejecta swept-up CSM would be asymmetric and would emit strongly thus generating the observed blue and red components. However, as discussed earlier the multi-component emission lines are assumed to be indicative of aspherical CSM. Additionally, the emission geometry shift seen in the spectropolarimetric observations is difficult to explain under a such scenario (see Fig.~\ref{fig:qu_plane}. The emission locations would have to remain in the region of the ejecta-CSM interaction and thus there is no reason to assume that the geometry would significantly change. While there is uncertainty on the amount and direction of the ISP, that alone cannot explain why the shift on the $Q$ -- $U$ plane would occur. Therefore we disfavour any highly aspherical explosion scenario for SN\,2018bsz.

\subsubsection{Disk-like CSM}
\label{subsec:CSM_disk}

\begin{figure}
    \centering
    \includegraphics[width=0.4\textwidth]{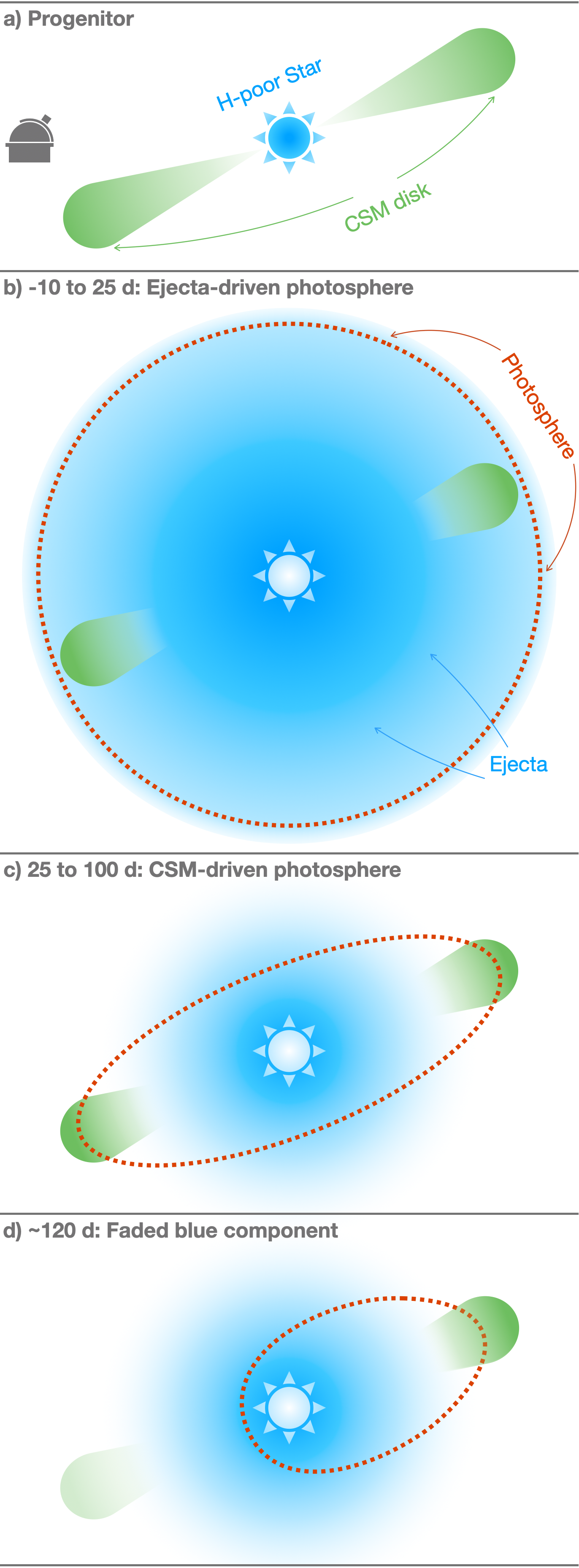}
    \caption{Illustration of the physical evolution of SN\,2018bsz assuming a scenario of disk-like CSM surrounding a H-poor star. Observer is assumed to be on the left as marked in the first panel. a) System before the explosion. b) Upon the explosion the ejecta quickly overtakes a significant part of the disk hiding the CSM emission lines. Only the faint blue component is possibly present. c) As the ejecta photosphere recedes, the CSM starts to re-emerge and multi-component emission becomes visible. At this stage the photosphere is CSM-driven. d) CSM on the near-side has completely re-emerged causing the photosphere to recede and the blue component to fade. Note that CSM disk is represented with cones and ejecta with a sphere for simplicity. The photosphere is marked by dashed red line.}
    \label{fig:disk-like_csm_schematic}
\end{figure}

To explain the peculiar spectral evolution of the Type IIn PTF11iqb and SN~1998S, \citet{Smith2015} presented a scenario with a disk-like CSM structure surrounding the SN explosion. The CSM disk is quickly enshrouded by the expanding ejecta hiding any CSM emission lines and giving rise to a typical ejecta spectrum. As the ejecta photosphere eventually starts to recede, the CSM disk re-emerges producing the prominent multi-component emission lines. Therefore, given the visual similarity of the H$\alpha$ profile of SN\,2018bsz in comparison to PTF11iqb and SN~1998S (see Fig. \ref{fig:Ha_comp}) it seems plausible to assume that a similar physical scenario could apply to all of them. As the SNe are obviously different, we have tailored the scenario to suit the properties of SN\,2018bsz better. The main difference is that we assume a H-poor progenitor star to explain the SESN-like spectra. The scenario is presented in Fig. \ref{fig:disk-like_csm_schematic} as a four-stage illustration visualising the sequence of events most important to understand the spectral evolution of SN\,2018bsz.

In panel a) we provide a schematic of the system before the SN explosion. Note that in the cutout the CSM disk is described with symmetric cones for simplicity. Panel b) refers to days from $\sim-10$ to $\sim25$~d i.e. the phase before the prominent multi-component H$\alpha$ appeared. At this stage the rapidly expanding ejecta has already overtaken at least a significant part of the CSM disk and the photosphere is ejecta-driven. As described by the 2D hydrodynamic models of \citet{McDowell2018}, in such situation the ejecta flows around and engulfs the disk hiding the CSM emission. The only exception is the blue component of H$\alpha$ which is possibly present as faint emission from the time of the first spectrum. As the CSM on the near-side of the explosion is travelling towards the observer the related blue component should appear before the red one. As it is likely accelerated to similar velocity as the ejecta, it is possible that some part of the CSM is visible from the beginning.  Note that while the ejecta is described as a sphere for simplicity, this does not need to be the case in reality. In panel c) we present the system at a stage when the blue and red emission components of the CSM -- most prominent for H$\alpha$ -- are visible (from $\sim25$ to $\sim100$~d). By this stage the ejecta-driven photosphere has receded enough for the CSM-disk to re-emerge from both sides of the explosion and to become geometrically visible for the observer. Based on the light curve plateau the photosphere seems to be held at least roughly in place at this epoch, indicating the influence of the CSM. As such the photosphere is now CSM-driven and stretching to follow it as shown in Fig. \ref{fig:disk-like_csm_schematic}. The schematic also provides a plausible explanation for the width of the red component: the longer travel distance through the SN ejecta likely results in more scatterings in comparison to the blue component. Finally, by $\sim120$~d presented in panel d) the blue component of H$\alpha$ has completely disappeared. As it fades simultaneously with the light curve break, they are both likely related to the evolution of the photosphere on the near-side of the SN. As discussed in Sect. \ref{subsec:CSM_constraints}, the CSM is no longer providing opacity thus resulting in a rapidly shrinking photosphere. However, as the red component is still prominent the photosphere has to still linger on the far-side of the explosion. This could possibly imply that the disk-like CSM is also \say{clumpy} and material is not evenly distributed in the CSM disk. The lingering red component could thus be explained by a larger amount of CSM on the far-side of the SN as suggested by \citet{Smith2015} to explain the observed difference in the brightness of the multiple H$\alpha$ components in PTF11iqb.

The schematic presented in Fig. \ref{fig:disk-like_csm_schematic} is a simplistic representation of how the scenario of a disk-like CSM can generate the main observables of the spectroscopic evolution of SN\,2018bsz. However, it does not touch one crucial feature -- the third H$\alpha$ component. While the component appears to be reasonably broad  ($\sim3000$~km/s), the width can be explained either by velocity dispersion or by electron scattering. The component could for instance be emitted from some CSM significantly further away from the SN. Under such an idea, the component would appear only when it has been shock-ionised which -- depending on the distance -- could occur after the red and blue components appeared. If we assume the central component appeared at 70~d post-maximum i.e up to 130~d after the explosion and assume a constant ejecta velocity of $15000$~km/s from the \ion{Ca}{II} absorption, the body of CSM would be found at $\sim11\times10^{15}$~cm. Similar values were for instance estimated by \citet{Yan2017} for the three iPTF SLSNe-I with late hydrogen emission (9 -- $40\times10^{15}$~cm). Whatever the cause for the central component is, the multi-component H$\alpha$ emission requires several emitting CSM regions.

Further supporting evidence for the enshrouded aspherical CSM is provided by the spectropolarimetric observations, which probe phases b) and c) in Fig. \ref{fig:disk-like_csm_schematic}, as the proposed scenario provides an intuitive explanation for the observed geometry shift of the emitting region (see Fig. \ref{fig:qu_plane} and related discussion in Sect. \ref{sec:spectropolarimetry}). At the time of the first FORS2 epoch at 10.2~d the spectrum is dominated by the \ion{C}{II} features, with only a faint presence of the CSM in form of the blue component of H$\alpha$. At this epoch it is clear that the photosphere is still ejecta-driven and any observed polarisation is related to the inherent geometry of the ejecta, i.e. the explosion itself. However, by 38.4~d the red and blue components of H$\alpha$ (and e.g. \ion{Ca}{II} H\&K) are prominent which -- under the discussed scenario -- indicates that the location of the photosphere is dictated by the CSM. Unless the photosphere at these two phases appeared to be similar on the sky, there is no reason to assume the geometry would be the same. Thus the scenario provides an natural explanation for the seen geometry shift and supports such scenario.

Under the scenario proposed in Fig. \ref{fig:disk-like_csm_schematic}, the CSM has to be located physically close to the progenitor star to be overtaken by the ejecta before the time of the first spectrum ($-8.1$~d). Given the known ejecta velocity (15000~km/s from \ion{Ca}{II} H\&K absorption) it is possible to estimate how close to the progenitor the bulk of the CSM had to be. The light curve of SN\,2018bsz shows a long $60$~d rise split in two phases as described in \citet{Anderson2018}. First the light curve rises slowly for $\sim40$~d forming a \say{plateau} until at about $20$~d pre-maximum the rise rate suddenly increases. Assuming both phases are related to the SN, the first spectrum is taken $\sim50$~d after explosion. As a result virtually all of the CSM have to be closer than $\sim6.5\times10^{15}$~cm ($430$~AU). As the value is just an upper limit in reality the bulk of CSM is like residing closer. In case the pre-maximum plateau is not related to the SN (i.e pre-explosion behaviour) the time from the explosion to the first spectrum is $\sim10$~d and the CSM would have to be closer than  $\sim1.3\times10^{15}$~cm ($87$~AU). 

If the CSM is created by a stellar eruption, these distances can be used to estimate the time of the eruption with respect to the SN. The only SLSN-I for which the CSM velocity has been directly estimated is iPTF16eh, $\sim3000$~km/s \citep[][]{Lunnan2018}. While the velocity is high in comparison to a typical stellar wind (few hundred km/s), similar velocity has been found for 1840 eruption of $\eta$~Car \citep[$3000$ -- $6000$~km/s;][]{Smith2008b}. Moreover, such velocities can be a result of pulsational pair instability \citep[PPI; e.g.][]{Woosley2007} eruptions \citep{Woosley2017}. Assuming the velocity of iPTF16eh, the ejected CSM would reach the distance of $\sim6.5\times10^{15}$~cm in $\sim250$~d. However, if the pre-max plateau is not related to the SN, the time would be reduced to $\sim50$~d. The latter estimate is close to the duration of the plateau itself \citep[$\sim40$~d;][]{Anderson2018}. This may imply that the plateau is caused by a mass eruption in the star just prior to the SN. Such pre-explosion behaviour has been discussed especially in relation to Type IIn SNe exhibiting non-terminal stellar eruption known as SN impostors \citep[see e.g.][]{Pastorello2019a}. While several Type IIn SNe have shown significant brightening just before the assumed SN explosion (e.g. SN\,2009ip; \citealt{Pastorello2013, Mauerhan2013, Fraser2013, Margutti2014}), the outbursts are typically fainter than $M=-16$ in optical bands. Therefore, if the pre-maximum plateau of SN\,2018bsz was caused by an eruption it would have to be exceptionally bright. 

The ejecta velocity and the highest measured HV H$\alpha$ velocity ($\sim9000$~km/s) can also be used to estimate the the largest possible viewing angle from the pole. Assuming that the CSM was accelerated to the ejecta velocity and we see it at a lower velocity only due to the projection, we find that the viewing angle is $\sim37\degree$. In case the \say{absolute} CSM velocity is lower the angle would be larger, but even if we assume a low CSM velocity of $10000$~km/s the angle would still be $\sim65\degree$. These values are all in the expected ballpark as no red component would be clearly seen if the disk was viewed edge-on and no clearly separate components would be seen face-on. 

Finally, it is of interest to discuss the observed \ion{He}{I} and \ion{Ca}{II} lines in context of the CSM scenario. As there is strong similarity between the H, He and Ca emission line profiles it seems natural to assume the lines originate in the same physical region i.e. in the CSM rather than the ejecta. While helium is not always detected in Type IIn SNe, there are several cases where prominent lines are seen alongside hydrogen throughout the spectral evolution \citep[i.e. SN\,2005ip and SN\,2006jd;][]{Stritzinger2012}. On the other hand, Type IIn SN\,2013L -- that also had multi-component hydrogen profiles -- exhibited \ion{Ca}{II} lines similar to H$\alpha$. While \citet{Andrews2017} stated that the \ion{Ca}{II} NIR triplet arises in the fast ejecta or at the location of the reverse shock, \citet{Taddia2020} argues that the similarity of the \ion{Ca}{II} and H$\alpha$ profiles means that the calcium emission must be coming from the same region as the H$\alpha$ (i.e. CSM). Thus, it is plausible that at least some helium and calcium would be found in the CSM with hydrogen. The key difference between the line profiles is the absence of the central component for \ion{Ca}{II} H\&K and \ion{He}{I} $\lambda5876$, clearly present for the NIR lines of both elements. This possibly implies that the emitting region of the central component has different physical conditions in comparison to the blue and red regions. For instance, perhaps a different temperature or density of the region could explain the absence of the bluer lines. Lastly, it is worth to remember that the \ion{He}{I} and \ion{Ca}{II} appear to be accompanied by blueshifted absorption. As this implies material flowing directly towards the observer and as Balmer lines do not exhibit clear absorption features, the \ion{He}{I} and \ion{Ca}{II} absorption lines likely originate in the ejecta. As such both calcium and helium appear to exist both in the CSM and in the ejecta.

\subsection{Progenitor}

As discussed in the previous sections, a disk-like CSM structure surrounding SN\,2018bsz at close vicinity seems to provide a reasonable description for significant part of the peculiar observables. Now of course the question is if that kind of a CSM is physically feasible. Given the spectroscopic similarity of SN\,2018bsz and SESNe, the progenitor scenarios discussed in the context of SESNe might be viable. In principle, there are two alternative scenarios proposed in the literature: single Wolf-Rayet (WR) stars and stars in binary systems. While the lack of of discovered progenitors seems to disfavour the bright WR stars as the progenitors \citep{Eldridge2013} it is not clear if they can be disregarded -- especially for the SLSNe-I. Even the only detected progenitor of a SESN -- that of Type Ib iPTF13bvn \citep{Cao2013, Folatelli2016} -- has been discussed both as a single WR star \citep{Groh2013} and as a star in a binary system \citep{Bersten2014}. Thus both scenarios could potentially be applicable to SN\,2018bsz as well. For a review of the SESNe progenitor discussion see e.g. \citet{Smartt2009} and \citet{VanDyk2018}. In the following we discuss the merits of both scenarios in the context of SN\,2018bsz.

\subsubsection{Single, rapidly rotating star}

The only way for a single star to create surrounding CSM structure is through some kind of mass loss i.e. wind or stellar eruptions. Such mechanisms are supported by discovery of several Type IIn SNe progenitors. One avenue to create highly aspherical CSM such as a disk or a torus, is for the progenitor star to be rapidly rotating. A famous example of such systems are the very rapidly rotating Be stars that form equatorial \say{decretion} disks of CSM via poorly understood mechanism \citep[for review see e.g.][]{Porter2003, Carciofi2011, Rivinius2013}. While it is not clear if such a H-rich CSM disk is likely to exist around a H-poor star -- as required for SN\,2018bsz -- Be stars clearly demonstrate that creating such a disk is physically possible for a single star. One caveat for such a scenario is the inhomogenous CSM distribution implied by the luminosities of the different H$\alpha$ components. As discussed by \citet{Smith2015}, creating a CSM disk with an azimuthal asymmetry is difficult for a single, rotating star and the system might have to be rather unique. In Sect. \ref{subsec:CSM_disk} we mention the possibility that the pre-maximum plateau of SN\,2018bsz could be caused by an high-velocity pre-explosion eruption. Such a violent ejection from a rotating star could possibly be able to create a aspherical CSM structure that satisfies the observational requirements.

\subsubsection{Binary system in mass transfer}

In addition to a single star, a binary system in mass transfer could also be the cause of the highly aspherical CSM \citep[see e.g.][]{Smith2002}. Under this scenario, the progenitor is assumed to be an evolved, massive star that has lost its H (and potentially He) shells to the the companion star during the binary evolution. The external disk-like CSM structure could arise from the mass transfer in the binary system. Such CSM nebula have been observed for one close-by binary system: RY Scuti. While RY~Scuti is the only binary system caught in the middle of a brief mass transfer episode \citep{Smith2002, Smith2011a, Grundstrom2007}, it clearly demonstrates that such aspherical and asymmetric CSM  (i.e. clumpy disks) can be created by binary interaction. Furthermore, \citet{Smith2015} compares the line profiles seen in RY~Scuti to the H$\alpha$ in PTF11iqb and concludes that they are very similar in shape -- with the difference that in RY Scuti the measured velocities are $<100$~km/s. In case an evolved massive star with little or no hydrogen were to explode inside CSM nebula similar to that of RY~Scuti, the spectral timeseries would likely be similar to SN\,2018bsz. 

One major question on such a scenario is which star acts as a donor. As for RY~Scuti the majority of the CSM is found within $\sim1000$~AU while the stars are measured to be separated by 0.66~AU \citep{Smith2002,Smith2011a}, it is clear that either of the stars could in principle be responsible for the CSM. The same conclusion applies also to SN\,2018bsz as we estimated that the CSM is concentrated within 430~AU. In case the progenitor star created the CSM structure, the star would have had to lose virtually all of its remaining hydrogen during the last mass transfer episode and explode before the CSM dissipated. Alternatively, the companion star could also be responsible. While the H-rich CSM composition would be easier to explain under this channel, the companion would have had to first accrete material from the progenitor and then start donating some back before the progenitor exploded. As both of the channels require a time-constrained sequence of events, it is difficult to determine which is more reasonable.

\subsubsection{Stellar Merger}

Common envelope mergers of massive stars are also discussed creating highly aspherical CSM \say{rings}. Most notably a merger scenario has been suggested to explain the triple ring CSM system seen around SN\,1987A \citep{Morris2007, Morris2009}, but the merger is also the favoured scenario to produce its blue supergiant (BSG) progenitor \citep{Menon2017}. In the merger scenario, the merged star sheds its excess angular momentum through equatorial outflow producing a ring-like structure around the star.  As the progenitor of SN\,1987A was a BSG -- i.e. H-rich star -- and as similar CSM structures have been seen around other BSGs \citep[see e.g.][]{Smartt2002, Smith2007a, Smith2013, Gvaramadze2015}, the 87A-like merger is unlikely directly applicable to SN\,2018bsz due to its H-poor nature. However, as the CSM configuration is similar to what we have envisioned for SN\,2018bsz, it is possible that some kind of a stellar merger could be responsible.

A key consideration of the merger scenario is that the time delay between the merger and the SN affects the observed spectral signatures. In case of SN\,1987A, the CSM rings ejected 20000~yr before the SN \citep{Crotts2000} are found relatively far ($\sim6$ -- $20\times10^{17}$~cm; \citealt{Sugerman2005, Tziamtzis2011}) and the photospheric spectra were dominated by the ejecta. However, as the merger process of SN\,1987A included a non-equatorial envelope ejection that eventually created the outer rings before the equatorial shedding \citep{Morris2007, Morris2009}, it is likely that if the star had exploded soon after the merger, the CSM emission would have been present. In fact, \citet{Pastorello2019} mentions the possibility that Type IIn-P SN\,2011ht \citep[spectroscopically IIn, but exhibits a light curve plateau;][]{Roming2012, Mauerhan2013a} could be a SN that occurred only $\sim300$~d after a common envelope ejection (potentially followed by a merger) that was faintly detected at the time \citep{Fraser2013a}. The CSM-dominated spectra of SN\,2011ht then arose from the ejected material. However, it is important to note that the nature of SN\,2011ht is under discussion and an eruption from a massive star has also been suggested \citep[e.g.][]{Mauerhan2013a, Smith2013a, Fraser2013}

For SN\,2018bsz we have estimated that the CSM is close to the SN ($\lesssim6.5\times10^{15}$~cm). In case a significant non-equatorial ejection related to a merger was present around SN\,2018bsz it would likely create features visible in the spectral timeseries. One possibility is that, assuming the ejected material is further away than the equatorial ring \citep[as implied by the larger distance of the outer rings in SN\,1987A;][]{Tziamtzis2011}, the ejection could be related to the delayed central H$\alpha$ component of SN\,2018bsz that would appear upon shock heating of the CSM. To conclude, while a merger is possibly applicable to SN\,2018bsz, the suggested scenario would have to successfully explain the combination of close-by H-rich CSM and H-poor explosion in a way that the SN itself is observable and not enshrouded by the CSM. 

While there are several open questions regarding the progenitor system of SN\,2018bsz, it seems physically feasible for both a single star and a binary system to produce the needed kind of CSM structure. Here we have only provided brief description of a few progenitor scenarios and it is important to note that alternative ways to produce inhomogenous CSM disk likely exist and SN\,2018bsz does not need to be explained by the ones presented here. 

\subsection{Implications for Type I SLSNe}
\label{subsec:implications}

In the previous sections we have demonstrated that while SN\,2018bsz exhibits several hallmark features of SLSNe-I it also shows clear signatures of CSM interaction setting it apart from the diverse population. However, SN\,2018bsz is also different from SLSN-II, despite the similar, late photospheric spectra (see Fig. \ref{fig:SLSN_spectral_comp}). As such SN\,2018bsz is spectroscopically between SLSNe-I and SLSNe-II, possibly indicating a continuum between the two classes. Furthermore, the long pre-maximum plateau \citep{Anderson2018} and significant dust formation at late times \citep{Chen2021} are highly atypical for SLSNe-I. Therefore, the key questions to ask are how does SN\,2018bsz fit in the midst of SLSNe-I and whether the CSM interaction alone can explain the peculiarity of SN\,2018bsz? 

While the five SLSNe-I with broad Balmer emission (see Fig. \ref{fig:SLSN_spectral_comp}) are diverse in observables, they do appear to have quite remarkable similarities. As already stated in Sect. \ref{subsec:spec_comp_SLSN}, PTF10aagc exhibits strong \ion{C}{II} features similar to those of SN2018bsz but the typical SLSN-I \ion{O}{II} \say{w}-shaped absorption feature is found at visibly lower velocity than typically in SLSNe-I. Furthermore, PTF10aagc also has broad Balmer lines blueshifted by $\sim-2000$~km/s first seen at $77$~d post-maximum. The three SN presented by \citet{Yan2017} are less similar to SN\,2018bsz than PTF10aagc. None of the three show evidence of the \ion{O}{II} absorption in their early spectra and only iPTF16bad clearly exhibits \ion{C}{II} lines. For iPTF15esb, we note that while \citet{Yan2017} states that they see clear Balmer emission at $73$~d, the spectrum at $30$~d does seem to have reasonably convincing detection of both H$\alpha$ and H$\beta$. The near zero-velocity emission appears to be accompanied by a blueshifted emission component -- reminiscent of SN\,2018bsz. The component is also present at $52$~d, but by $73$~d it has faded. \citet{Yan2017} demonstrated that these features can be explained by a combination of \ion{C}{II} and \ion{Si}{II} emission (the two H$\alpha$ components) and \ion{Fe}{II}  absorption (H$\beta$), but after the detection of multiple components of H$\alpha$ in SN\,2018bsz one has to wonder if the features could actually be caused by hydrogen. If so, it is important note that the light curve of iPTF15esb appears to have a break at $\sim50$~d, right where the disputable blue component appears to be fading -- just as in SN\,2018bsz. However, the light curve of iPTF15esb also exhibits a double peak before the break. The other two SNe, iPTF13ehe and iPTF16bad, have poorer spectral coverage and thus the times when the Balmer lines first appeared are less constrained: For iPTF13ehe the H$\alpha$ is first detected at $\sim+252$~d but the previous spectrum was taken at $-5$~d and for iPTF16bad it was detected at $\sim+97$~d with previous spectrum at $3$~d. For iPTF16bad \citet{Yan2017} state that the H$\alpha$ emission appears to move from blueshift of $\sim-400$~km/s to redshift of $\sim500$~km/s between at $125$~d and $242$~d, possibly indicating a presence of multiple components. Similarly, for iPTF15esb the shift occurred from blueshift of $\sim-1000$~km/s at $73$~d to redshift of $\sim400$~km/s at 102~d. For iPTF13ehe such shift was not seen but the late detections of H$\alpha$ appeared to be blueshifted by $\sim400$~km/s instead. %

\citet{Yan2017} promoted a scenario where an episode of violent mass loss occurred $\gtrsim10$~yr before the explosion and the Balmer lines appeared when the SN shock finally reached the ejected CSM. A similar CSM shell could also be present around SN\,2018bsz, but it would have to be aspherical as implied by the multi-component emission lines. As the H$\alpha$ components are found at high velocities ($-9000$~km/s and $3000$~km/s) and are significantly broader than in the iPTF SNe \citep[$\sim4000$~km/s;][]{Yan2017}, the main difference between the SNe seems to be the distance to the CSM assuming that earlier interaction provides more kinetic energy to the CSM. This is also supported by the distance estimates as radius of $9$ -- $40\times10^{15}$~cm found by \citet{Yan2017} is higher than the upper limit of the CSM ($\sim6.5\times10^{15}$~cm) for SN\,2018bsz. Thus, a similar CSM structure could exist around SN\,2018bsz as well, but based on the photometric, spectroscopic and spectropolarimetric evolution it was close enough to be overtaken by the ejecta as discussed in Sect. \ref{subsec:CSM_disk}. However, interaction with more distant CSM could explain the late appearance of the central component (see Sect. \ref{subsec:CSM_conf}). Alternatively, \citet{Moriya2015} proposed that the late H emission of iPTF13ehe could be created by hydrogen stripped from a companion star. However, as having both blueshifted and redshifted emission lines are not expected when stripping a companion star, it is unlikely applicable to SN\,2018bsz. On the other hand, aspherical CSM hidden inside the photosphere could possibly explain the emission from the literature SNe as well. If one assumes that the disks for the iPTF SNe were more face-on than in case of SN\,2018bsz, the multi-component nature of the emission lines would be less clear and the lines likely be blended close to zero velocity. The apparent velocity shifts could then be caused by evolution of the blended components. 

Even if the similarity of the iPTF SNe and SN\,2018bsz is more debatable, PTF10aagc can easily be explained with the same scenario as SN\,2018bsz. As the H emission lines are found only at a blueshift of $\sim-2000$~km/s, the only significant difference to SN\,2018bsz would be the configuration of the CSM which would be more concentrated on the near side of the explosion. Additionally the disk could possibly be a bit more face-on in comparison, explaining the lower velocity. Furthermore, the two SNe also share the strong \ion{C}{II} features and comparatively low velocity \ion{O}{II} absorption in comparison to SLSNe-I. However, it is unclear whether these features are somehow a consequence of the CSM interaction or if they are inherently related to the explosion itself.

Now the important question is if a scenario similar to what we have presented for SN\,2018bsz could also be applicable to SLSNe-I more generally. However, as significant majority of SLSNe-I do not exhibit H emission lines, any scenario involving H-rich CSM seems disfavoured as one would expect the H lines to eventually appear. Even if not all SLSNe-I are observed late enough to be certain that no H emission appears at a epoch comparable to PTF10aagc (77~d) for instance, a large number are \citep[see e.g.][]{Quimby2018}. Therefore, a H-rich CSM disk hidden inside the photosphere seems unlikely applicable to a significant portion of SLSNe-I. As such SN\,2018bsz seems to be an extension of the SLSNe-I population made spectroscopically different due to the especially prominent CSM interaction. However, detailed non-local thermodynamic equilibrium radiative transfer calculations -- that take into account e.g. presence of shocks due to interaction and complicated 3D geometry -- are required to investigate if hydrogen features from interaction with H-rich CSM might not be seen in some circumstances \citep[see e.g.][]{Chatzopoulos2013}. While it is now accepted that interaction plays a role in many SLSNe as either the driving mechanism or a minor contributor to another dominant mechanism \citep[see e.g.][]{Yan2017,Lunnan2018,Hosseinzadeh2021}, a key remaining question revolves around understanding the physics behind H-poor interaction where models are not available in the literature. 

\section{Summary \& Conclusions}
\label{sec:conclusions}

We have presented an in depth spectroscopic analysis of the nearby Type I SLSN, SN\,2018bsz. Its photometric properties have shown several peculiar features such as a pre-maximum plateau \citep{Anderson2018} and late time dust formation \citep{Chen2021}, and here we demonstrate that the spectroscopic and spectropolarimetric evolution are unique as well. While the SN demonstrates early similarity to SLSNe-I with \ion{O}{II} absorption and \ion{C}{II} P~Cygni lines followed by typical SESNe features (e.g. Ca, Mg, Fe, O), the multi-component H$\alpha$ emission emerging at $\sim30$~d post-maximum is highly unusual for the class. The H$\alpha$ profile is at first characterised by two components -- one at $\sim-7500$~km/s and second at $\sim3000$~km/s. The blue component is visibly fainter and narrower ($\mathrm{FWHM}\sim5000$~km/s) and can be described with a Gaussian profile, while the red is a brighter and broader ($\mathrm{FWHM}\gtrsim10000$~km/s) Lorentzian. The blue component also appears to be present from the time of the first spectrum at $-8.1$~d as a faint emission in the absorption trough of \ion{C}{II} $\lambda6580$. At $\lesssim75$~d a third component found at zero-velocity appears. The central component can be described by a Gaussian component and it is found to be the narrowest of the three ($\mathrm{FWHM}\sim3000$~km/s). At $\sim100$~d the blue component starts to fade and by $121.3$~d it is no longer detected, resulting in a skewed H$\alpha$ profile with sharp drop in the blue side and a long red tail. Similar multi-component emission profiles are also seen in other hydrogen lines including Pa$\beta$ but also in lines of \ion{Ca}{II} and \ion{He}{I}. 

As SN\,2018bsz is a Type I SLSN based on its general spectral evolution, and as similar multi-component emission lines have been seen in some Type IIn SNe (e.g. SN~1998S, PTF11iqb and SN\,2013L), the H emission likely originates in CSM with several emitting regions. Here we have demonstrated that a asymmetric, disk-like CSM structure can explain the observed spectroscopic evolution. Upon the explosion the CSM would be overtaken by the ejecta, allowing only the blue component to be seen. As the ejecta-driven photosphere recedes, the CSM re-emerges and the blue and red emission components become visible. At this phase, the photosphere is CSM-driven. Later on, the blue component starts to fade, implying that the CSM on the near-side of the explosion has completely re-emerged and recombined. Consequently the photosphere recedes rapidly causing the break in the optical light curves and the general spectral evolution towards the nebular phase. Based on the first appearance of H$\alpha$ ($-8.1$~d) we can constrain the distance of the CSM to be $<\sim6.5\times10^{15}$~cm ($430$~AU). If the pre-max plateau is not related to the SN explosion the distance is $<\sim1.3\times10^{15}$~cm ($87$~AU). Assuming the eruption velocity of SLSN-I iPTF16eh ($\sim3000$~km/s), the CSM would reach the distance in $\sim50$~d, possibly implying the plateau of $\sim40$~d is caused by a mass eruption just before the explosion.

Spectropolarimetric observations obtained during both the ejecta- and CSM-dominated phases (10.2~d and 38.4~d) confirm the viability of the proposed scenario. We observe a large shift on the Stokes $Q$ -- $U$ plane, which is independent of the ISP and argues that the SN underwent radical changes in its projected geometry in a matter of four weeks. Although the exact polarisation values depend on the ISP, which is hard to estimate, two different limiting solutions were examined. Assuming that the SN is almost unpolarised at the first epoch, results in continuum polarisation of $P \sim 1.8$\% at the second epoch, arguing for a highly aspherical CSM. This is fully consistent with the scenario of an asymmetric disk-like CSM, highly inclined with respect to our line of sight. The other limiting ISP case, where polarisation is higher ($P \sim 1.4$\%) at 10.2 days and reduces to $P \sim 0.7$\% at 38.4 days, is less favoured as it is less theoretically motivated and agrees less with the spectroscopic findings, but the truth may lie somewhere in between.

In comparison to literature Type I SLSNe, SN\,2018bsz appears to be unique. Only a few exhibit late hydrogen emission, but only in SN\,2018bsz it is multi-component. SLSNe-I PTF10aagc is the most similar to SN\,2018bsz as both of them exhibit non-zero velocity H emission as well as strong \ion{C}{II} features and comparatively low velocity \ion{O}{II} absorption in comparison to the population of Type I SLSNe. As such they might be the first examples of a subclass of SLSNe-I with aspherical CSM. More Type I SLSNe with similar multi-component emission lines need to be discovered to understand their relation to the class of SLSNe and to investigate the diversity of CSM surrounding SLSNe. 

\begin{acknowledgements}
We thank the anonymous referee for comments that helped improve this paper.

Based on (in part) observations collected at the European Organisation for Astronomical Research in the Southern Hemisphere, Chile, as part of ePESSTO (the advanced Public ESO Spectroscopic Survey for Transient Objects Survey). ePESSTO observations were obtained under ESO program ID 1103.D-0328 (PI: Smartt). 

Based on (in part) observations collected at the European Organisation for Astronomical Research in the Southern Hemisphere under ESO programme 2101.D-5023,

M.P, G.L. and P.C. are supported by a research grant (19054) from VILLUM FONDEN.

S.B. would like to thank their support from Science Foundation Ireland and the Royal Society (RS-EA/3471)

M.B. acknowledges support from the Swedish Research Council (Reg. no. 2020-03330).

E.C. acknowledges additional support from the National Agency for Research and Development (ANID) / Scholarship Program / Doctorado Nacional grant 2021 - 21211203. 

T.-W. C. acknowledges the EU Funding under Marie Sk\l{}odowska-Curie grant H2020-MSCA-IF-2018-842471.

MF is supported by a Royal Society - Science Foundation Ireland University Research Fellowship

L.G. acknowledges financial support from the Spanish Ministerio de Ciencia e Innovaci\'on (MCIN), the Agencia Estatal de Investigaci\'on (AEI) 10.13039/501100011033, and the European Social Fund (ESF) "Investing in your future" under the 2019 Ram\'on y Cajal program RYC2019-027683-I and the PID2020-115253GA-I00 HOSTFLOWS project, from Centro Superior de Investigaciones Cient\'ificas (CSIC) under the PIE project 20215AT016, and the program Unidad de Excelencia María de Maeztu CEX2020-001058-M.

MG is supported by the EU Horizon 2020 research and innovation programme under grant agreement No 101004719.

T.E.M-B. acknowledges financial support from the Spanish Ministerio de Ciencia e Innovaci\'on (MCIN), the Agencia Estatal de Investigaci\'on (AEI) 10.13039/501100011033 under the PID2020-115253GA-I00 HOSTFLOWS project, and from Centro Superior de Investigaciones Cient\'ificas (CSIC) under the PIE project 20215AT016.

MN is supported by the European Research Council (ERC) under the European Union’s Horizon 2020 research and innovation programme (grant agreement No.~948381) and by a Fellowship from the Alan Turing Institute.

F.O. acknowledges the support of the GRAWITA/PRIN-MIUR project: "The new frontier of the Multi-Messenger Astrophysics: follow-up of  electromagnetic transient counterparts of gravitational wave sources" and the support of HORIZON2020: AHEAD2020 grant agreement n.871158.

\end{acknowledgements}

\bibliographystyle{aa} 
\bibliography{bib} 

\begin{appendix} 
\onecolumn
\centering
\section{ISP corrected F, P, Q, U, $\theta$}

\begin{figure}[h]
\centering
     \begin{subfigure}[h]{0.87\textwidth}
         \centering
         \includegraphics[width=\textwidth]{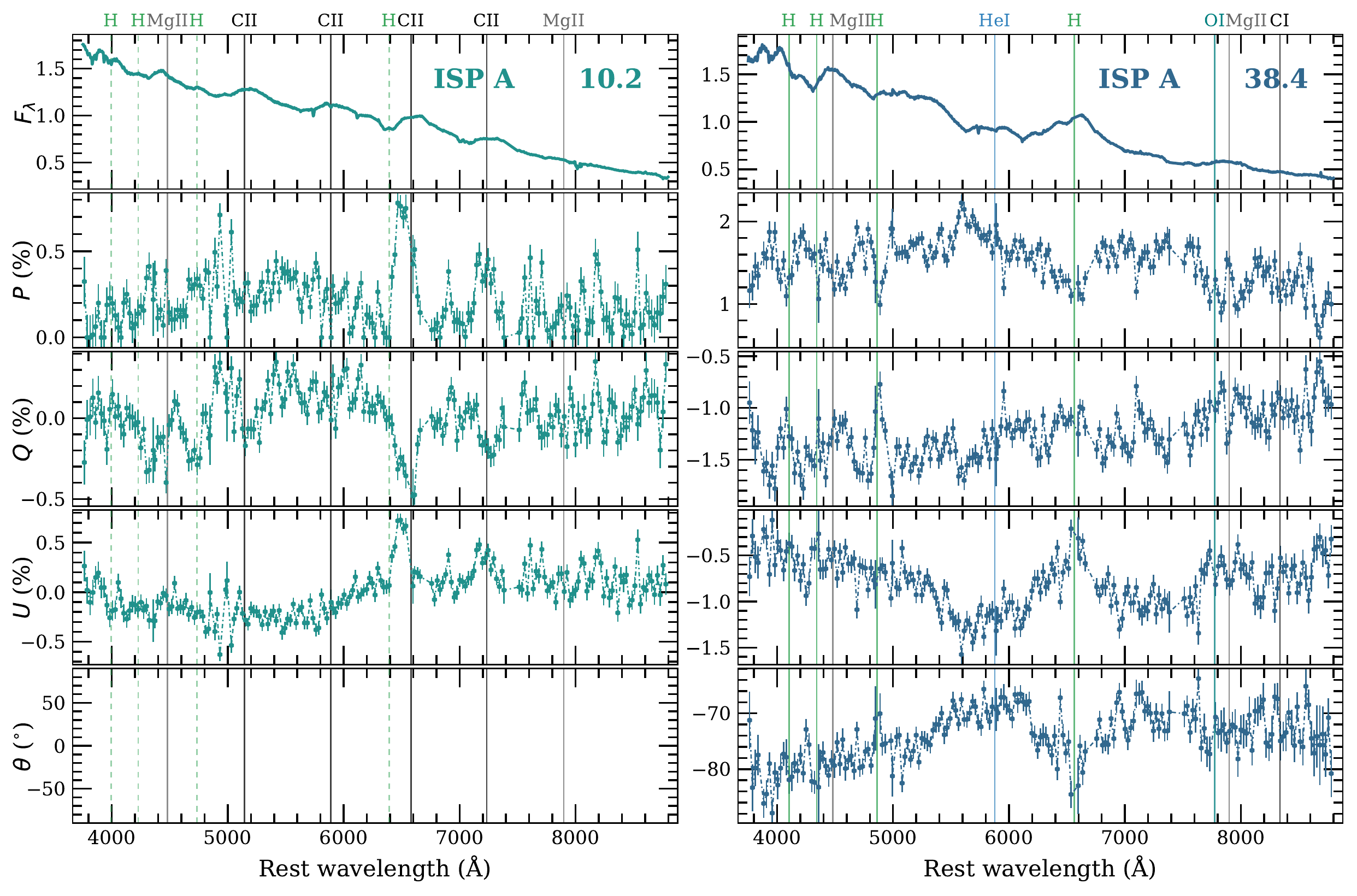}
     \end{subfigure} 
     \begin{subfigure}[h]{0.87\textwidth}
         \centering
         \includegraphics[width=\textwidth]{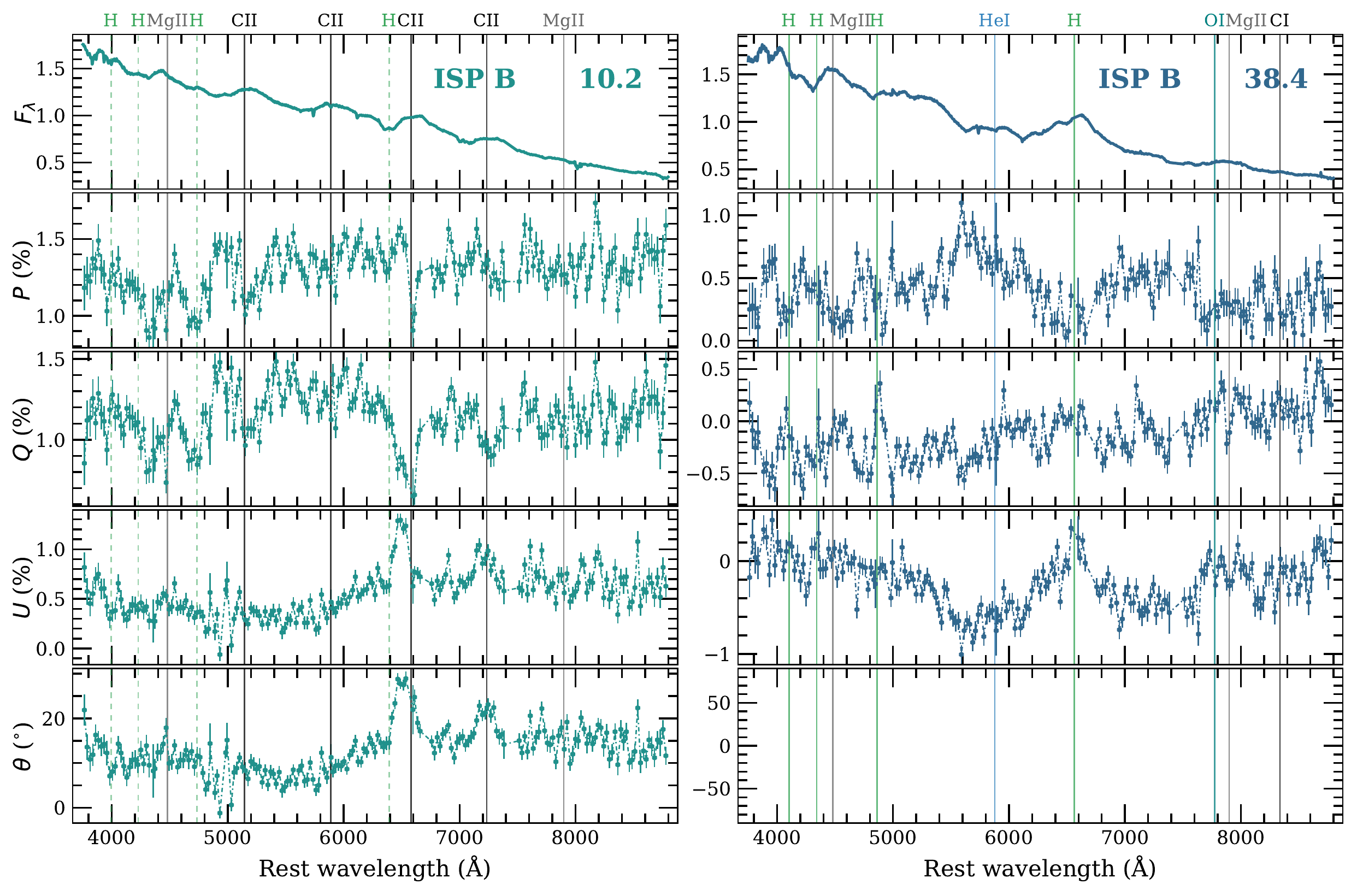}
    \end{subfigure}
    \caption{Flux spectrum, polarisation degree $P$, normalised Stokes $Q$ and $U$ parameters and polarisation angle $\theta$ of SN\,2018bsz at 10.2~d (left) and 38.4~d (right) corrected for ISP A (top) and ISP B (bottom). Polarisation angles have not been provided at 10.2~d for ISP A and at 38.4~d for ISP B as ISP solutions are found over the data points (see Fig. \ref{fig:qu_plane}) and the $\theta$ values are thus erratic.}
    \label{fig:FPQUX_ISP_corr}
\end{figure}
\end{appendix}
\end{document}